\documentclass[printer]{aa}
\pdfoutput=1

\usepackage{graphicx}
\usepackage[intlimits,sumlimits]{amsmath}
\usepackage{deluxetable}
\usepackage{times,epsfig} 
\usepackage{amssymb}
\usepackage{mathrsfs}  

\usepackage{natbib}
\bibpunct{(}{)}{;}{a}{}{,} 
\usepackage{caption}

\usepackage{amsmath}
\usepackage[english]{babel}
\usepackage{longtable}
\usepackage{afterpage}
\usepackage{url}
\usepackage{hyperref}
\usepackage{xcolor}
\definecolor{dark-red}{rgb}{0.4,0.15,0.15}
\definecolor{dark-blue}{rgb}{0.15,0.15,0.4}
\definecolor{medium-blue}{rgb}{0,0,0.5}
\hypersetup{
    colorlinks, linkcolor={dark-red},
    citecolor={dark-blue}, urlcolor={medium-blue}
}
\usepackage{txfonts}

\newcommand{\beqa}{\begin{eqnarray}} 
\newcommand{\eeqa}{\end{eqnarray}}

\newcommand{\bsub}{\begin{subequations}}
\newcommand{\esub}{\end{subequations}}
\newcommand{\beal}{\begin{align}}
\newcommand{\ealn}{\end{align}}

\newcommand{\msun}{M$_{\odot}$}
\newcommand{\rsun}{R$_{\odot}$}

\newcommand{\Msun}{{\ensuremath{\mathrm{M}_{\odot}}}}
\newcommand{\Rsun}{{\ensuremath{\mathrm{R}_{\odot}}}}

\authorrunning{C.~Fremling et al.}
\titlerunning{PTF12os and iPTF13bvn}

\begin{document}

\title{PTF12os and iPTF13bvn:}
\subtitle{Two stripped-envelope supernovae from low-mass progenitors in NGC 5806}

\author{C.~Fremling\inst{1} \and J.~Sollerman \inst{1} \and F.~Taddia \inst{1} \and M.~Ergon \inst{1} 
\and M.~Fraser \inst{2} 
\and E.~Karamehmetoglu \inst{1} 
\and S.~Valenti \inst{3} 
\and A.~Jerkstrand \inst{4} 
\and I.~Arcavi \inst{3,5} 
\and F.~Bufano \inst{6}
\and N.~Elias~Rosa \inst{7} 
\and A.~V.~Filippenko \inst{8}
\and D.~Fox \inst{9} 
\and A.~Gal-Yam \inst{5} 
\and D. A.~Howell \inst{3,10}
\and R.~Kotak \inst{4} 
\and P.~Mazzali \inst{11,12}
\and D.~Milisavljevic \inst{13} 
\and P.~E.~Nugent\inst{8,14} 
\and A.~Nyholm \inst{1} 
\and E.~Pian \inst{15,16}
\and S.~Smartt \inst{4}}

\institute{Department of Astronomy, The Oskar Klein Center, Stockholm University, AlbaNova, 10691 Stockholm, Sweden \and
Institute of Astronomy, University of Cambridge, Madingley Road, Cambridge, CB3 0HA, UK \and
Las Cumbres Observatory Global Telescope Network, 6740 Cortona Dr., Suite 102, Goleta, CA 93117, USA\and
Astrophysics Research Centre, School of Mathematics and Physics, Queen's University Belfast, Belfast BT7 1NN, UK \and
Benoziyo Center for Astrophysics, The Weizmann Institute of Science, Rehovot 76100, Israel \and
INAF-Osservatorio Astrofisico di Catania, Via Santa Sofia, 78, 95123, Catania, Italy \and
INAF-Osservatorio Astronomico di Padova, Vicolo dell'Osservatorio 5, 35122 Padova, Italy \and
Department of Astronomy, University of California, Berkeley, CA 94720-3411, USA \and
Department of Astronomy and Astrophysics, Pennsylvania State University, University Park, PA 16802, USA \and
Department of Physics, Broida Hall, University of California, Santa Barbara, CA 93106-9530, USA\and
Astrophysics Research Institute, Liverpool John Moores University, Liverpool L3 5RF, UK\and
Max-Planck-Institut f\"ur Astrophysik, Karl-Schwarzschild-Str. 1, D-85748 Garching bei M\"unchen, Germany \and
Harvard-Smithsonian Center for Astrophysics, 60 Garden St, Cambridge MA, 02138, USA\and
Lawrence Berkeley National Laboratory, 1 Cyclotron Road, MS 50B-4206, Berkeley, CA 94720, USA\and
INAF, Institute of Space Astrophysics and Cosmic Physics, via P.Gobetti 101, 40129 Bologna, Italy\and
Scuola Normale Superiore, Piazza dei Cavalieri 7, I-56126 Pisa, Italy}

\date{Received; Accepted}

\abstract
{We investigate two stripped-envelope supernovae (SNe) discovered in the nearby galaxy NGC 5806 by the (intermediate) Palomar Transient Factory [(i)PTF]. These SNe, designated PTF12os/SN~2012P and iPTF13bvn, exploded within $\sim520$ days of one another at a similar distance from the host-galaxy center.
We classify PTF12os as a Type IIb SN based on our spectral sequence; iPTF13bvn has previously been classified as Type Ib having a likely progenitor with zero age main sequence (ZAMS) mass below $\sim17$ \Msun. Because of the shared and nearby host, we are presented with a unique opportunity to compare these two SNe.}
{Our main objective is to constrain the explosion parameters of iPTF12os and iPTF13bvn, and to put constraints on the SN progenitors. We also aim to spatially map the metallicity in the host galaxy, and to investigate the presence of hydrogen in early-time spectra of both SNe.}
{We present comprehensive datasets collected on PTF12os and iPTF13bvn, and introduce a new automatic reference-subtraction photometry pipeline (FPipe) currently in use by the iPTF.  We perform a detailed study of the light curves (LCs) and spectral evolution of the SNe. The bolometric LCs are modeled using the hydrodynamical code \textsc{hyde}. 
We analyze early spectra of both SNe to investigate the presence of hydrogen; for iPTF13bvn we also investigate the regions of the Paschen lines in infrared spectra. We perform spectral line analysis of helium and iron lines to map the ejecta structure of both SNe. We use nebular models and late-time spectroscopy to constrain the ZAMS mass of the progenitors. We also perform image registration of ground-based images of PTF12os to archival \textit{HST} images of NGC~5806 to identify a potential progenitor candidate.}
{We find that our nebular spectroscopy of iPTF13bvn remains consistent with a low-mass progenitor, likely having a ZAMS mass of $\sim12$~\Msun. Our late-time spectroscopy of PTF12os is consistent with a ZAMS mass of $\sim15$~\Msun. We successfully identify a source in pre-explosion \textit{HST} images coincident with PTF12os. The colors and absolute magnitude of this object are consistent between pre-explosion and late-time \textit{HST} images, implying it is a cluster of massive stars. Our hydrodynamical modeling suggests that the progenitor of PTF12os had a compact He core with a mass of $3.25^{+0.77}_{-0.56}$~\msun\ at the time of the explosion, which had a total kinetic energy of $0.54^{+0.41}_{-0.25}\times10^{51}$ erg and synthesized $0.063^{+0.020}_{-0.011}$~\msun\ of strongly mixed \,\element[][56]{Ni}. Spectral comparisons to the Type IIb SN~2011dh indicate that the progenitor of PTF12os was surrounded by a thin hydrogen envelope with a mass lower than $0.02$~\Msun.  
We also find tentative evidence that the progenitor of iPTF13bvn could have been surrounded by a small amount of hydrogen prior to the explosion. This result is supported by possible weak signals of hydrogen in both optical and infrared spectra.} 
{}

\maketitle
\clearpage
\section{Introduction}
Prior to their final fate as core-collapse (CC) supernovae (SNe) releasing $\sim10^{51}$ erg of kinetic energy, the progenitor stars of Type IIb and Ibc SNe have had their envelopes stripped of hydrogen. These SN types are therefore commonly referred to as stripped-envelope (SE) supernovae.

In the case of SNe IIb, the stripping is partial, and early-time spectra show clear signatures of Balmer lines. At later times the spectra of SNe IIb instead closely resemble those of Type Ib SNe, which by definition do not show significant hydrogen signatures at any time \citep[e.g.,][]{1997ARA&amp;A..35..309F}. Type Ic SNe likely experience even stronger stripping, resulting in a loss of their entire (or nearly entire) \ion{He}{} envelope prior to explosion.

Two main mechanisms have been suggested to produce SE~SNe: either binary mass transfer \citep[e.g.,][]{1985ApJS...58..661I,Yoon:2010aa,2011A&amp;A...528A.131C,2012MNRAS.424.2139D,2015PASA...32...15Y} or strong line-driven winds from massive Wolf-Rayet (WR) stars \citep[e.g.,][]{conti76,Groh:2013aa,2014ARA&amp;A..52..487S}.

Binary mass transfer is an appealing mechanism for producing the partial stripping seen in SNe IIb; when the envelope of the donor star decreases in radius below the Roche limit, the mass transfer naturally stops. 
Simulations of binary evolution show that the final amount of stripping of the exploding star depends mainly on the initial masses of the two stars in the binary system, the spatial configuration of the system, and the metallicity of the stars.
Within the parameter space explored by \cite{Yoon:2010aa} and others, both SNe~IIb and Ib are produced in a bimodal fashion, but not at the observed rates (e.g., \citealp{2011MNRAS.412.1441L}; see also \citealp{2011A&amp;A...528A.131C}).

It is also possible to produce both complete and partial stripping in a single-star scenario \citep[e.g.,][]{Groh:2013aa,2014ARA&amp;A..52..487S}. In this case the final amount of hydrogen depends strongly on the initial mass and metallicity of the isolated star, with higher metallicity allowing stronger line-driven winds and more mass loss. In this picture, SNe occurring in galaxies with low metallicity could end up as partially stripped SNe~IIb, and SNe occurring in higher-metallicity environments could end up as completely stripped SNe~Ib. 

In the literature there is evidence \citep[e.g.,][]{2012ApJ...759..107K} for SNe IIb tending to reside in lower-metallicity hosts compared to SNe~Ib. 
However, to produce completely stripped SN progenitors from single stars, progenitors with high zero age main sequence (ZAMS) masses of 30~\msun\, and beyond are typically needed. Such progenitors give rise to very large ejecta masses of $\sim 10$~\msun\, as they explode. This is not consistent with the observed ejecta-mass range of 3.6--5.7~\msun\ derived for most SNe~Ibc by \cite{2015A&amp;A...574A..60T}. 
In a binary scenario, a low ejecta mass is more easily produced since less-massive stars can end up completely stripped by the mass transfer. Similar low values for ejecta masses of SNe~Ib/c were also deduced by \cite{Cano:2013aa} and \cite{2014arXiv1406.3667L}.

Observational evidence for binarity in the progenitors of SNe~IIb exists; for the well-known Type IIb SN~1993J, a binary companion was directly detected by \cite{Maund:2009}; see also \cite{2014ApJ...790...17F}. Further, the progenitor of the well-observed Type IIb SN~2011dh has been shown by \cite{2014A&amp;A...562A..17E,2015A&amp;A...580A.142E} to most likely be a composite of a nearly completely stripped compact He core with a mass of 3.3--4.0 M$_{\sun}$ surrounded by a thin H-rich envelope extending out to 200--300~\rsun. Binary evolution modeling can readily reproduce these properties at the time of explosion, while single-star models encounter difficulties in reproducing the relatively low ejecta mass in combination with the observed \ion{H}{} shell \citep{2011A&amp;A...528A.131C,2013ApJ...762...74B}. Recently, \cite{2014ApJ...793L..22F} have also claimed that the remaining flux in {\it Hubble Space Telescope} (\textit{HST}) imaging obtained $\sim1160$~d past the explosion is consistent with a suitable binary companion to SN~2011dh; however, \cite{2015MNRAS.454.2580M} disagree with this conclusion.

The Type Ib SN iPTF13bvn was initially suggested to have been the result of the explosion of a single massive WR star based on model fits to the early photometry as well as on the colors and absolute magnitude of the progenitor candidate identified in pre-explosion \textit{HST} images \citep[e.g.,][]{2013ApJ...775L...7C,Groh:2013aa}. However, hydrodynamical modeling and nebular $r$-band photometry indicate that the ZAMS mass was likely not more than $17$ \msun \,\citep{2014A&amp;A...565A.114F}. This result was confirmed by the nebular-phase observations of \cite{2015A&amp;A...579A..95K}. A single-star model with this ZAMS mass should not produce any kind of SE~SN. This indicates that a binary was also the origin for this system. The observed colors in the pre-explosion \textit{HST} images of iPTF13bvn have since also been reassessed and reproduced by evolutionary modeling of a binary system \citep{2014AJ....148...68B,2015MNRAS.446.2689E}. Furthermore, the observables of iPTF13bvn are consistent with other SNe Ib in the literature \citep{2015A&amp;A...574A..60T}, which indicate that most SNe Ib have progenitors of similar nature. Independent photometry and spectroscopy of iPTF13bvn was also collected by \cite{2014MNRAS.445.1932S}, showing consistent data and similar conclusions. 

In this paper we present the first comprehensive dataset on PTF12os (SN~2012P), a SN~IIb that occurred in the same nearby galaxy, NGC 5806, as the already well-studied Type Ib SN iPTF13bvn. We also supplement previously published data on iPTF13bvn with more photometry and spectra, and perform a comprehensive comparison of these two SNe. In addition, we compare their host environments via spectroscopic metallicity measurements of nearby H$\mathrm{\alpha}$ regions. Our analysis is aided by comparisons with SN~2011dh.  

This paper is organized as follows.
In Sect.~\ref{sec:observations} we discuss our observations and reduction procedures. A description of the reference-subtraction code that has been used to reduce most of our imaging is presented in detail. Section~\ref{sec:host} gives long-slit metallicity measurements of H$\mathrm{\alpha}$ regions in NGC 5806 to determine the metallicity at the positions of iPTF13bvn and PTF12os and the metallicity gradient of the host galaxy.  Extinction estimates for the two SNe are given in Sect.~\ref{sec:extinction}, and in Sect.~\ref{sec:prog_id} we discuss the result of astrometric identification of a progenitor candidate for PTF12os using \textit{HST} pre-explosion images of NGC 5806. 
In Sect.~\ref{sec:lc} we present the multiband LCs of the SNe and construct pseudobolometric LCs of PTF12os and iPTF13bvn, which are then compared with SN~2011dh to get a first handle on the properties of the SN progenitors and explosions. In Sect.~\ref{sec:spectra} we report our spectroscopy of the two SNe; we investigate the presence of early-time H in PTF12os, search for similar indications in the early spectra of iPTF13bvn, and perform spectral-line velocity measurements of the He and Fe lines to estimate the photospheric expansion velocities. The latter are used in Sect.~\ref{sec:hydro} together with the hydrodynamical model {\sc hyde} \citep{2015A&amp;A...580A.142E} to constrain the explosion parameters, such as the synthesized Ni and ejecta masses as well as the He-core mass of the progenitor of PTF12os. To allow direct comparison, we have recalculated the explosion parameters for  iPTF13bvn and SN~2011dh using the same code. 
In Sect.~\ref{sec:spectra} we also use late-time spectroscopy ($>200$~d past the explosions) of both iPTF13bvn and PTF12os to constrain the amount of oxygen in the ejecta of each SN by comparing our data to the nebular models of \cite{jerkstrand2014}. 
Finally, Sect. \ref{sec:conclusions} contains a summary of this work, some discussion, and our conclusions.

\label{sec:intro}

\section{Observations and Data Reduction}
\label{sec:observations}
\subsection{Discovery and imaging}
The discoveries\footnote{All dates in this paper are given in Universal Time (UT).} of PTF12os (SN~2012P\footnote{SN~2012P was also independently discovered by amateur astronomers \citep{2012CBET.2993....1D}.}) \citep[2012 Jan. 10.48;][]{2012ATel.3881....1A} and iPTF13bvn \citep[2013 June 16.24;][]{2013ATel.5152....1A,2013ATel.5140....1A} in NGC~5806 were made with the Palomar Oschin 48-inch (P48) Schmidt telescope \citep{Law:2009aa}. PTF12os was discovered 4.0~d past the explosion (+4~d\footnote{Throughout this paper, we adopt the convention that a plus sign followed by a time implies a time past the estimated explosion date.}), $t_{\rm exp}=$~2012 Jan. 6.50$^{+0.5}_{-1.3}$, as estimated from the best fit to our hydrodynamical model grid (Sect.~\ref{sec:hydro}). The explosion date of iPTF13bvn is very well constrained ($t_{\rm exp}=$~2013 June 15.67; \citealp{2013ApJ...775L...7C}) based on a power-law fit to the earliest P48 data points, which sets the discovery at +0.6~d.

\cite{2013ApJ...775L...7C} reported early-time photometry on iPTF13bvn obtained with the P48, the robotic Palomar 60-inch telescope \citep[P60;][]{2006PASP..118.1396C}, and the Las Cumbres Observatory Global Telescope network \citep[LCOGT;][]{2013PASP..125.1031B} up to 20~d past the discovery. Later follow-up data obtained with the same telescopes, along with the addition of nebular-phase photometry up to $\sim 240$~d past discovery from the Nordic Optical Telescope (NOT), were reported by \cite{2014A&amp;A...565A.114F}.

In this paper we provide final reductions in table format (Tables~\ref{tab:13bvnphot} and \ref{tab:13bvnphot2}) of the previously reported photometry from the P60, NOT, and LCOGT of iPTF13bvn \citep{2013ApJ...775L...7C,2014A&amp;A...565A.114F}, along with previously unpublished late-time photometry obtained with the P60, NOT, and the Palomar 200-inch telescope (P200) through $\sim 350$~d past discovery.

We also present P48, P60, NOT, Liverpool Telescope (LT), Gran Telescopio Canarias (GTC), and New Technology Telescope (NTT) multiband observations of PTF12os through $\sim 210$~d past discovery (Table~\ref{tab:12osphot}) along with ultraviolet (UV) photometry obtained with {\it Swift} \citep{2004SPIE.5165..262R} for both PTF12os (Table~\ref{tab:12osswift}) and iPTF13bvn (Table~\ref{tab:13bvnswift}). 

\subsection{Imaging reductions and photometry}
The reductions of our multicolor LCs of PTF12os and iPTF13bvn from data obtained with the P48, P60, P200, NOT, LT, and GTC are based on our in-house reference image-subtraction pipeline currently performing real-time automatic reductions of iPTF P60 data. This pipeline, while operating in automatic mode, uses mosaiced Sloan Digital Sky Survey \citep[SDSS;][]{2014ApJS..211...17A} images for the subtraction of the host-galaxy contribution from each science frame taken with the P60. This allows us to quickly obtain host-subtracted magnitudes and colors of newly discovered transients, a valuable aid when deciding on follow-up strategy for a particular transient. The code can in principle operate on any imaging dataset as long as suitable data for producing a reference are available. In this paper it is used to obtain publication-quality host-subtracted magnitudes for both iPTF13bvn and PTF12os from all of our imaging data. 

To reduce P48 data, we utilize deeply stacked P48 references generated from images obtained during the years before the explosion of PTF12os. For data from the P60, LT, GTC, NOT, and the P200, deep stacks of P60 images are used as the references. For iPTF13bvn, the P60 dataset obtained on PTF12os is used to generate the reference images, and vice versa. Since the two SNe happened more than 500~d apart and they both peaked at $\sim 16$ mag in the $r$ band, any possible remaining flux from the SNe in the reference frames is negligible. 

The top layer of the pipeline is written in {\sc MATLAB}\footnote{We also use the astronomy \& astrophysics package for {\sc MATLAB} \citep{2014ascl.soft07005O}.}, operating on images that are bias-subtracted and flat-fielded. The basic steps performed by this pipeline are described in sequential order below. When relevant, we also give some details pertaining to the quality of the reductions of our data on PTF12os and iPTF13bvn presented in this paper.

\subsubsection{The Fremling Automated Pipeline, {\sc FPipe}}
\begin{itemize}

\item
\textbf{Quality control of the science data and bad-pixel masking:}
Before any further processing and image subtraction, the pipeline checks for the presence of a World Coordinate System (WCS) solution and its reliability. If no WCS is present, an astrometric solution is attempted. The approximate seeing and the number of stars present in each frame are also measured using {\sc SExtractor} \citep{1996A&amp;AS..117..393B}. Any frames for which it is not possible to determine a robust WCS are discarded. 
If there is a known map of bad pixels for the CCD in use (e.g., the P60), these are also fixed by linear interpolation in this step.

\item
\textbf{Sky subtraction:}
Following the quality control, the sky background is removed from the images by masking out all detected sources (point-like or otherwise, using {\sc SExtractor} with a low detection threshold), fitting a second-order polynomial surface with a first-order cross term to the remaining data in two iterations with 3$\sigma$ clipping, and then subtracting.

\item
\textbf{Reference image generation:}
If a reference image is not manually specified, or if SDSS references are not used, a stacked reference is generated from suitable reference data as follows. Initially, images with seeing worse than $1.8''$ and images where there are a smaller number of detected stars than one third of the average in the dataset are filtered out. The remaining frames are sky subtracted, registered, matched in intensity, and combined using the average of all images with 3$\sigma$ clipping for each pixel. The sky subtractions are performed as described above and the registrations are performed as described below, using the image with the best seeing as the reference onto which the rest of the images are stacked. 

\item 
\textbf{Image registration:}
To register the science images to the reference image, the centroids of common point sources in the reference and science frames are first identified using {\sc SExtractor}. The geometric transformation is subsequently determined with different complexity depending on the number of common point sources. If there are fewer than 7 common point sources, the transformation allows for shifting, rotation, and scaling of the science image. If there are 7 to 15 common sources we also allow for shearing. With more than 15 common sources, which is typically the case, a second-order polynomial transformation with 12 parameters is determined, and the polynomial transformation is finally applied using Lanczos resampling. When operating on P60 references and science frames, the standard deviation in the distance between the centroids in the reference and a registered science frame is typically below 0.05 pixels. Frames with fewer than 3 point sources in common with the reference are discarded.

\item
\textbf{Point-spread-function (PSF) modeling:}
To obtain a model for the PSF in the reference frame and in each science frame, we use {\sc SExtractor} in combination with {\sc PSFex} \citep{2013ascl.soft01001B}. The current version of the pipeline uses a nonparametric PSF model, as measured from the raw data by cleaning and stacking the isolated point sources. We assume that the PSF is constant across the CCD\footnote{For P48 data this assumption is not true. To handle this, we cut out a small portion around the transient within which the PSF can be assumed to be constant for these reductions.}, an assumption that we have found to work very well for datasets solely obtained by the P60 or data from the P60 in combination with SDSS references, the main use of this pipeline. We find no evidence for a spatial dependence on the subtraction residuals in our subtracted frames.

\item
\textbf{PSF matching:}
Once the PSFs of the reference and each science frame have been determined, we convolve the reference image with the PSF of the science frame, as well as the science frame with the PSF of the reference, for each individual subtraction. This method, also known as the common PSF method (CPM), has been previously proposed by \cite{2008ApJ...680..550G}.

\hspace{0.4cm}We find that this method is well suited for subtractions where the reference image is not necessarily obtained with the same telescope as the science observations, since there are no parameters to be tuned except the size of the box region used to measure the PSF. However, both the reference and the science frames are somewhat degraded owing to convolution being performed on both images. Finally, the PSF models themselves are also convolved with each other to obtain the final PSF in each subtraction for later use when performing PSF photometry.

\item
\textbf{PSF photometry:}
To determine the counts of a point source, we perform weighted least-squares fitting of the convolved nonparametric PSF model to the data. The fit is weighted by the photon noise in the images, so that areas with increased signal from the source receive higher weights in the fit. Before the fit is performed, the PSF model is centered on the point source using Discrete Fourier Transform cross-correlation.

\item
\textbf{Zero-point determinations and intensity-scale matching of the reference and science data:}
Whenever there is SDSS coverage, or other local standard stars with known magnitudes are available for the field, we perform PSF photometry as described above on the locations of these stars in both the science and reference frames for each subtraction. Sources for which the quality of the fit to the PSF model is below a certain threshold are excluded, and the remaining sources are matched to their known magnitudes to determine the zero points (ZPs) of the frames\footnote{In practice, the result is that only sources that are detected in the science frame are used when determining the ZP in both the reference and the science frame.}. We do not apply any color terms. The ZP information is used to scale the counts of the convolved reference frame to the convolved science frame, resulting in a common ZP as determined from the science frame and data that are ready to be subtracted.

\hspace{0.4cm}For our Sloan-filter data of iPTF13bvn and PTF12os, we use SDSS magnitudes to determine the ZPs. Typically at least 20 useable SDSS stars are present in each science frame, resulting in very precise ZP determinations, with standard deviations of the mean for the ZP typically $< 0.01$ mag\footnote{For the data from the LT that we present in this paper, the ZP is in some cases difficult to determine in the science frames because of their low signal-to-noise ratio (S/N), resulting in ZP uncertainties up to 0.05 mag for some epochs.}.

\hspace{0.4cm}To set the ZPs in our $B$-band images we use $g$- and $r$-band magnitudes of SDSS stars within the fields with the magnitude conversions described by \cite{2005AJ....130..873J}. 

\hspace{0.4cm}For our P48 $R$-band data, we use SDSS $r$-band magnitudes to determine the ZPs. The P48 uses a Mould $R$-band filter, which is not identical to the SDSS $r$ band. However, we find that this method gives LCs that are consistent within approximately $\pm0.05$ mag compared to our P60 $r$-band LCs that have been reduced and calibrated in the same way. Thus, we can make the assumption that the ZPs we find in this way converts our Mould $R$-band data into Sloan $r$-band data, at least to a first approximation. Since this accuracy is sufficiently good for the science performed in this paper, we have not applied any color-corrections when determining the ZPs or corrections for the filter differences and spectral shapes (S-corrections) when determining the magnitudes of the transients in any of the imaging filters used, unless otherwise stated\footnote{We caution that for other fields, color-corrections could be needed when determining the ZPs. For other SNe, S-corrections could also be more important than what we find for the SNe in this paper. Convolving the normalized transmission curves of the filters mounted on the P60 along with standard Sloan filter-profiles on the spectra of PTF12os and iPTF13bvn typically result in S-corrections smaller than than 0.1 mag. We also find that our data obtained from other telescopes are consistent with the P60 LCs at this level or better, when we have overlap between the LCs.}.

\item
\textbf{PSF photometry of the transient, error determination, and detection limits:}
After subtracting the scaled reference image from the science frame, we finally perform a PSF model fit at the expected location of the transient. If the quality of the fit is above the detection threshold, the result is used to determine the magnitude of the transient based on the flux and the ZP of the science frame. To determine the uncertainty of the measured flux we insert artificial transients with the measured flux, scattered in a circular pattern around one PSF size away from the real transient, in the unsubtracted science frame. The subtraction is redone for each artificial transient and we measure the uncertainty as the standard deviation in the flux of 35 artificial sources.
If the quality of the initial fit is below the detection threshold, a limiting magnitude is determined by inserting artificial sources of increasing magnitude until fewer than $66.7\%$ of the inserted sources are detected. The initial threshold has been tuned so that the results from this procedure represent $3\sigma$ detection limits, by comparison to results from aperture photometry.

\item
\textbf{Final uncertainty in the measured magnitude:}
The final error in the determined magnitude of the transient is taken as the statistical uncertainty determined from the artificial sources added in quadrature to the standard deviation of the mean when doing the ZP determination from the detected point sources with known magnitudes in the science frame. Typically, for iPTF13bvn and PTF12os, the standard deviation of the mean of the ZP in our data is below 0.01 mag, as mentioned above; consequently, the final error is generally dominated by the statistical uncertainty when measuring the flux of the transients.
\end{itemize}

\subsubsection{LCOGT photometry}
For the images of iPTF13bvn obtained by LCOGT, we estimate the galaxy contribution by fitting and subtracting a low-order surface, and then performing PSF-fitting photometry (Valenti et al., in prep.). The Sloan filter data were calibrated against a minimum of 10 SDSS \citep{2014ApJS..211...17A} stars in the field. The Johnson-Cousins {\it UBVRI} filter data were calibrated against Landolt standard stars \citep{1992AJ....104..340L} observed during photometric nights. We find that this procedure gives consistent LCs in the bands where we also have reference-subtracted P60 data ($g, r, i$). Thus, the host-galaxy contribution at the location of iPTF13bvn appears to be negligible after the galaxy is subtracted in this way.

\subsubsection{Reduction of {\it Swift} UVOT photometry}
Both PTF12os and iPTF13bvn were observed with the UV Optical Telescope onboard {\it Swift} \citep[UVOT;][]{2004ApJ...611.1005G,2005SSRv..120...95R}. PTF12os was observed in 6 epochs from 2012 Jan. 14 to Jan. 29 (+8.7~d to +23.5~d), and iPTF13bvn in 10 epochs from 2013 June 17 to 2013 July 23 (+2.0~d to +37.8~d). Photometry was obtained as described by \cite{2009AJ....137.4517B}, and an aperture with radius 5$''$ was used. 

To estimate the host-galaxy contribution at the location of the SNe we use the deepest {\it Swift} observation obtained when observing PTF12os as the reference frame for iPTF13bvn, and vice versa. The flux at the position of the transients was measured with the same aperture size in the reference frames as in the science frames and the flux measured in the reference was subtracted from each detection to obtain host-subtracted magnitudes. In this paper we only include data in the $U$, $UVW1$, and $UVM2$ filters, even though more bands were observed. We report our host-subtracted {\it Swift}-UVOT photometry as AB magnitudes in Table~\ref{tab:12osswift} and Table~\ref{tab:13bvnswift}, and it is also shown in Fig.~\ref{fig:lc_full}. For one epoch of observations of iPTF13bvn, there were no significant detections in these filters after the host contribution was subtracted.

\subsection{Spectroscopic observations and reductions}
We obtained 19 epochs of low-resolution optical spectra of PTF12os, starting before peak luminosity at $4$~d past discovery up until $211$~d past discovery, when the SN has entered the nebular phase. These are listed in Table~\ref{tab:spec12os} along with the telescopes and instruments that were used. The full spectral sequence is shown in Fig.~\ref{fig:spec12os}.

Optical and near-infrared (NIR) spectra of iPTF13bvn were obtained starting within 24~hours after discovery. Spectra until $16$~d past discovery were published by \cite{2013ApJ...775L...7C}, whereas \cite{2014A&amp;A...565A.114F} presented 6 additional later-time optical spectra obtained between $18$ and $86$~d after discovery. In this paper we provide our complete dataset, consisting of 26 epochs of optical spectra and 4 NIR spectra, including the previously published data. The earliest spectrum of iPTF13bvn \citep{2013ATel.5142....1M}, obtained with the SALT telescope, is also included in our analysis. We have obtained several new spectra in the nebular phase, one using the NOT and the Andalucia Faint Object Spectrograph (ALFOSC) at $250$~d past discovery, one with the Deep Extragalactic Imaging Multi-Object Spectrograph \citep[DEIMOS;][]{2003SPIE.4841.1657F} on Keck~2 at $344$~d past discovery, one with the ESO Very Large Telescope using the FORS2 spectrograph at $346$~d past discovery and one with the Intermediate dispersion Spectrograph and Imaging System (ISIS) at the William Herschel Telescope (WHT) at $376$~d past discovery. We also present an additional NIR spectrum that was not published by \cite{2013ApJ...775L...7C}, obtained with the Folded-port InfraRed Echellette spectrograph \citep[FIRE;][]{2013PASP..125..270S} on Magellan-Baade at $78$~d past discovery. 
Our spectral data on iPTF13bvn are listed in Table~\ref{tab:spec13bvn}. The optical spectra are shown in Fig.~\ref{fig:spec13bvn} and the NIR spectra are in Fig.~\ref{fig:spec13bvnir}.

All spectra were reduced using standard pipelines and procedures for each telescope and instrument. For our nebular spectrum of PTF12os obtained $211$~d past discovery, we use spectra of the underlying \ion{H}{II} region from several years after the explosion to subtract the background continuum.
All spectral data and corresponding information is available via WISeREP\footnote{\href{http://www.weizmann.ac.il/astrophysics/wiserep/}{http://www.weizmann.ac.il/astrophysics/wiserep/}} \citep{Yaron:2012aa}.

\begin{figure*}
\centering
\includegraphics[width=17cm]{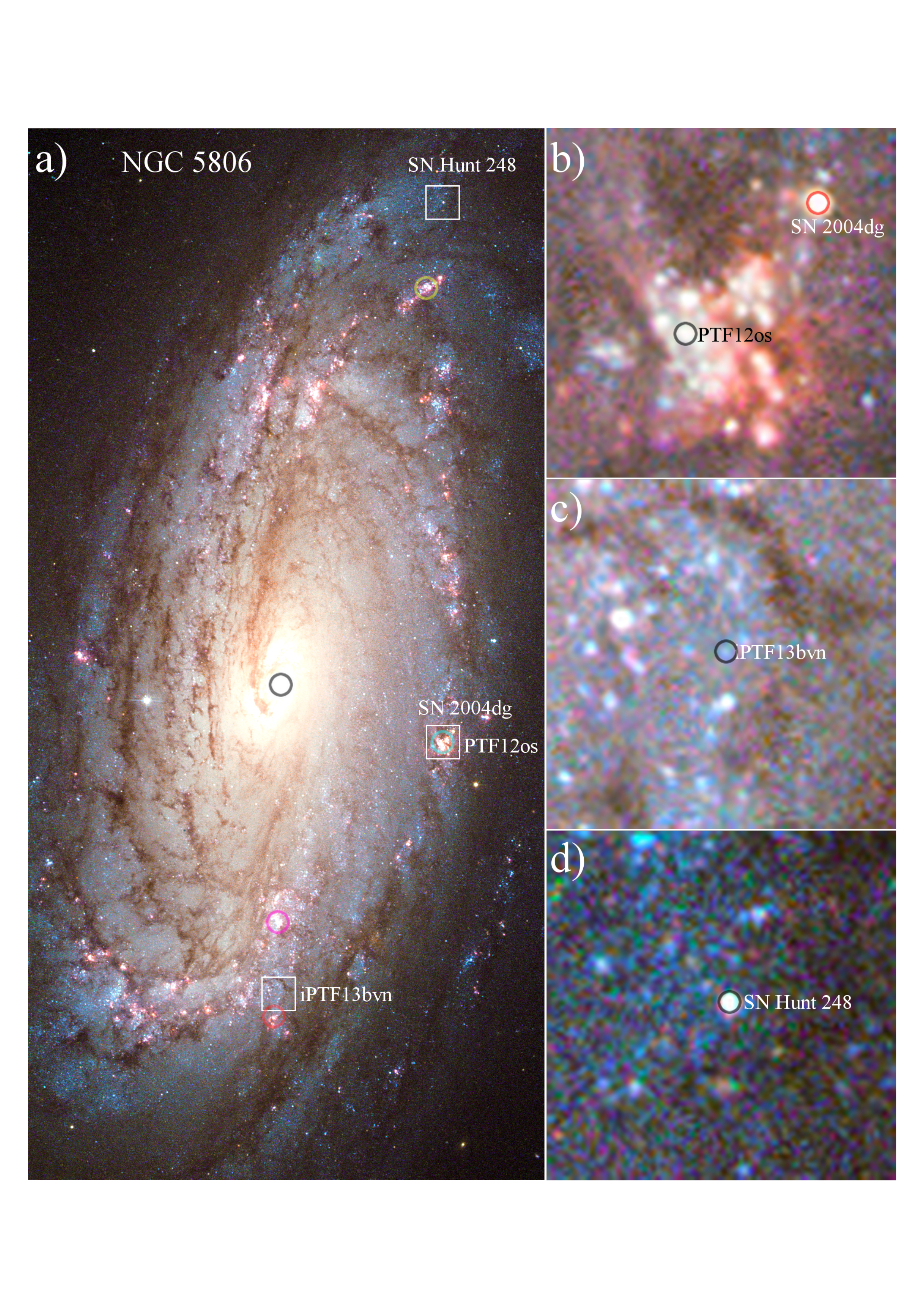}  
\vspace{-2cm}
 \caption{NGC~5806 imaged by \textit{HST} using WFC filters F435W, F555W, and F814W during 2004 when SN 2004dg occurred in the galaxy. Regions with strong H$\alpha$ emission (F658N) are shown in red (Credits: ESA/NASA/Andre van der Hoeven). Panel (a) shows the locations of PTF12os, PTF13bvn, SN 2004dg, and SN~Hunt~248 marked by white boxes. Bright \ion{H}{II} regions of which we have obtained spectra are indicated by colored circles. Panel (b) displays a close-up view of the location of the PTF12os progenitor candidate (black circle). SN 2004dg, which exploded close to the position of PTF12os, is indicated by a red circle. Panel (c) shows a close-up view of the location of iPTF13bvn, with the progenitor candidate marked by a black circle. Panel (d) highlights the location of SN~Hunt~248, which is marked by a black circle. North is up and east is to the left in all panels.}
 \label{fig:galaxy}
\end{figure*}

\section{Host-galaxy properties}
\label{sec:host}
NGC~5806 is a nearby spiral galaxy having SAB(s)b morphology. The spectroscopic redshift of the galaxy from the SDSS is $z=0.00449$. We adopt the distance modulus $\mu=32.14\pm0.20$~mag \citep{2013AJ....146...86T}, corresponding to a distance of $26.8_{-2.4}^{+2.6}$~Mpc\footnote{When the uncertainties are taken into account, this value is consistent with the kinematic distance of $23.9\pm1.7$~Mpc.}\textsuperscript{,}\footnote{An older and more uncertain distance modulus of $\mu=31.76\pm0.38$~mag \citep{Tully:2009} has previously been used extensively in the literature for work on iPTF13bvn. However, we point out that the updated distance modulus estimate of $\mu=32.14\pm0.2$~mag by \cite{2013AJ....146...86T} is very close to the median value and standard deviation ($\mu=32.09\pm0.2$~mag) of all the distance modulus estimates reported on NED.}. We note that since PTF12os and iPTF13bvn occurred in the same galaxy, we do not have to worry about systematic uncertainties in the relative distances of the two SNe. 
The above distance results in a $B$-band absolute magnitude for NGC~5806 of $M_B=-20.12$~mag\footnote{Total $B$-band magnitude from NED, not corrected for extinction.}. Adopting the values from NED, the major and minor diameters of NGC~5806 are 185\farcs40 and 94\farcs554, respectively, and the position angle is 170$^{\circ}$. The morphological T-type is 3.0 according to the Third Reference Catalogue of Bright Galaxies \citep[RC3;][]{1991rc3..book.....D}.

A stack of \textit{HST}/WFC images taken in filters F658N, F435W, F555W, and F814W during 2004 is shown in Fig.~\ref{fig:galaxy}, with the locations of the known transients in the galaxy marked. In addition to iPTF13bvn and PTF12os which we investigate in detail here, the Type IIP SN 2004dg \citep{Smartt:2009} and the SN impostor SN~Hunt~248 \citep{2015MNRAS.447.1922M,2015A&amp;A...581L...4K} (discovered in 2014) also occurred in this galaxy. SN~2004dg is present in the image stack shown in Fig.~\ref{fig:galaxy}.

 \begin{figure}
\centering
\includegraphics[width=9cm]{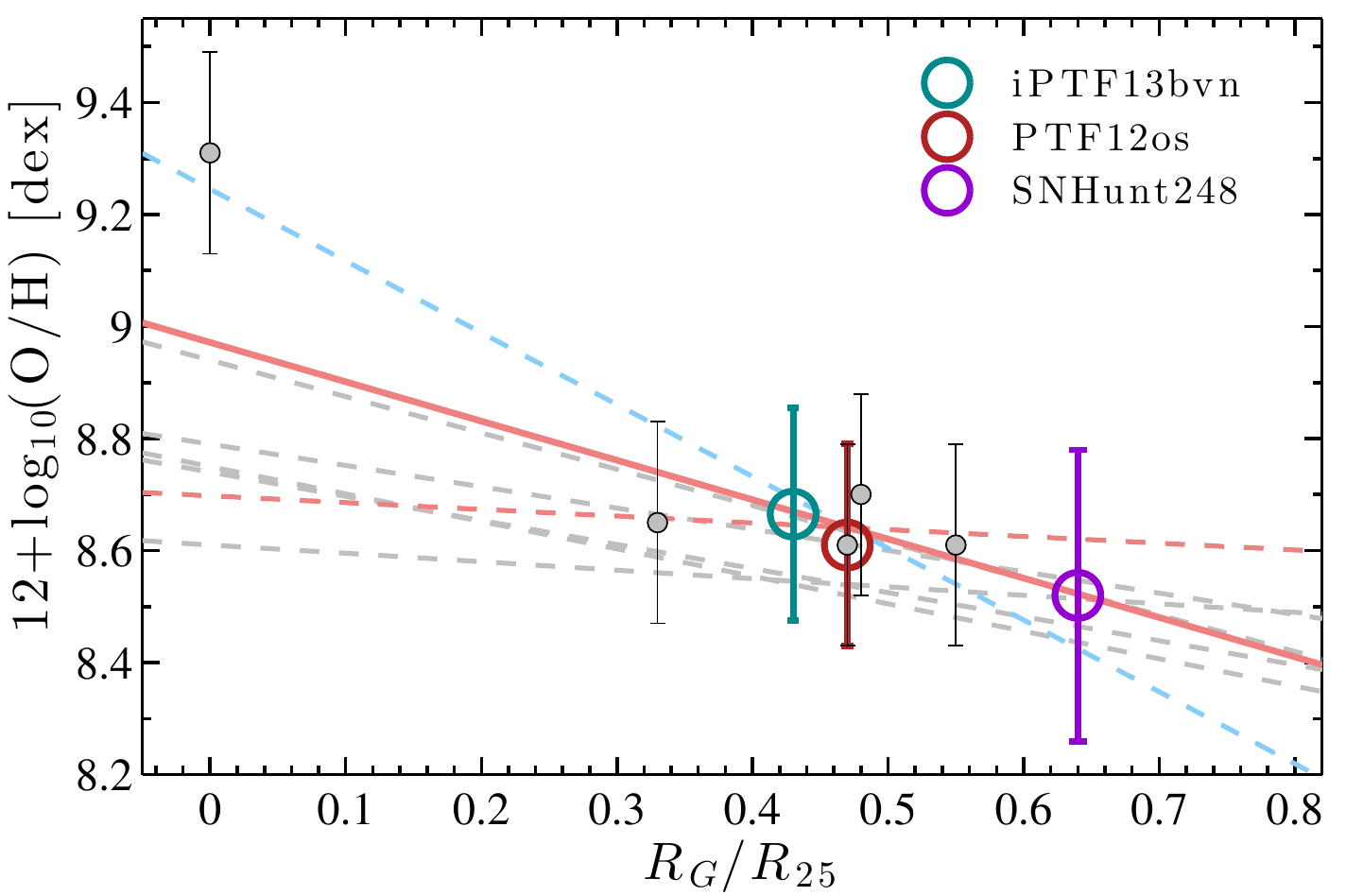}
\caption{Metallicity measurements (solid circles), metallicity estimates (open circles), and gradient estimate (solid red line) for NGC~5806 compared to the spiral-galaxy sample (gray dashed lines) by \cite{Gusev:2012aa}. The dashed blue line shows a first-order polynomial fit which includes the central oxygen metallicity measured for NGC~5806 using an SDSS spectrum, and the dashed red line shows the same fit with the metallicity datapoint at the center of the galaxy excluded.}
\label{fig:metgrad}
\end{figure}

\subsection{Metallicity estimates}
\label{sec:metallicity}
We have mapped the metallicity of NGC~5806 via a spectroscopic programme conducted at the NOT\footnote{Proposal ID 48-408, PI C. Fremling, conducted in service mode.} within which we performed spectroscopy of \ion{H}{II} regions in the galaxy using ALFOSC in long-slit mode. This dataset was also supplemented by two WHT spectra (see Fig.~\ref{fig:galaxy}).

Our metallicity measurements are based on strong-line diagnostics using the N2 method \citep[][]{2004MNRAS.348L..59P}, which utilizes the ratio of the 
[\ion{N}{ii}]~$\lambda$6584 and H$\alpha$ lines to estimate the abundance of oxygen in the line-emitting region. We estimate deprojected galactocentric radii ($r_G/r_{25}$) by following the procedures described by \cite{2013A&amp;A...558A.143T}.

The location of PTF12os is coincident with a strong \ion{H}{II} region, indicated by a blue circle in panel (a) of Fig.~\ref{fig:galaxy} (see also Sect.~\ref{sec:prog_id} for details on the progenitor identification). At this position we obtained the N2 metallicity estimate $12+\mathrm{log_{10}(O/H)}~=~8.61\pm0.18$~dex using a NOT long-slit spectrum, and we measure the deprojected galactocentric radius $r_G/r_{25}=0.47$. The same metallicity value was obtained at the position of the closest strong \ion{H}{II} region to SN~Hunt~248. This \ion{H}{II} region is marked by a yellow circle in panel (a) of Fig.~\ref{fig:galaxy}, and it is located at $r_G/r_{25}=0.55$. SN~Hunt~248 itself is situated to the north of this \ion{H}{II} region at $r_G/r_{25}=0.64$.

For the location of iPTF13bvn (see Fig.~\ref{fig:galaxy} and Sect.~\ref{sec:prog_id}), we have estimated the deprojected galactocentric radius $r_G/r_{25}=0.43$, similar to that of PTF12os. With WHT/ISIS we obtained spectra of two bright \ion{H}{II} regions close to this location, located at $r_G/r_{25}=0.48$  (red circle in Fig.~\ref{fig:galaxy}) and $r_G/r_{25}=0.33$ (pink circle in Fig.~\ref{fig:galaxy}). For these regions we estimated a N2 metallicity of $12+\mathrm{log_{10}(O/H)}=8.70\pm0.18$~dex and $12+\mathrm{log_{10}(O/H)}=8.65\pm0.18$~dex, respectively.

To estimate the metallicity gradient in NGC~5806 we use the equation
\begin{equation}
12+\mathrm{log_{10}(O/H)}= 12+\mathrm{log_{10}(O/H)}_0 + C_{\mathrm{O/H}}\times(r_G/r_{25}),
\end{equation}
where $12+\mathrm{log_{10}(O/H)}_0$ is the central oxygen abundance and $C_{\mathrm{O/H}}$ is the gradient (in dex~$R_{25}^{-1}$). 
This equation fitted to the above data, including the central metallicity which we measure as $12+\mathrm{log_{10}(O/H)}=9.31\pm0.18$ using an SDSS spectrum taken at the center of the galaxy, gives $12+\mathrm{log_{10}(O/H)}= 9.25-1.28\,(r_G/r_{25})$~dex. This is a steeper gradient than the average for spirals (e.g., \citealp{2004A&amp;A...425..849P}; \citealp{Gusev:2012aa}). However, we note that in the above fit we are using the central oxygen abundance, without accounting for the possibility of AGN activity at the center of the galaxy. We see no clear evidence of an AGN in the spectrum, but a weak effect on the lines used in the abundance calculation cannot be excluded. If only \ion{H}{II} regions in the disk are used, and the central abundance is extrapolated, we find the gradient in the disk of NGC~5806 to be $C_{\mathrm{O/H}}=-0.12$~dex~$R_{25}^{-1}$ and an extrapolated central oxygen abundance of $12+\mathrm{log_{10}(O/H)_0}=8.70$~dex. The average of the two different gradients is
\begin{equation}\label{eq:grad}
12+\mathrm{log_{10}(O/H)_{NGC~5806}}= 8.97-0.70\,(r_G/r_{25}),
\end{equation}
and this is what we use as the best estimate for the metallicity gradient of NGC~5806. We show our measurements and gradients, along with the gradients of the sample of spiral galaxies studied by \cite{Gusev:2012aa}, in Fig.\ref{fig:metgrad}. The average gradient we find for NGC~5806 is in good agreement with this sample.

Given the metallicity gradient of NGC~5806 (Eq.~\ref{eq:grad}), the distance of iPTF13bvn from the center of the galaxy implies a metallicity of $12+\mathrm{log_{10}(O/H)}=8.67\pm0.19$~dex at the position of the SN. At the position of SN~Hunt~248 we find $12+\mathrm{log_{10}(O/H)}=8.52\pm0.26$~dex. To estimate the uncertainties we have accounted for the systematic error of 0.18~dex from the N2 method, as well as the difference between the two gradient fits and the average gradient at the position of each object. 

In conclusion, it appears that both PTF12os and iPTF13bvn occurred in regions of very similar metallicity, where the oxygen abundance is close to the solar value ($12+\mathrm{log_{10}(O/H)}= 8.7$~dex; \citealp{2009ARA&amp;A..47..481A}). The environment of SN~2011dh was also found to be roughly of solar metallicity by \cite{2011ApJ...741L..28V}.
However, a possible caveat here is that we do not have a direct measurement exactly at the position of iPTF13bvn, since the SN did not occur in an \ion{H}{II} region which would allow the metallicity to be estimated using the N2 method.

\section{Extinction}
\label{sec:extinction}
Throughout this paper all reddening corrections are applied using the \cite{1989ApJ...345..245C} extinction law with $R_V=3.1$. For the Milky Way (MW) color excess we adopt $E(B-V)_\mathrm{MW}=0.0437$~mag toward NGC~5806\footnote{We note that in \cite{2013ApJ...775L...7C} the foreground (MW) extinction toward NGC~5806 was mixed up with the local extinction found at the position of iPTF13bvn. This mixup was also propagated to \cite{2014A&amp;A...565A.114F}.} \citep{2011ApJ...737..103S}.

To estimate the host-galaxy color excess of PTF12os and iPTF13bvn we perform optical color comparisons to SN~2011dh\footnote{We use the LC data of SN~2011dh from \cite{2014A&amp;A...562A..17E,2015A&amp;A...580A.142E}.}, after correcting for the MW contributions. The multiband light curves (Sect.~\ref{sec:lc}) were interpolated to evenly spaced dates in order to give one value for each night of observations, and then we used an iterative procedure that minimizes the following expression:
\begin{equation}
\Delta_{\mathrm{C}}=\sum^{n=4}_{n=1}{\sum^{t=30~\mathrm{d}}_{t=10~\mathrm{d}}{[(C(n,E,t)_{\mathrm{SN}}}} - (C(n,t)_{\mathrm{dh}})]^2,
\end{equation}
where the sum over $n$ is performed for $C(n)=B-g,g-r,r-i,$ and $i-z$ colors. $C(n,E,t)_{SN}$ is also a function of $E(B-V)$ and time; it is the extinction-corrected color in a chosen set of filters as a function of time for the object for which the extinction is being computed. Similarly, $C(n,t)_{\mathrm{dh}}$ is the corresponding extinction-corrected color for SN~2011dh.

If we assume that the extinction of SN~2011dh is known, we can now compute the host-galaxy color excess we need for our objects to get the best match to the colors of SN~2011dh. Here we adopt a total extinction $E(B-V)=0.07_{-0.04}^{+0.07}$~mag for SN~2011dh \citep{2014A&amp;A...562A..17E}. 
Effectively, this procedure minimizes the total difference in the $B-g$, $g-r$, $r-i$, and $i-z$ colors measured between 10~d and 30~d after explosion for a SN compared to SN~2011dh. These specific colors were chosen since we have the best-quality data in terms of cadence and uncertainties in these bands for both SNe simultaneously. For iPTF13bvn we find $E(B-V)_\mathrm{host}=0.08^{+0.07}_{-0.04}$~mag, and for PTF12os we find $E(B-V)_\mathrm{host}=0.29^{+0.08}_{-0.05}$~mag. The main assumption here is that these SNe have the same intrinsic colors (temperature) as SN~2011dh.

To estimate the total uncertainty intervals of these extinction values we have added in quadrature the statistical error resulting from the method described above to the error in the extinction of SN~2011dh. The statistical errors were derived via Monte-Carlo simulations by randomly applying $1\sigma$ errors on all of the photometric data points used in the calculation and iterating the calculation a large number of times. 
The final errors are dominated by the uncertainty in the extinction estimate for SN~2011dh.
Assuming another value for the extinction of SN~2011dh directly reflects in 
the calculated extinction values as a simple 1:1 relation.

We also checked the host-galaxy color excess of PTF12os by measuring the \ion{Na}{I}~D equivalent width (EW) in the spectrum taken on 2012 Jan. 25. Using the relation suggested by \cite{2003fthp.conf..200T},
\begin{equation}\label{eq:NaID}
E_{(B-V)}=-0.01+0.16\times\mathrm{EW(\ion{Na}{I} D)},
\end{equation}
we find $E(B-V)_\mathrm{host}=0.265$~mag. This method is not very reliable since it is based on low-resolution spectra, and the \ion{Na}{I}~D feature is not resolved \citep[see, e.g.,][]{2011MNRAS.415L..81P,2012MNRAS.426.1465P}, but the presence of significant \ion{Na}{I}~D absorption confirms that the host extinction is not negligible. We note that two different slopes for the \ion{Na}{I} D to $E_{(B-V)}$ relation were found by \cite{2003fthp.conf..200T}. We have adopted the shallower slope in Eq.~\ref{eq:NaID}. Using the relation derived by \cite{2011MNRAS.415L..81P} we find $E(B-V)_\mathrm{host}=0.66$~mag, which is consistent with what would be found by adopting the steeper slope found by \cite{2003fthp.conf..200T}. However, the scatter is very high in these relations (e.g. 0.3 mag in \citealp{2011MNRAS.415L..81P}), thus we consider the results from these relations as roughly consistent with each other as well as with our color-based method.

For iPTF13bvn, a host-galaxy color excess $E(B-V)_\mathrm{host}=0.0278$~mag was previously derived from the \ion{Na}{I}~D absorption from high-resolution spectroscopy \citep{Cao:2013aa}. A larger value of $E(B-V)_\mathrm{host}=0.17\pm0.03$~mag was later suggested by \cite{2014AJ....148...68B} based on a preliminary $B-V$ color comparison to the Carnegie Supernova Project sample. \cite{2014AJ....148...68B} estimate $E(B-V)_\mathrm{host}=0.07$--0.22~mag from another high-resolution spectrum. We note that the result from our color comparison to SN~2011dh is roughly consistent with all of these values when the uncertainties are taken into account.

\begin{figure}
\centering
\includegraphics[width=9cm]{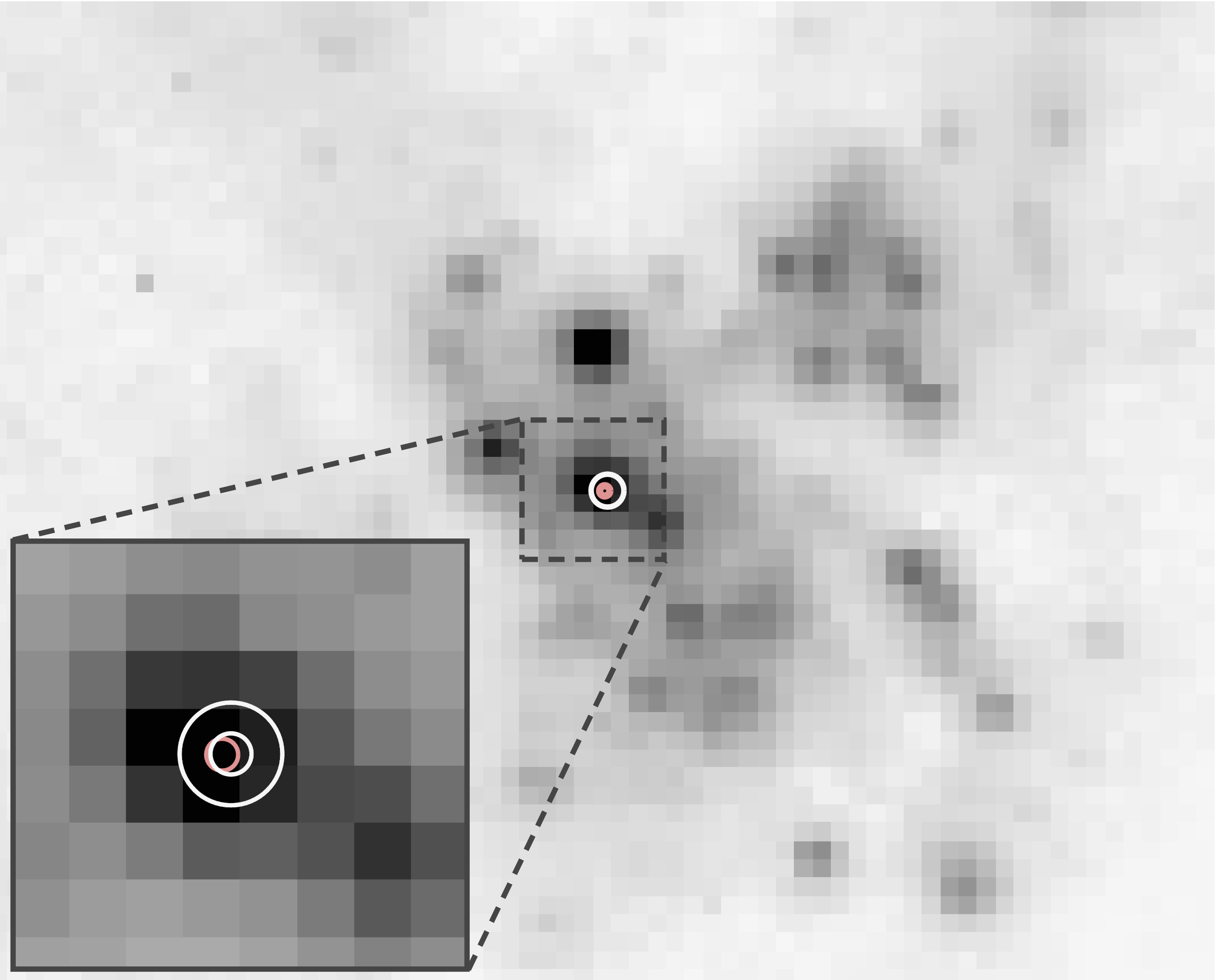}
\caption{Progenitor identification of PTF12os found by registering 8 ground-based NTT/EFOSC $i$-band images of PTF12os to a stacked pre-explosion \textit{HST} ACS F814W image. The field of view shown here is $4\farcs0\times2\farcs8$. North is up and east is to the left. The white circles show the average position of the centroid of PTF12os after image registration with a radius reflecting the statistical error from 7 image registrations. The smaller circle represents a 2$\sigma$ uncertainty and the larger circle represents a 5$\sigma$ uncertainty. The 2$\sigma$ uncertainty is only shown in the zoomed inset. The pink circles indicate the centroid of the possible progenitor measured in the \textit{HST} image, with a radius representing the 2$\sigma$ uncertainty in its centroid.}
\label{fig:prog_12os}
\end{figure}

\section{Progenitor identifications and photometry}
\label{sec:prog_id}

NGC~5806 has been comprehensively imaged with \textit{HST}, both before and after the explosions of PTF12os and iPTF13bvn. We give a summary of the currently publicly available archival \textit{HST} images covering the region of PTF12os in Table~\ref{tab:hst}.

Moreover, images of iPTF13bvn were obtained in 2013 when this SN was present\footnote{GO-12888, PI S. Van Dyk, F555W. These images do not cover the explosion site of PTF12os.}, and they were used by multiple authors to confirm the progenitor candidate identification that was previously proposed from ground-based imaging (see Sect.~\ref{sec:prog_id}). NGC~5806 was also reobserved at two epochs\footnote{2015 June 26 (see Table.~\ref{tab:hst}) and 2015 June 30, GO-13822, PI G. Folatelli, F225W and F814W.} in 2015 in order to search for the binary companion of iPTF13bvn.

In this paper we use these observations to perform accurate astrometric registrations to locate a progenitor candidate of PTF12os and to constrain the mass of this candidate via its photometry (Sect.~\ref{sec:prog_12os}). For iPTF13bvn we do not perform any new analysis, but provide a summary of previous work (Sect.~\ref{sec:prog_13bvn}).

\begin{figure*}
\begin{center}
\includegraphics[width=18cm]{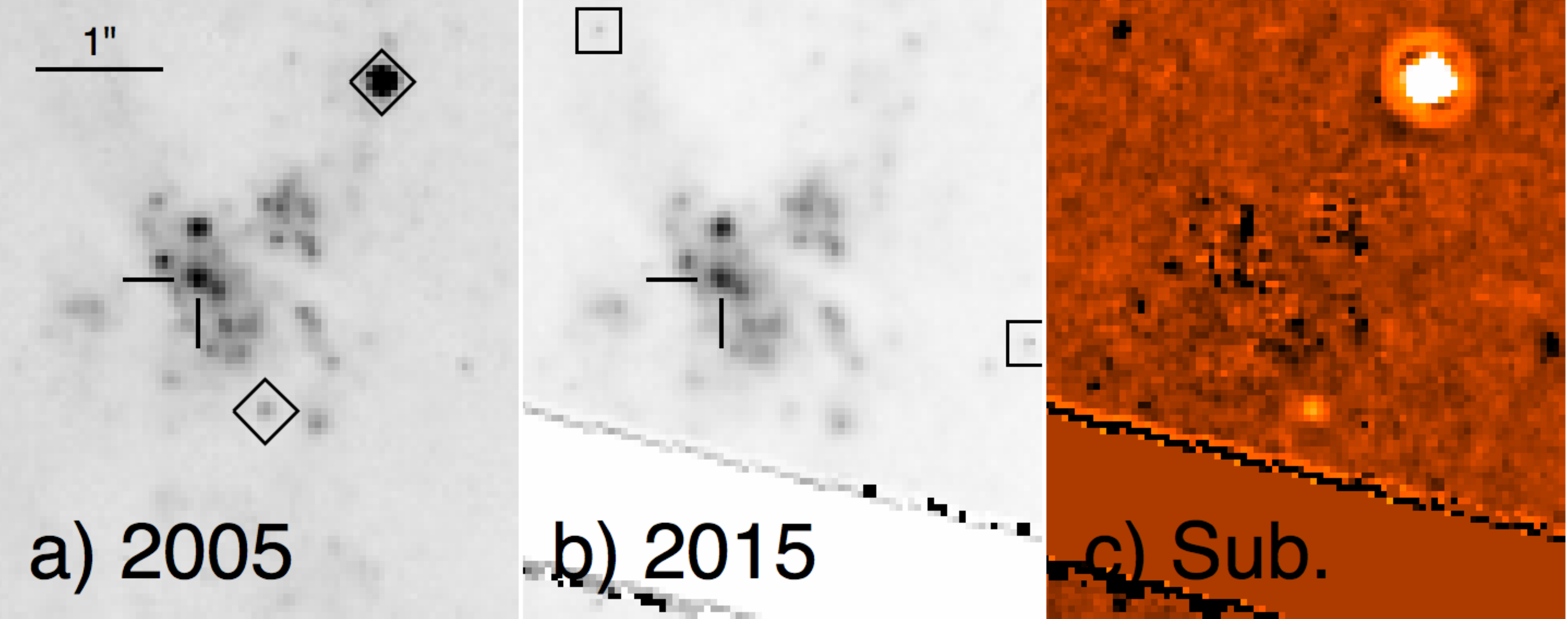}
\caption{Difference imaging using \textit{HST} data at the position of PTF12os. Panel (a) shows the 2005 ACS F555W image, panel (b) the 2015 WFC3 F555W image transformed to the same pixel grid as the 2005 image, and panel (c) the difference image. Sources brighter in 2005 are indicated with diamonds in panel (a) and appear brighter in the difference image. Sources brighter in 2015 are marked with squares in panel (b) and appear darker in the difference image. North is up and east is left, and the scale is indicated in panel (a). The diagonal band at the bottom of the 2015 image is the gap between the two WFC3 chips.}
\label{fig:F555_sub}
\end{center}
\end{figure*}

\subsection{PTF12os}
\label{sec:prog_12os}
A progenitor candidate for PTF12os was suggested by \cite{2012ATel.3884....1V}. To independently determine the location of the progenitor in NGC 5806, we obtained 11 $i$-band images of PTF12os on 2012 Feb. 16 (+42~d) with the NTT at the La Silla Observatory in Chile. Using the 8 frames with the best seeing (0\farcs8--1\farcs0), we perform astrometric registrations to a stacked (filters $F814W$, $F555W$, and $F435W$) archival \textit{HST} (ACS/WFC) 
image from 2005 \citep{Smartt:2009} using 11 common point sources.

For the registrations we use a procedure based on standard {\sc MATLAB} functions following the same principle as  described by \cite{2014A&amp;A...565A.114F}; the centroids of common point sources are derived from fits of two-dimensional (2D) Gaussians that allow for rotation, and the geometric transformation is determined as a second-order polynomial transformation with 6 free parameters and applied using bicubic interpolation.

After we performed the 8 registrations we applied 3$\sigma$ clipping to the values found for the position of the progenitor. This resulted in one calculation being excluded, leaving us with 7 values for the progenitor position. As the final position we then used the mean of these values and as the uncertainty we used the standard deviation in the mean, giving us a final uncertainty of 0.179 ACS/WFC pixels or 8.9~mas in right ascension ($\alpha$) and 0.176 pixels or 8.8~mas in declination ($\delta$).

We find that it is possible to constrain the progenitor location to the central source of a bright \ion{H}{II} region (see Fig.~\ref{fig:prog_12os} and Fig.~\ref{fig:galaxy}) located at $\alpha=~14\degr59\arcmin59\farcs082$, $\delta=~+01\degr53\arcmin23\farcs67$ (J2000.0) in the \textit{HST} image. The centroid of this source, again determined by fitting a 2D Gaussian with rotation in the \textit{HST} image, is offset by only 0.15 pixels or 7.7~mas from the progenitor location that we obtained from our registrations. Furthermore, the 2$\sigma$ uncertainty in the progenitor position that we find almost completely overlap with the 2$\sigma$ uncertainty in the centroid of the \textit{HST} source (Fig.~\ref{fig:prog_12os}). Thus, we conclude that they are coincident. This is the same source that was first reported as a potential progenitor candidate by \cite{2012ATel.3884....1V}.

\textit{HST} photometry of this progenitor candidate is shown in Table~\ref{tab:hst}. The ACS/WFC photometry based on the images obtained in 2005 transformed to $BVI$ filters via a blackbody (BB) fit to the \textit{HST} filters and by performing synthetic photometry using standard $BVI$ filter profiles are $B=23.30\pm0.01$~mag, $V=22.94\pm0.01$~mag, and $I=22.47\pm0.01$~mag. 
After corrections for the distance modulus of NGC 5806 and the total extinction we get absolute magnitudes $M_B=-10.18\pm0.20$~mag, $M_V=-10.24\pm0.20$~mag, and $M_I=-10.29\pm0.20$~mag, as well as colors $B-V=0.06^{+0.06}_{-0.09}$~mag and $V-I=0.05^{+0.07}_{-0.11}$~mag. In the error bars for the colors we include the uncertainty in the extinction estimate.

PSF-fitting photometry was performed on the WFPC2 data using the {\sc hstphot} package \citep{2000PASP..112.1383D}, and on the ACS and WFC3 images using {\sc dolphot}, a modified version of {\sc hstphot}. In all cases, images were masked using the data-quality files, the sky background measured, and library PSFs fitted to sources detected in each frame. The magnitudes were calibrated to the {\it HST} Vegamag system using standard ZPs, and in the case of WFPC2 and WFC3 the data were corrected for charge-transfer losses (the ACS images have already been corrected for this effect at the pixel level).

We find that the source coincident with PTF12os is consistently brighter in the WFPC2 images. This is likely caused by the lower resolution of these data leading to blending with nearby sources. We therefore do not include the WFPC2 data in our further analysis. The two epochs of ACS imaging both include F814W data, and  we find a comparable magnitude for the progenitor in these two epochs. The magnitudes we measure are consistent with those reported by \cite{2012ATel.3884....1V}.

We also find consistent magnitudes between the ACS F435W image taken on 2005 March 10 and the WFC3 F438W image taken after the SN had faded on 2015 June 26. The fact that the source is still present at late times, coupled with the relatively bright absolute magnitudes, clearly suggest that this is a cluster rather than a single stellar progenitor of PTF12os.

The F555W late-time magnitude measured in the post-explosion WFC3 image is $\sim0.4$ mag fainter than what was measured in 2005. While it is appealing to interpret this as a deficit of flux owing to the disappearance of a progenitor, it is difficult to reconcile this with the lack of change in F435W/F438W. To further investigate this, we performed difference imaging using the 2005 and 2015 F555W images. The latter was registered to the former, before the two images were convolved to match their PSFs and scaled using the {\sc hotpants} package\footnote{http://www.astro.washington.edu/users/becker/v2.0/hotpants.html}. The result of this subtraction is shown in Fig. \ref{fig:F555_sub}. No source can be seen at the position of PTF12os, suggesting no major change in flux between the two epochs. The flux at the position of PTF12os in this pre-explosion image cannot have been dominated by a single massive SN progenitor. In the precursor study by \cite{2015ApJ...811..117S}, it was also found that the source located at the position of PTF12os is still at approximately the same brightness 3 years after the SN occurred.

Under the assumption that the source coincident with PTF12os is a cluster, we fitted the pre-explosion photometry using the {\sc chorizos} package \citep{2004PASP..116..859M}. Starburst99 models \citep{1999ApJS..123....3L} were fitted to the measured \textit{HST} ACS F435W, F555W, and F814W magnitudes from 2005, along with the narrow-band F658N magnitude from 2004. The metallicity was set to be solar, while the extinction law was fixed to $R_V=3.1$. The extinction was allowed to vary within the range estimated in Sect.~\ref{sec:extinction}. The best fit is achieved for a cluster age $\sim 5.5$~Myr. 

This would imply that the progenitor of PTF12os was relatively massive. Using the STARS models, this corresponds to a ZAMS mass of $\gtrsim25$~\msun\ \citep{2009MNRAS.400.1019E}, since single stars less massive than this should not yet have exploded. This value is significantly higher than the $<15$~\msun\ estimate of the progenitor mass from our nebular spectroscopy found in Sect.~\ref{sec:nebular}. However, we are cautious about attributing too much weight to the cluster age derived here given the limited (4-band) photometric data. For comparison, the host cluster age and progenitor mass derived for SN 2004dj by \cite{2009ApJ...695..619V} relied on measurements in $\sim 20$ different bandpasses. It is also quite possible that the cluster contains multiple stellar populations of differing ages, with the SN progenitor being part of an older population and with the flux in the optical being dominated by a younger population.

\begin{deluxetable}{lllcc}
\tabletypesize{\scriptsize}
\tablewidth{0pc}
\tablecaption{\textit{HST} imaging of NGC~5806.\,\,\,\,\,\,\,\,\,\,\,\,\,\,\,\,\,\,\,\,\,\,\,\,\,\,\,\,\,\,\,\,\,\,\,\,\,\,\,\,\,\,\,\,\,\,\,\,\,\,\,\,\,\,\,\,\,\,\,\,\,\,\,\,\,\,\,\,\,\,\,\,\,\,\,\,\,\,\,\,\,\,\,\,\,\,\,\,\,\,\,\,\,\,\,\,\,\,\,\,\,\,\,\,\,\,\,\,\,\,\,\,\,\,\,\,\,\,\,\,\,\,\,\,\,\,\,\,\,\,\,\,\,\,\,\,\,\,\,\,\,\,\,\,\,\,\,\,\,\,\,\,\,\,\,\,\,\,\,\,\,\,\label{tab:hst}}
\tablehead{
\colhead{UT Date} &
\colhead{Instrument}&
\colhead{Filter}&
\colhead{Exposure [s]}&
\colhead{Mag ($1\sigma$)}}
\startdata
2001-07-05\tablenotemark{a} & WFPC2/WFC & F450W & 2 $\times$ 230 & 22.204 (0.003)  \\
-           & -                 & F814W & 2 $\times$ 230 & 21.385 (0.003)  \\
2004-04-03\tablenotemark{b} & ACS/WFC   & F658N    & 2 $\times$ 350 & 21.728 (0.053) \\
-          & -            & F814W  & 1 $\times$ 120  & 22.317 (0.033)  \\
2005-03-10\tablenotemark{c} & ACS/WFC   & F435W & 2 $\times$ 800 &23.319 (0.012) \\
-          & -            &  F555W & 2 $\times$ 700 & 23.061 (0.011) \\
-          & -            & F814W & 2 $\times$ 850 & 22.436 (0.009) \\
2015-06-26\tablenotemark{d} & WFC3/UVIS & F438W & 2 $\times$ 1430 &  23.441 (0.014)  \\
-        & -            & F555W & 2 $\times$ 1430 & 23.481 (0.010)    \\
\enddata
\vspace{-0.3cm}
\tablenotetext{a, b, c, d}{SNAP-9042, PI S. Smartt; SNAP-9788, PI L. Ho; GO-10187, PI S. Smartt; GO-13684, PI S. Van Dyk.}
\end{deluxetable}

\begin{figure*}[ht]
\centering
\includegraphics[width=18cm]{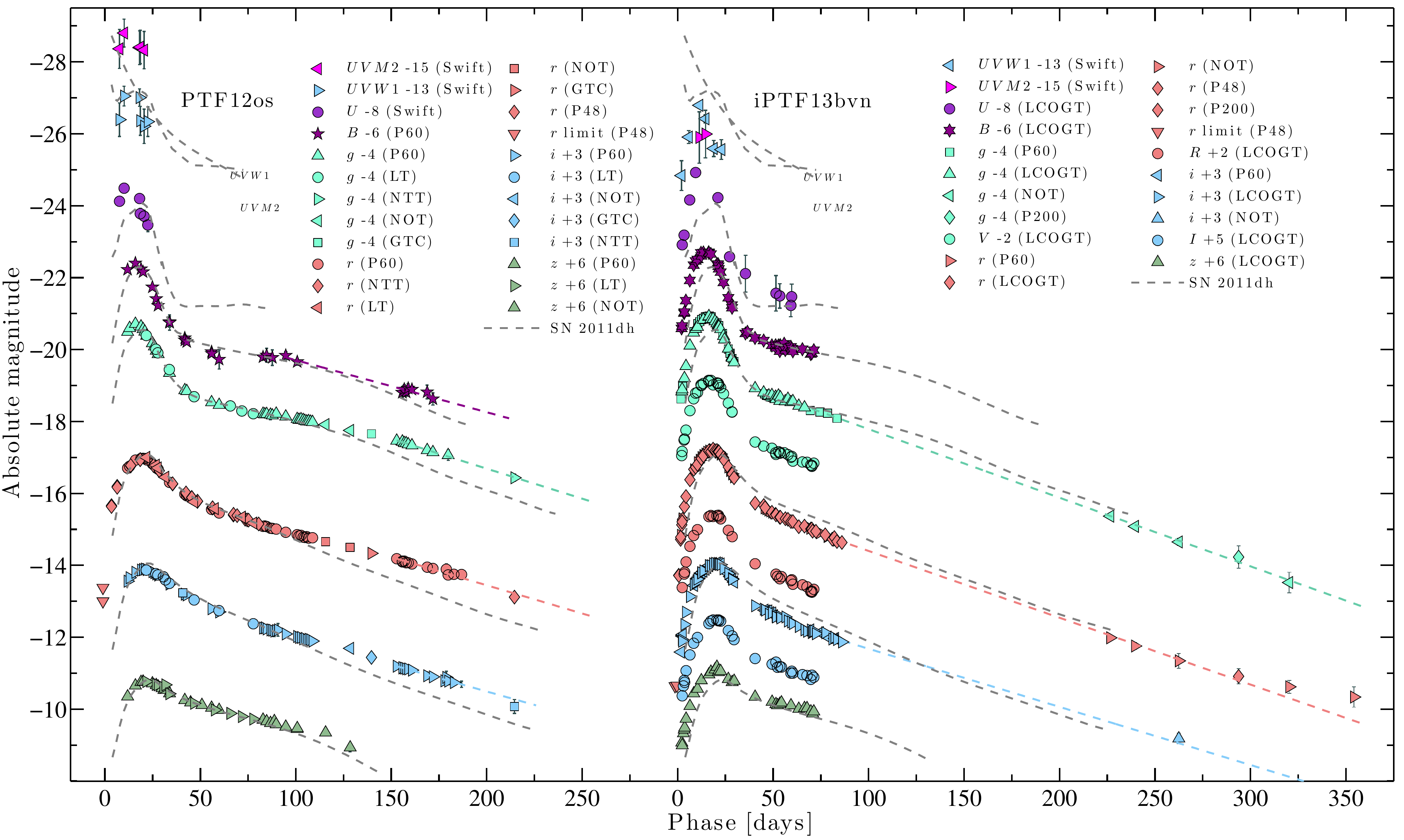}
\caption{Multicolor LCs of PTF12os (left) and iPTF13bvn (right) along with SN 2011dh (dashed black lines). The dashed colored lines represent first-order linear fits to the LC tails to estimate the late-time decline rates of the SNe (see Sect.~\ref{sec:lc}). Corrections for extinction and distance have been applied according to Sect.~\ref{sec:host} and Sect.~\ref{sec:extinction}. An S-correction based on a standard Johnson $U$-band filter has been applied to the {\it Swift} $U$-band data.}
\label{fig:lc_full}
\end{figure*}

\subsection{iPTF13bvn}
\label{sec:prog_13bvn}
A probable progenitor candidate for iPTF13bvn, located at $\alpha=~15\degr00\arcmin00\farcs147$, $\delta=+01\degr52\arcmin53\farcs19$ (J2000.0), was first identified in pre-explosion \textit{HST} images by \cite{Cao:2013aa}. In \cite{2014A&amp;A...565A.114F} we used post-explosion \textit{HST} (WFC3) images\footnote{Obtained on 2013 Sep. 2.37, GO-12888, PI S. Van Dyk.} of iPTF13bvn to confirm this result, also illustrated in Fig.~\ref{fig:galaxy} here. With these data it was possible to rule out other sources to a $5\sigma$ level. This result was yet again confirmed by \cite{2015MNRAS.446.2689E}, who reanalyzed the same pre- and post-explosion \textit{HST} images.

In \cite{Cao:2013aa} the apparent magnitudes for this source were first reported. However, a reanalysis by \cite{2015MNRAS.446.2689E} resulted in somewhat different values; $F435W=25.80\pm0.12$~mag, $F555W=25.80\pm0.11$~mag, and $F814W=25.88\pm0.24$~mag.

After correcting for the reddening and distance modulus, we find the absolute magnitudes\footnote{Due to our choice of $\mu=32.14\pm0.2$~mag, these absolute magnitudes are higher compared to previous studies. However, the upper limits considered by e.g. \cite{2015MNRAS.446.2689E} are consistent within our uncertainties. In \cite{2016arXiv160405050E} the higher distance modulus is also fully taken into account}. $M_{F435W}=-6.86\pm0.23$~mag, $M_{F555W}=-6.74\pm0.23$~mag, and $M_{F814W}=-6.47\pm0.31$~mag. By fitting a BB to these magnitudes and performing synthetic photometry using standard $BVI$ filter profiles, we find
$M_B=-6.83\pm0.23$ mag, $M_V=-6.78\pm0.23$ mag, and $M_I=-6.73\pm0.31$ mag, as well as color estimates 
$B-V=-0.05^{+0.17}_{-0.16}$~mag and 
$V-I = -0.05^{+0.27}_{-0.26}$~mag, 
where the error bars on the colors include the uncertainty in the extinction estimate. 

These colors and absolute magnitudes could be consistent with a single WR star \citep{Eldridge:2013aa,Groh:2013aa2}. However, previous studies \citep{2014A&amp;A...565A.114F,2014AJ....148...68B,2015MNRAS.446.2689E,2015A&amp;A...579A..95K} conclude that a binary system is more likely for the progenitor of iPTF13bvn, based on the low ejecta and oxygen masses. In the binary model for the iPTF13bvn system (20~\msun\ primary and 19~\msun\ secondary initial masses in a very close binary) by \cite{2014AJ....148...68B}, the source in the pre-explosion \textit{HST} images was predicted to be dominated by light from the primary, which implies that the system should be significantly fainter after the SN has faded. Further, based on the binary model grid used by \cite{2015MNRAS.446.2689E}, the \textit{HST} photometry was shown to possibly be consistent with a wide range of initial separations and masses (10--20~\msun\ for the SN progenitor). However, in the recent study by \cite{2016arXiv160405050E}, based on the \textit{HST} observations obtained on 2015 June 26 (+740~d), it is shown that the object at the position of iPTF13bvn has dimmed, and is now fainter than in the pre-explosion images, confirming that the progenitor identification was correct. A similar conclusion was also made by \cite{2016arXiv160406821F}. While \cite{2016arXiv160405050E} concluded that the observations were likely still dominated by light from the SN at the time of these observations, the mass range of the progenitor could be narrowed down significantly to 10--12~\msun. A range that is consistent with our nebular models of iPTF13bvn, discussed in Sect.~\ref{sec:nebular}.

\section{Light curves}\label{sec:lc}
The observed light curves (LCs) in Johnson/Cousins/Bessel $UBVRI$, SDSS $griz$ and {\it Swift} $UVW1$ and $UVM2$ filters for PTF12os and iPTF13bvn along with SN 2011dh are shown in Fig.~\ref{fig:lc_full}. All of the data in Fig.~\ref{fig:lc_full} have been corrected for distance and reddening according to Sect.~\ref{sec:host} and Sect.~\ref{sec:extinction}.

\begin{figure}
\centering
\includegraphics[width=8.5cm]{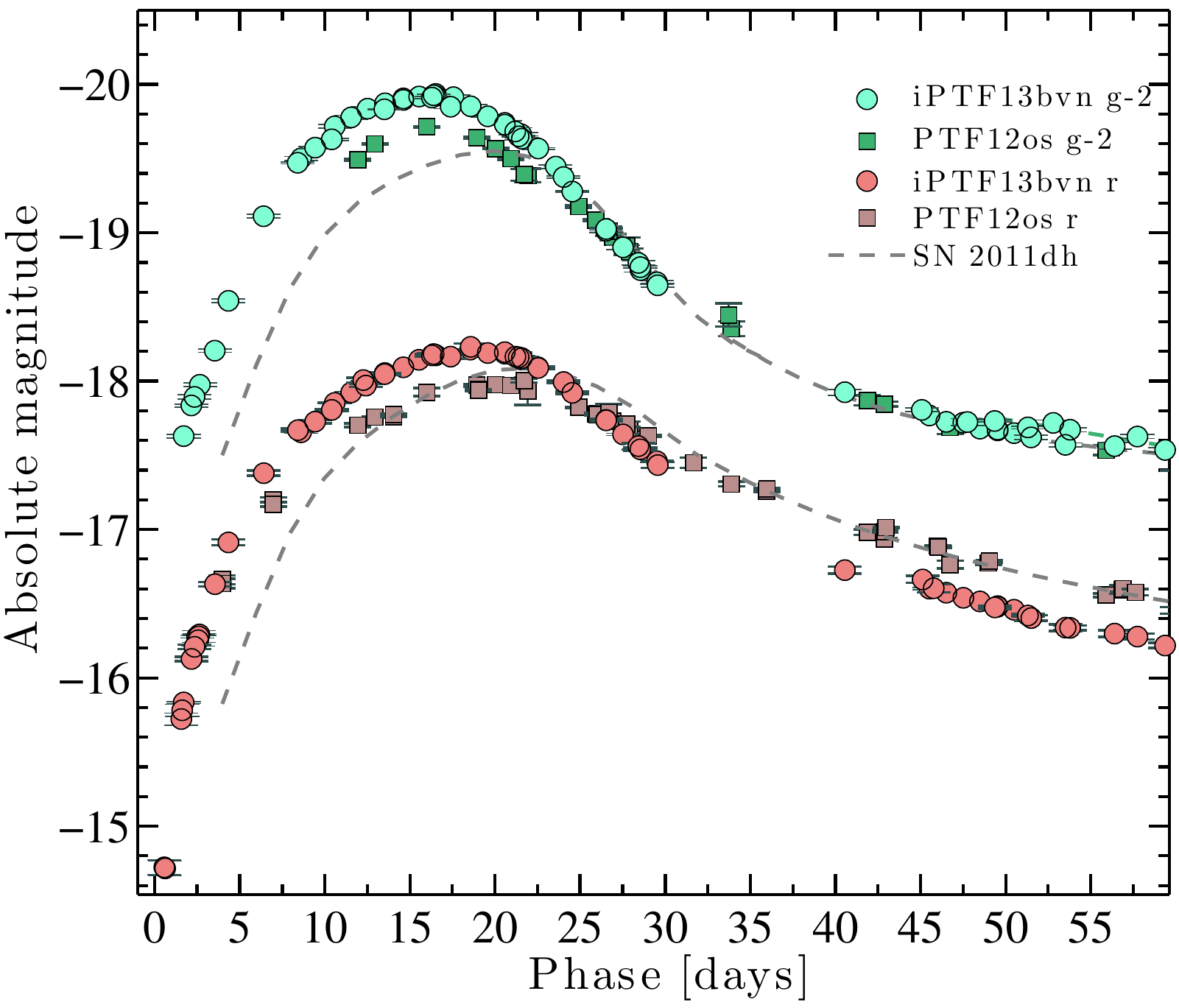}
\hspace{0.2cm}
\caption{Early-time $g$- and $r$-band LCs of PTF12os (squares) and iPTF13bvn (circles) along with SN 2011dh (dashed black lines). Corrections for extinction and distance have been applied according to Sect.~\ref{sec:host} and Sect.~\ref{sec:extinction}.}
\label{fig:lc_early}
\end{figure}

At first glance, the LCs of these three SNe appear surprisingly similar across all photometric bands, but a more detailed look does bring out some minor differences between them. PTF12os peaks in the $g$ band at approximately +16~d, iPTF13bvn at +16.5~d, and SN~2011dh at +20~d (see Fig.~\ref{fig:lc_early} for close-up view of the $g$- and $r$-band LC peaks). The widths of the LC peaks are also slightly different. In the $r$ and $i$ bands especially, the LCs of iPTF13bvn appear to be somewhat narrower than those of both PTF12os and SN~2011dh. The width of the LC peak in $r$ is $\sim 35$ days for iPTF13bvn and 50 days for PTF12os and SN~2011dh. This width was measured by finding the point on the rising part of the LC that is 1.5 mag fainter than the maximum, and then measuring the time until the same magnitude is reached on the declining part of the (interpolated) LC (again, see Fig.~\ref{fig:lc_early}). Using the same measure a similar difference is observed in the $i$ band. However, it is somewhat difficult to isolate the peaks in the LCs of SN~2011dh and PTF12os in the $i$ and $z$ bands. It is also difficult to assess when the LC peaks end and the LCs start to decline in a more linear fashion. For iPTF13bvn, the LC peaks are somewhat more easily discernible in both the $i$ and $z$ bands. 

Another minor difference can be seen in the apparent lack of a plateau phase immediately following the peak (+50~d) in bands bluer than the $r$ band for iPTF13bvn compared to PTF12os. PTF12os shows a temporarily slower decline between approximately +50~d and +100~d in $g$ and a virtually constant magnitude in $B$. For PTF12os we measure the $g$-band decline rate in this phase to be $0.011\pm0.002$~mag~d$^{-1}$, based on a linear (first-order polynomial) fit. 
For SN~2011dh we find a very similar decline rate of $0.013\pm0.001$~mag~d$^{-1}$. For iPTF13bvn we measure 
$0.015\pm0.002$~mag~d$^{-1}$. In the fit for iPTF13bvn, data points between +50~d and +80~d were used, since after +80~d there is a gap in our photometric coverage until +210~d. However, our late-time data points between +210~d and +300~d indicate it is likely that the same decline rate as measured between +50~d and +80~d continued up until 300~d past the explosion in the $g$ band. For the $B$ band we measure $0.000\pm0.002$~mag~d$^{-1}$ for PTF12os, $0.010\pm0.001$~mag~d$^{-1}$ for SN~2011dh, and $0.010\pm0.003$~mag~d$^{-1}$ for iPTF13bvn. For SN~2011dh this $B$-band decline rate continued until around +125~d, whereafter the decline rate increased significantly. This behavior is consistent with our data on PTF12os. However, for iPTF13bvn we lack coverage in the $B$ band past +75~d.

\begin{figure*}[ht]
\centering
\includegraphics[width=18cm]{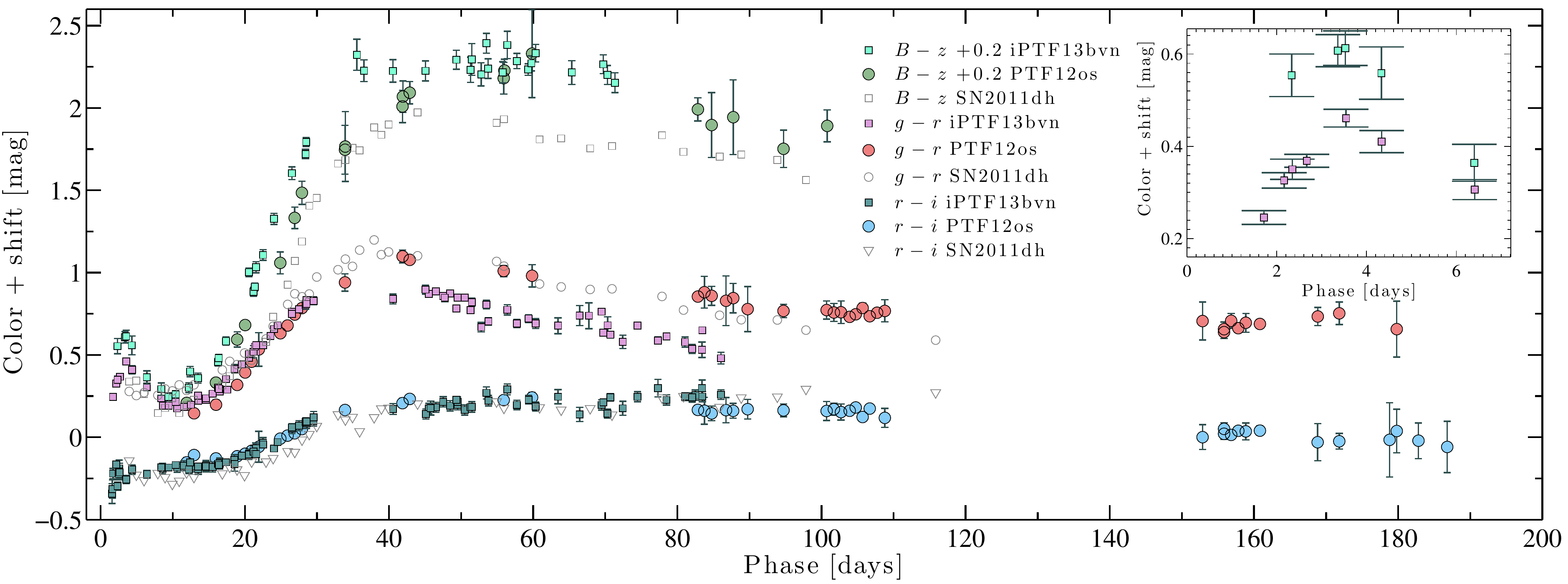}
\caption{Color evolution of PTF12os (solid colored circles) and iPTF13bvn (solid colored squares) along with SN 2011dh (open black markers). The photometry has been corrected for the extinction as described in Sect.~\ref{sec:extinction}. The error bars represent only the photometric uncertainty; we do not include the uncertainties in the extinction here.}
\label{fig:color}
\end{figure*}

In the $U$ band, the data for both iPTF13bvn and PTF12os are not of sufficiently high quality to be able to discern whether there is a similar plateau as seen in the SN~2011dh data past +50~d. However, the photometric points for both PTF12os and iPTF13bvn are consistent with the photometry of SN~2011dh within the uncertainties. In the redder bands ($riz$), there is a similar trend as in the bluer bands in the decline rates between +50~d and +100~d, with PTF12os showing marginally slower decline rates. In the $riz$ bands we respectively measure $0.018\pm0.001$~mag~d$^{-1}$, $0.018\pm0.001$~mag~d$^{-1}$, and $0.012\pm0.001$~mag~d$^{-1}$ for PTF12os; $0.022\pm0.002$, $0.020\pm0.002$, and $0.016\pm0.003$~mag~d$^{-1}$ for SN~2011dh; and $0.022\pm0.001$, $0.021\pm0.002$, and $0.011\pm0.003$~mag~d$^{-1}$ for iPTF13bvn, based on similar first-order polynomial fits as described above. 

While the widths of the LC peaks are somewhat different for iPTF13bvn and SN 2011dh, and iPTF13bvn lacks the plateau phase (at least in $g$) as discussed above, the decline rates become almost identical past +150~d in all of the bands where our data allow this comparison to be made ($gri$; see the dashed lines in Fig.~\ref{fig:lc_full}). For PTF12os, both the LC peaks and the late-time decline rates are very similar to those of SN 2011dh, but owing to the slower declines immediately after the peak the fluxes in the $g$, $r$, and $i$ bands are somewhat higher past +150~d compared to both SN~2011dh and iPTF13bvn. 

To roughly estimate the post $+100$~d decline rates of PTF12os and iPTF13bvn in the individual bands (the dashed and colored lines in Fig.~\ref{fig:lc_full}), we performed first-order polynomial fits to the photometric data. For PTF12os we fit the data obtained between +150~d and +220~d in $gri$ to avoid the plateaus in $g$ and $r$. Photometric data between +70~d and +300~d are used for iPTF13bvn. Based on these fits we find the late-time decline rate in $r$ to be $0.019\pm0.001$~mag~d$^{-1}$ for iPTF13bvn, and for PTF12os we measure a decline of $0.016\pm0.002$~mag~d$^{-1}$. In $g$ we measure $0.019\pm0.001$~mag~d$^{-1}$ for iPTF13bvn and $0.017\pm0.001$~mag~d$^{-1}$ for PTF12os. In the $i$ band we measure $0.016\pm0.005$~mag~d$^{-1}$ for iPTF13bvn and $0.017\pm0.001$~mag~d$^{-1}$ for PTF12os. These decline rates are very similar to the decline rates of SN~2011dh in the corresponding bands at +200~d (again, see Fig.~\ref{fig:lc_full}). We note that we are by necessity fitting different time intervals for PTF12os and iPTF13bvn, since there is a large gap in the coverage of iPTF13bvn past +80~d and we lack photometry past +220~d for PTF12os. This makes direct comparisons of the decline rates difficult. However, from Fig.~\ref{fig:lc_full}, it is clear that PTF12os, overall, declined less from peak until at least +220~d compared to both SN~2011dh and iPTF13bvn, since the LC peaks are generally very similar in PTF12os while the late-time $gri$ photometry is significantly brighter. 

In the UV regime, the flux is comparable for PTF12os and SN 2011dh in both the $UVW1$ and $UVM2$ bands. For SN~2011dh a smooth decline was observed in the $UVM2$ band, while an initial decline followed by a peak was observed in the $UVW1$ band\footnote{We note that the peak observed in the $UVW1$ band for SN~2011dh is likely a result of leakage from redder wavelengths.}. Our data on PTF12os are not of sufficiently high quality to determine whether this SN shows similar behavior. The first point in the $UVW1$ band is not early enough, compared to the epochs where the initial decline was observed for SN~2011dh. Furthermore, the uncertainties of our $UVM2$-band data are too large to disentangle between a peak or a smooth decline in the observed LC. Compared to PTF12os and SN~2011dh, iPTF13bvn appears to have had a significantly lower UV flux. For this SN the first datapoint in the $UVW1$ band was obtained very early, at around +2~d, but we still do not see the initial decline followed by a rise as for SN~2011dh. This could be interpreted as iPTF13bvn having a more compact progenitor compared to SN~2011dh, and hence a faster UV cooling tail following shock breakout (see also Sect.~\ref{sec:earlyphot}). In the  $UVM2$ band we only have two useable data points, and thus we cannot determine the LC shape. Finally, one should note that the UV regime is very sensitive to extinction corrections, and especially to the shape of the extinction law, which we simply assume here to be the same as for the MW (Sect.~\ref{sec:extinction}) for both objects. This, or alternatively line blocking by metal lines, could also explain why the UV flux appears to be different in iPTF13bvn compared to PTF12os (and SN~2011dh). If the extinction law is too steep in the UV, the considerable extinction corrections we are using for PTF12os could result in an overestimated UV flux.

\subsection{Color evolution and blackbody fits}
\label{sec:colors}
A comparison of the color evolution in $B-z$, $g-r$, and $r-i$ for PTF12os, iPTF13bvn, and SN~2011dh is shown in Fig.~\ref{fig:color}. For these calculations the LCs have been interpolated to the dates of the bluest filter in each color. Since the multiband LCs are very similar, as previously discussed, all three SNe also appear similar in terms of the observed colors and color evolution. We note that this is in part by construction; we are matching the colors of PTF12os and iPTF13bvn to the colors of SN~2011dh between +10~d and +30~d, to estimate the extinction (Sect.~\ref{sec:extinction}). However, other methods to calculate the extinction for both objects are consistent with the values that we have adopted, and the color evolution for each SN relative to the values between +10~d and +30~d is in any case not affected by our assumptions about the extinction.

The most apparent differences between these three SNe appear in the $B-z$ and $g-r$ colors. SN~2011dh temporarily shows a bluer color in $B-z$ by around 0.5~mag compared to PTF12os and iPTF13bvn, between +50~d and +70~d. In $g-r$ the color is almost identical among the SNe up until +30~d, after which the color of iPTF13bvn plateaus at a $\sim0.25$~mag bluer color of $g-r\approx0.8$~mag, indicating a significantly lower flux in the $g$ band, since the $r-i$ color continues to be similar also after +30~d. However, compared to the sample of SNe~Ibc  in \cite{2015A&amp;A...574A..60T}, the $g-r$ color and time evolution in both PTF12os and iPTF13bvn appear consistent with those of other SNe~Ib. The spread in the observed $g-r$ color in that sample after +30~d is significantly larger than the difference observed between PTF12os and iPTF13bvn. 

\begin{figure}
\centering
\includegraphics[width=9cm]{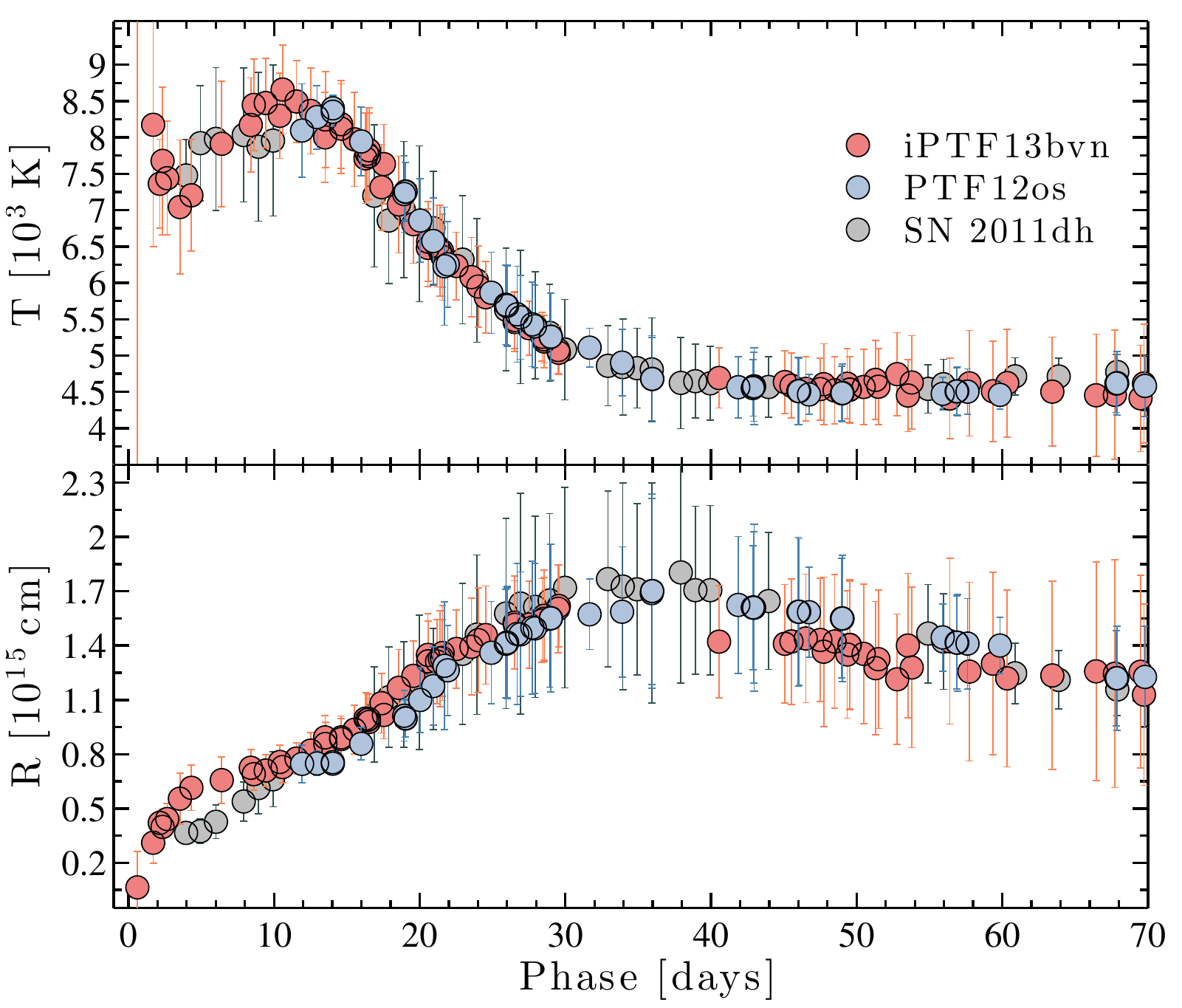}
\caption{Blackbody temperature (top panel) and radius (bottom panel) of iPTF13bvn (red circles), PTF12os (blue circles), and SN~2011dh (gray circles).}
\label{fig:bb}
\end{figure}

\begin{figure*}[ht]
\centering
\includegraphics[width=18cm]{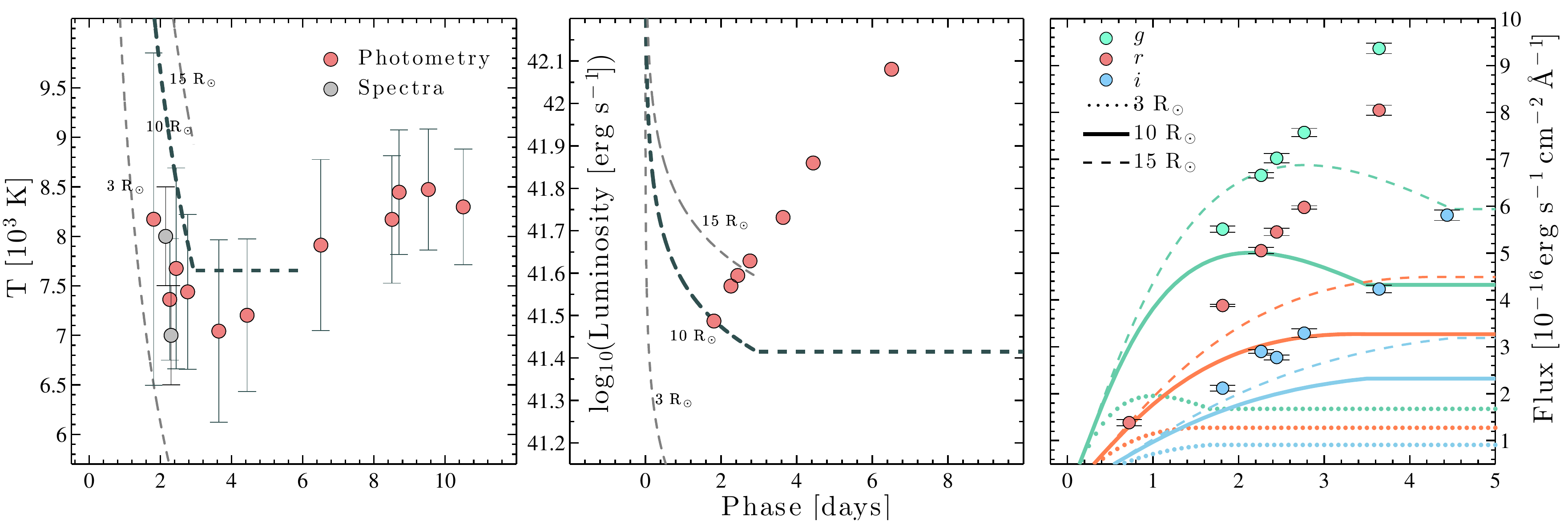}
\caption{Early BB temperature evolution of iPTF13bvn (left panel). Early bolometric luminosity calculated using the temperature evolution from the left panel and the Stefan-Boltzmann law (center panel). Early $g$, $r$, and $i$-band flux of iPTF13bvn (right panel). In each panel we also show the relevant quantity predicted by the PN1 models for the cooling emission after shock breakout for progenitor radii of 3, 10, and 15~\rsun. A model with a radius of 10~\rsun\ is consistent with the observed data. In the left and center panels the 3~\rsun\ and 15~\rsun\ models are shown as thin dashed gray lines and the optimal model with a radius of 10~\rsun\ is a thicker dashed black line. In the right panel the 3~\rsun\ model LCs are displayed as dotted lines and the 15~\rsun\ model LCs are thin dashed lines in matching colors to the photometry. The 10~\rsun\ model LCs are shown as thicker solid lines.}
\label{fig:early_lc_fit}
\end{figure*}

Another noteworthy difference is seen in the very early $g-r$ color of iPTF13bvn; up until approximately +4~d, a clear phase of gradual reddening in this color is observed (see the inset in Fig.~\ref{fig:color}). Similar behavior is seen to a lesser extent in the other colors. 
This could be evidence of a significant contribution from the cooling phase following shock breakout in the SN explosion, and it 
is discussed in more detail in Sect.~\ref{sec:earlyphot}. 

We have determined the BB parameters of PTF12os, iPTF13bvn, and SN~2011dh by fitting BBs to the spectral energy distribution (SED) derived from the $griz$ LCs (see Fig.~\ref{fig:bb}). For these BB calculations we have interpolated the flux in the other bands to the dates of the $g$-band measurements. We correct for the distance and extinction according to Sect.~\ref{sec:host} and Sect.~\ref{sec:extinction}.

Both in terms of the BB temperature and the radius, all three SNe appear to be very similar, except for the very early temperature evolution where iPTF13bvn shows signs of cooling in the earliest data. The BB temperature peaks at $\sim 8500$~K for all three SNe, and the radii peak at $\sim 1.7\times10^{15}$~cm. 

We note that the peak radius and uncertainties generally become smaller for SN~2011dh if the $r$ band is excluded from the BB fits. This indicates that the SED significantly deviates from a BB for SN~2011dh in $r$, which could possibly be explained by the relatively strong signature from hydrogen in the early spectra (Sect.~\ref{sec:spectra}). We also note that the radius found for SN~2011dh by \cite{2014A&amp;A...562A..17E} via fitting the $VIJHK$ bands peaks at $\sim 1.4\times10^{15}$~cm. That estimate should be more robust, since it covers a much larger wavelength range, and deviations from a BB in small parts of this range should not have as large an effect on the final result. A similar decrease in the BB radii of PTF12os and iPTF13bvn would be expected, given the similarity of all other LC bands. However, we do not have the needed wavelength coverage to perform that comparison.

\begin{figure*}
\centering
\includegraphics[width=17.999cm]{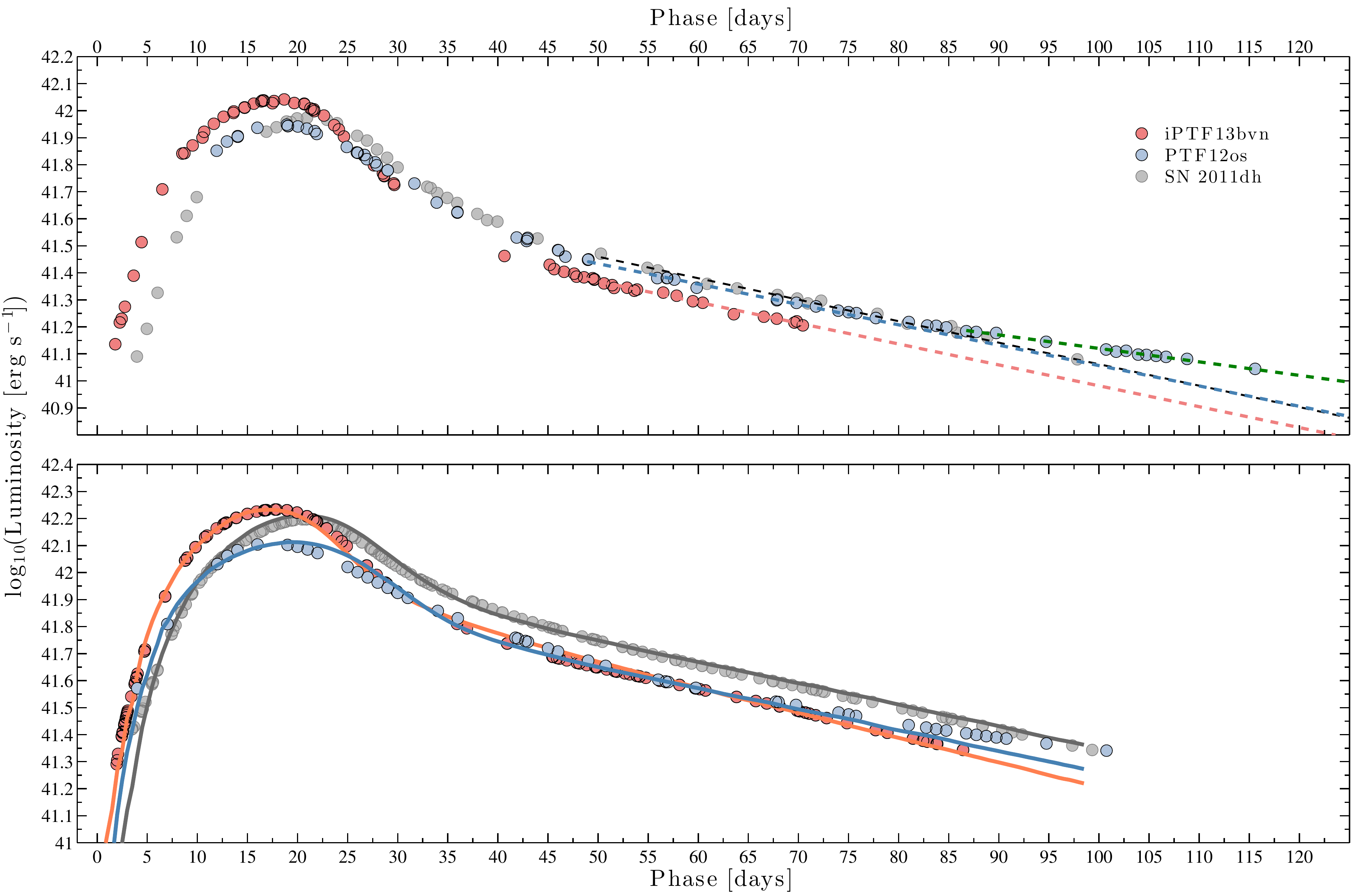}
\caption{Bolometric light curves of iPTF13bvn (red circles), SN~2011dh (gray circles), and PTF12os (light-blue circles). The top panel shows quasibolometric {\it Bgriz} LCs calculated as described in Sect.~\ref{sec:lc}. In the top panel we also show linear fits to the LC points between +53~d and +75~d for the three SNe (dashed lines, PTF12os in blue, iPTF13bvn in red and SN~2011dh in black). For PTF12os an additional fit to the LC points after +85~d is shown as a green dashed line, which clearly exhibits a slower decline. The bottom panel, following the same color scheme as in the top panel, illustrates the bolometric LCs of the SNe, with corrections for the UV and IR contributions based on observations of SN~2011dh applied to PTF12os and iPTF13bvn. The bottom panel also shows our best-fitting hydrodynamical models for each SN as solid lines of matching color.}
\label{fig:bol_lcs}
\end{figure*}

\subsection{Early-time photometry}
\label{sec:earlyphot}
For iPTF13bvn we have very early-time $r$-band photometry (at +0.6~d) from the P48, along with color information starting already from +1.7~d. In the early color evolution we observe a cooling trend up to around +4~d in $g-r$ (see the inset of Fig.~\ref{fig:color}). Our BB fits to the $griz$ filters (Fig.~\ref{fig:bb}) show temperatures declining from 8500~K at around +2~d to 7000~K at +4~d. Similar evolution is found when fitting BB SEDs to the earliest two spectra (see Sect.~\ref{sec:spectra}). In Fig.~\ref{fig:early_lc_fit} we show the early BB temperature evolution, an estimated early bolometric LC, and the early $gri$ photometry in detail.

One possible interpretation of these data is that we are at early times observing significant flux from the cooling phase following shock breakout \citep{Piro:2012aa}. Based on this assumption we have performed fitting of models with different radii to the early $gri$-band luminosity, the temperature evolution, and the early bolometric luminosity estimated from the BB temperature and radius using the Stefan-Boltzmann law. Our models are calculated according to
\citet[][their Eqs. 1 and 2, henceforth abbreviated the PN1 model]{Piro:2012aa},
and we also let the explosion date of iPTF13bvn be a free parameter in the fitting procedure.

We find that it is possible to fit the early temperature evolution with a model that is also consistent with the early photometry in the individual bands (the right panel of Fig.~\ref{fig:early_lc_fit}), by using a progenitor radius of 10~\Rsun\ and an explosion date of $t_{\rm exp}=$~2013 June 15.57. We note that after around +2~d, the luminosities in the individual bands start to greatly exceed what can be produced from a model based purely on shock-breakout cooling while still fitting the observed temperature (and estimated bolometric luminosity) simultaneously. We interpret this as the heating from radioactive nickel gradually becoming the dominant energy source.

For these calculations we have adopted the ejecta mass and total kinetic energy from our hydrodynamical model described in Sect.~\ref{sec:hydro} and an opacity $\kappa=0.2$~cm$^2$g$^{-1}$. 
The radius we find here is significantly larger than what was found by \cite{2013ApJ...775L...7C}. This difference can be largely explained by a different approach; in \cite{2013ApJ...775L...7C} it was assumed that the luminosity from the cooling phase should reach the plateau phase at a luminosity limited by the first photometric point observed by the P48 at +0.6~d. Here we assume that the cooling is still significant at least up until around +2~d. We are also using explosion parameters (ejecta mass, total kinetic energy) from hydrodynamical modeling, while \cite{2013ApJ...775L...7C} adopted typical values for an SE~SN. We note that the explosion date we find using this method is within 0.1~d of the explosion date estimate by \cite{2013ApJ...775L...7C}, but around 0.3~d later than the explosion date derived from our hydrodynamical modeling. We also note that a radius on the order of 10~\rsun\ immediately prior to the explosion of an SE~SN is expected in a binary scenario \citep[see, e.g.,][]{Yoon:2010aa}.

From hydrodynamical simulations, \cite{2014AJ....148...68B} have argued that the early-time observations of iPTF13bvn could be consistent with a significantly larger radius, possibly even as large as $150$~\Rsun, if the explosion is assumed to have happened somewhat later than our estimate. That calculation was based only on the P48 $r$-band data. Using only the $r$ band and later explosion dates, we can also fit significantly larger radii. However, these models then become inconsistent with all of the remaining observations; too high temperatures compared to the observations are predicted, resulting in bad fits to the bolometric LC. However, there is evidence that analytical shock-cooling models tend to predict unrealistically high temperatures when the radius becomes large \citep[e.g.,][]{2012ApJ...757...31B}. Lower temperatures in our models with larger radii and later explosion times could make them more consistent with the observed LCs. Thus, we do not consider the above radius estimate of $10$~\Rsun\,as particularly robust, but it does seem unlikely that the progenitor would have had a significantly smaller radius. In the binary modeling by \cite{2016arXiv160405050E} a radius on the order of 50~\rsun\ is found for the progenitor of iPTF13bvn.

Finally, for PTF12os it is not possible to put any meaningful constraints on the progenitor radius based on a similar analysis, since the early P48 coverage does not start until around +5~d and we only have color information after $+10$~d. We stress the importance of obtaining early-time color information (ideally before +2~d) for constraining the radius of more SE~SNe, following the above recipe or using hydrodynamical calculations.

\begin{figure*}
\centering
\includegraphics[width=18cm]{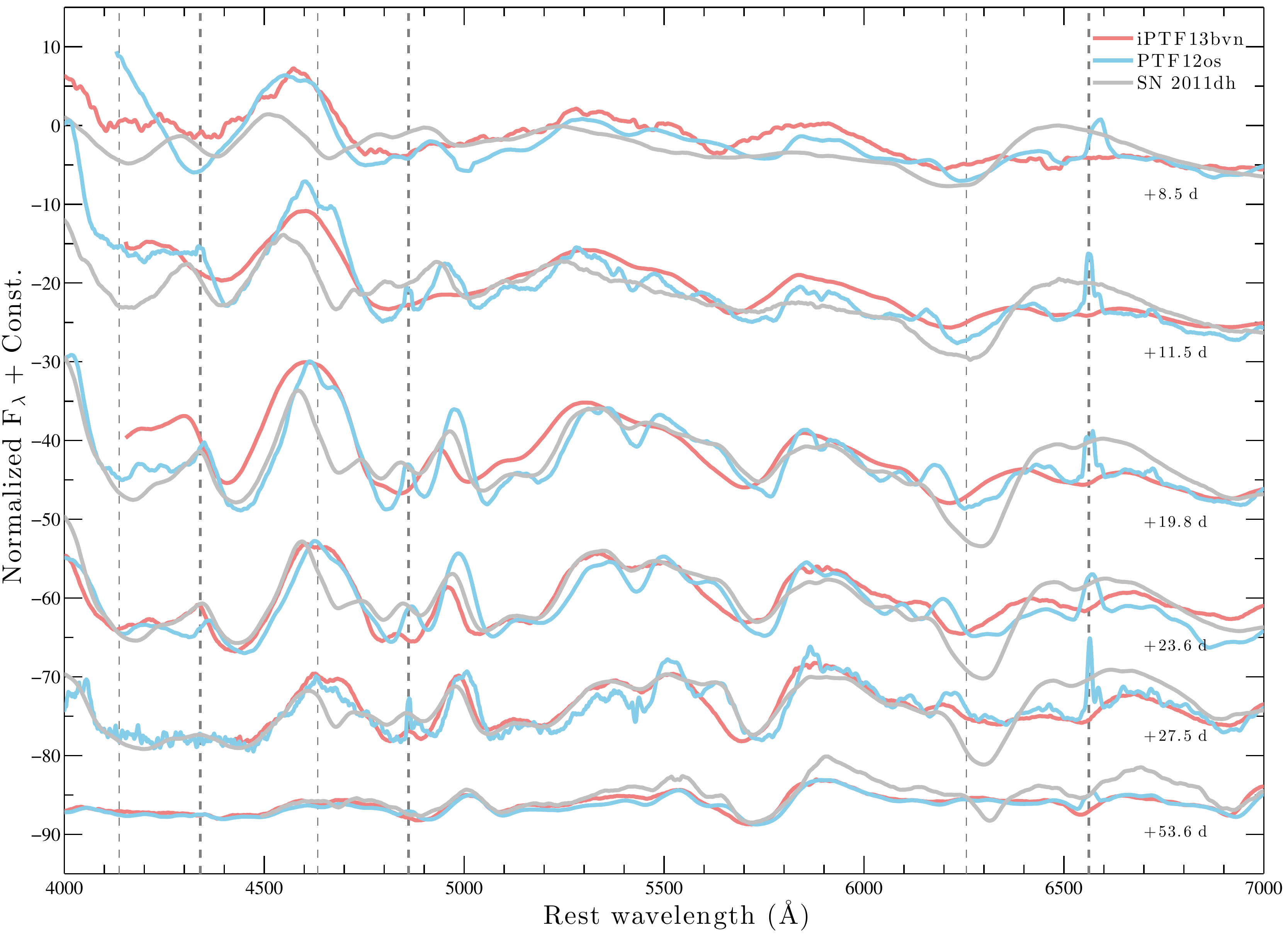}
\caption{Spectral comparison of PTF12os (blue), iPTF13bvn (red), and SN~2011dh (gray). The spectra have been normalized using the median and shifted by a constant for each epoch. The thick dashed lines show the rest wavelength of H$\alpha$, H$\beta$, and H$\gamma$, and the thin dashed lines mark the corresponding wavelengths shifted by 14,000 km~s$^{-1}$, which coincides very well with absorption minima in the spectra of PTF12os for all three lines. The spectra have been smoothed to improve their visibility. The date labels refer to the date of the spectral observations of PTF12os for each epoch. The spectra of the other SNe have been chosen as close as possible to these epochs.}
\label{fig:earlyspec}
\end{figure*}

\subsection{Quasi-bolometric light curves}
\label{sec:bolometric}
We have constructed quasi-bolometric LCs by integrating the flux in the $Bgriz$ bands. For these calculations we have corrected for the distance and extinction according to Sect.~\ref{sec:host} and Sect.~\ref{sec:extinction}. To determine the integrated fluxes we have interpolated the magnitudes in the other bands to the dates of the $r$-band measurements. We then converted the magnitudes into fluxes for each observed $r$-band epoch at the effective wavelengths of the filters and performed linear interpolation of these data at an even 1~\AA~spacing, followed by integration between the effective wavelengths of the $B$ and $z$ bands. 

We show a comparison of the $Bgriz$ quasi-bolometric LCs of PTF12os, iPTF13bvn, and SN~2011dh in Fig.~\ref{fig:bol_lcs}. Qualitatively, these LCs are again very similar for these three SNe, with only minor differences, as expected from the very similar multiband photometry. Similar to what was seen for the $g$ band, PTF12os and iPTF13bvn rise to peak in 16--18~d, while SN~2011dh rises to peak in $\sim 20$~d. The widths of the LC peaks also differ slightly, with the LC peak of iPTF13bvn being somewhat narrower, as was also seen in some of the individual LC bands. The bolometric LC peak width correlates with the ejecta mass of the explosion \citep{1977ApJS...33..515F}; thus, this indicates that the ejecta mass could be slightly lower in iPTF13bvn compared with PTF12os and SN~2011dh, which should have very similar ejecta masses if only the LC peak widths are considered (but see Sect.~\ref{sec:hydro} for a more comprehensive analysis based on our hydrodynamical models, which also account for the observed expansion velocities). The ejecta mass of SN~2011dh was estimated to be $\sim 2$~\msun, from previous hydrodynamical modeling \citep{2014A&amp;A...562A..17E}.

The quasi-bolometric peak luminosities are remarkably similar, with iPTF13bvn peaking at the highest luminosity of $10^{42.05}$~erg~s$^{-1}$, SN~2011dh peaking at $10^{42}$~erg~s$^{-1}$, and PTF12os at $10^{41.95}$~erg~s$^{-1}$. The peak luminosity is strongly correlated with the amount of radioactive nickel ejected in the explosions. Thus we find that these three SNe should have very similar nickel masses. In \cite{2014A&amp;A...565A.114F}, the nickel mass of iPTF13bvn was found to be $\sim 0.05$~\msun, which is on the lower end of what is observed for typical SNe~Ib/IIb. However, we have here adopted a distance modulus that is higher by 0.38~mag. The nickel mass of SN~2011dh \citep[$\sim0.075$~\msun,][]{2014A&amp;A...562A..17E} is typical for a SN~IIb.

In the linear decline phase of the quasi-bolometric LCs, starting at approximately +50~d for all three SNe, the decline rates are very similar, when measured as linear (first-order polynomial) fits between +50~d and +70~d. We find $0.019\pm0.001$~mag~d$^{-1}$, for iPTF13bvn, $0.019\pm0.002$~mag~d$^{-1}$ for PTF12os, and $0.020\pm0.003$~mag~d$^{-1}$ for SN~2011dh. 
After around +80~d, we observe a break in the LC of PTF12os toward a slower decline of $0.012\pm0.001$~mag~d$^{-1}$, as measured by a linear fit between +85~d and +110~d. This could indicate a more efficient trapping of $\gamma$-rays in the SN ejecta of PTF12os, as this decline rate is close to the expected decay rate of $\sim 0.01$~mag~d$^{-1}$ from the energy deposition of radioactive \element[][56]{Co} assuming complete trapping. This behavior is not seen in the LC of SN~2011dh, for which the same decline as measured between +50~d and +70~d continues for a significantly longer time, at least until around +300~d. However, \cite{2015MNRAS.450.1295W} show that the late-time decline rates ($\gamma$-rays trapping efficiencies) among SE~SNe can vary significantly. For iPTF13bvn, our photometric coverage in the $B$ and $z$ bands stops after +70~d, and thus we are not able to comment on the late-time $Bgriz$ quasi-bolometric LC behavior. However, in Sect.~\ref{sec:lc} we found that the $gri$-band decline rates are very similar for iPTF13bvn and SN~2011dh also at much later times.

When constructing a quasi-bolometric LC of iPTF13bvn using methods similar to ours, \cite{2014MNRAS.445.1932S} found a rather steep post-peak decline rate of $\sim0.03$~mag~d$^{-1}$ between +40~d and +70~d. This is a steeper decline compared with most other SNe~IIb/Ib \citep{2015MNRAS.450.1295W}. However, the decline rate found by \cite{2014MNRAS.445.1932S} depended heavily on a single bolometric epoch at $~+70$~d. Our quasi-bolometric LC of iPTF13bvn shows a slower decline rate, similar to that of SN~2011dh between +50~d and +70~d. Our result is based on multiple epochs of $Bgriz$ data during this phase, of which several are at around +70~d.
Finally, we want to point out that although the quality of our photometry has allowed us to scrutinize minor differences in the multiband LCs, color properties, and quasi-bolometric LCs of these three SNe, the bottom line must still be that PTF12os, iPTF13bvn, and SN~2011dh are astonishingly similar photometrically.

\section{Spectroscopy}
\label{sec:spectra}
\begin{figure}
\centering
\includegraphics[width=16cm]{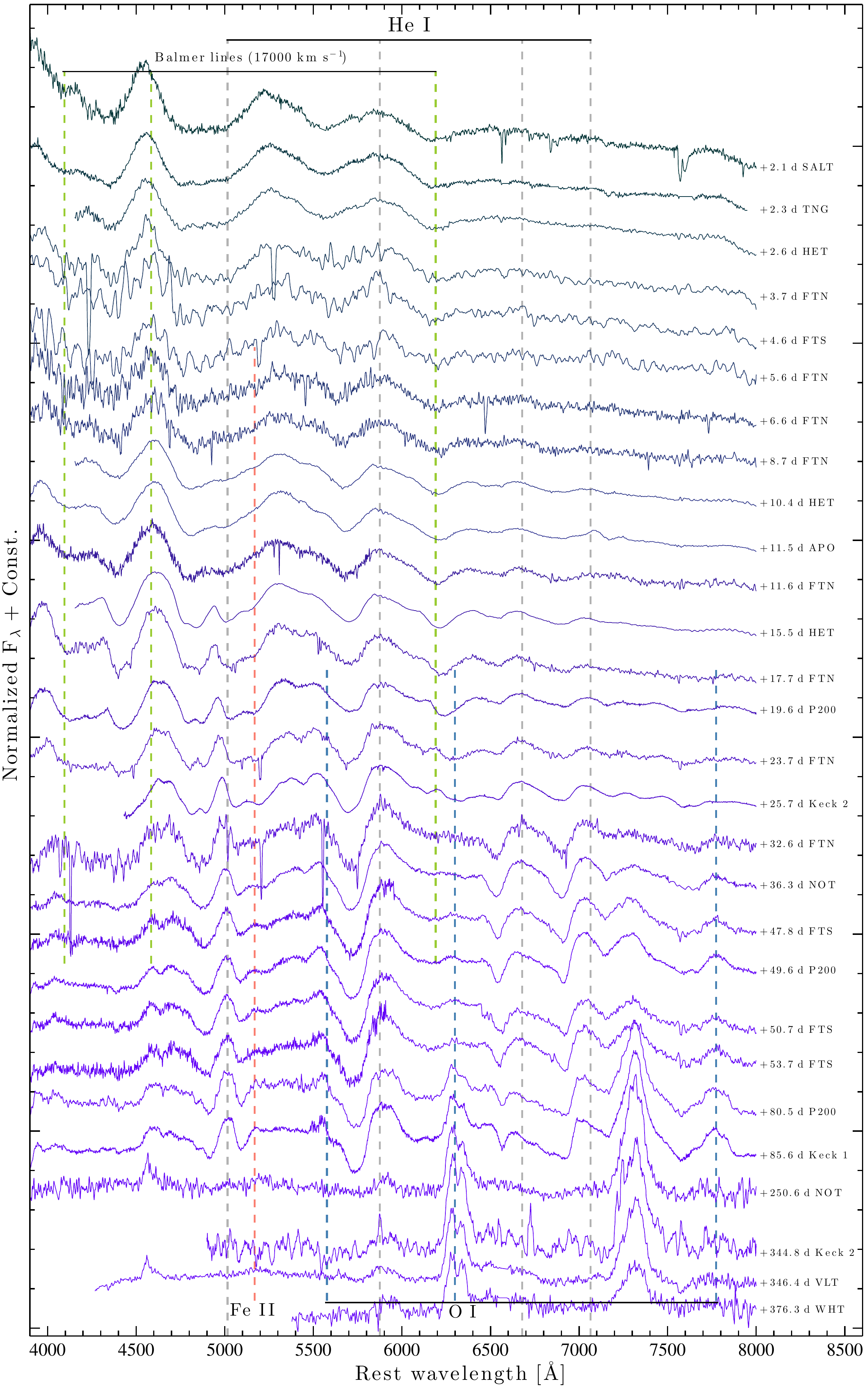}  
 \caption{\label{fig:spec13bvn}Spectral sequence of iPTF13bvn. Thick dashed lines mark the central wavelength of the marked emission lines at rest, except for the Balmer lines, which are marked by green dashed lines at $17, 000$ km~s$^{-1}$. Telluric features near the wavelengths 6860~\AA\, and especially 7600~\AA\, are present in some of the spectra.}
\end{figure}

In Sect.~\ref{sec:lc} we showed that PTF12os and iPTF13bvn are photometrically very similar. Only minor differences, both in terms of the filtered LCs and quasi-bolometric LCs, can be identified. Spectroscopically, the picture is similar, although somewhat more complex. PTF12os (Fig.~\ref{fig:spec12os}) and iPTF13bvn (Fig.~\ref{fig:spec13bvn}) qualitatively show the same spectral features and evolve in a similar way (see also Fig.~\ref{fig:earlyspec}). However, the spectra of iPTF13bvn generally show somewhat broader features, indicating faster expansion velocities. The spectra of PTF12os are more affected by host-galaxy contamination from the strong underlying H$\alpha$ region (Sect.~\ref{sec:prog_12os}).

The very early spectra (until +10 d) of iPTF13bvn are dominated by broad features, indicating high expansion velocities. We can identify emission from the \ion{He}{I} 5876~\AA\,line in the earliest spectrum of the SN obtained at +2.15~d. Its absorption minimum indicates a velocity of 17,000--18,000 km~s$^{-1}$, consistent with what was measured in the spectrum at +2.3~d 
by \citet{2014A&amp;A...565A.114F}. For PTF12os, our spectroscopic coverage starts at +8.5~d, and we can therefore not make a comparison of the very early expansion velocities. However, the \ion{He}{I} 5876~\AA\,line shows a significantly slower velocity of only $7000$~km~s$^{-1}$ compared to around 12,000~km~s$^{-1}$ for iPTF13bvn in spectra obtained around +8~d.

\begin{figure}
\centering
\includegraphics[width=9cm]{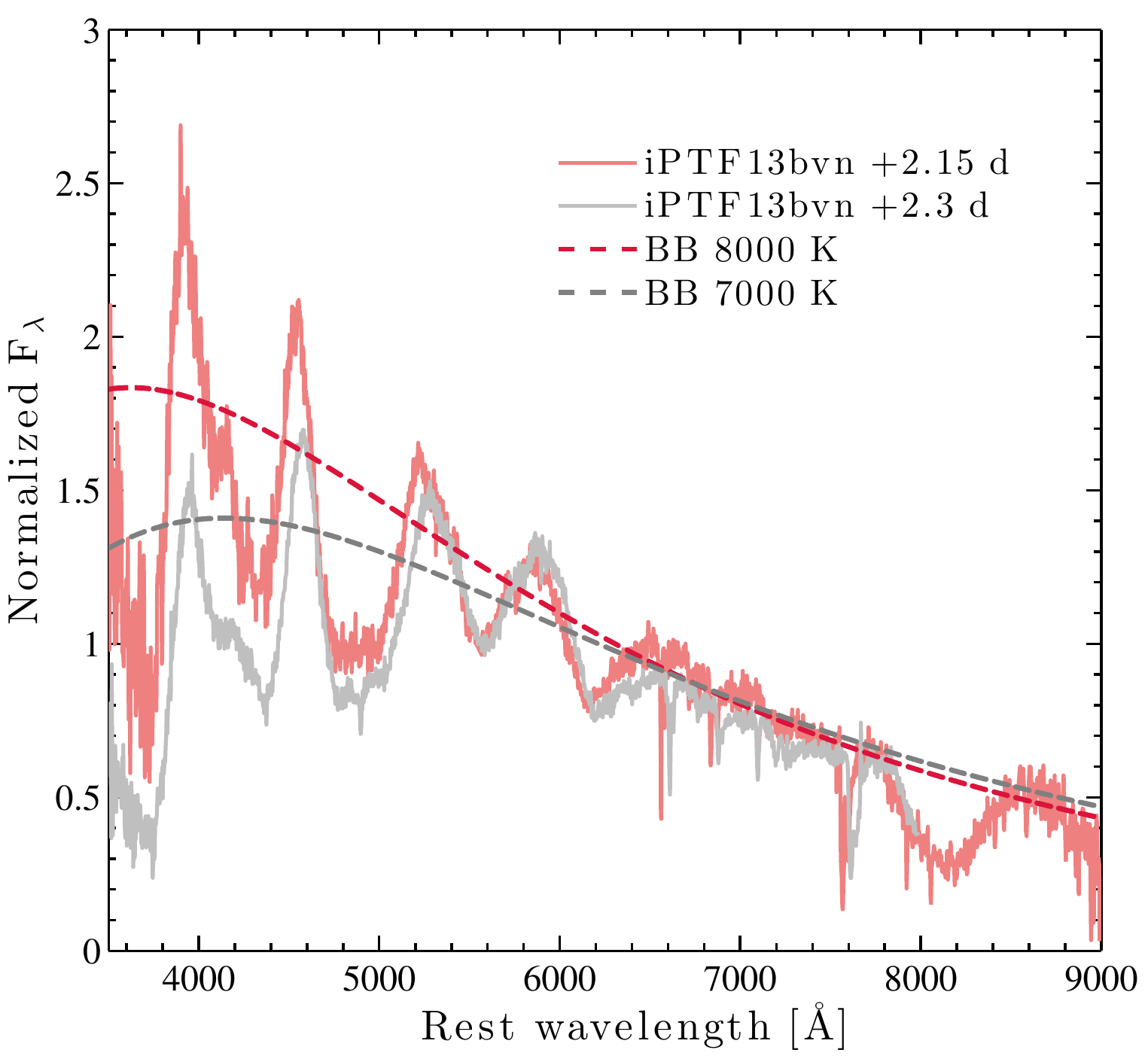}
\caption{Early-time spectra of iPTF13bvn obtained at +2.15~d (red) and +2.3~d (gray) together with blackbody fits (dashed lines, 8000~K in red and 7000~K in gray). The spectra have been normalized by their median values in a region around 6000~\AA.}
\label{fig:bb_spec}
\end{figure}

The iPTF13bvn spectrum at +2.15~d shows a significantly bluer continuum compared to the later spectra obtained at +2.3~d and +2.6~d.
Blackbody fits to these three very early-time spectra yield temperatures of $8000\pm500$~K at +2.15~d, $7000\pm500$~K at +2.3~d, and $7500\pm500$~K at +2.6~d\footnote{We note that since the spectra are actually not smooth BBs, these fits come with large systematic uncertainties.}
(Fig.~\ref{fig:bb_spec}). This is consistent with the cooling trend observed in the optical LCs of the SN during this phase (Sect.~\ref{sec:earlyphot}). Note also that a decline in temperature from around 
9000~K to 7000~K is predicted to occur between +2~d and +3~d in the semianalytical model for the early-time photometry we used to estimate the radius of the progenitor of this SN (Sect.~\ref{sec:earlyphot} and Fig.~\ref{fig:early_lc_fit}).

Based on spectral comparisons with other SNe~Ib, iPTF13bvn has been classified as a spectroscopically normal SN~Ib both in the optical and the NIR \citep{2013ApJ...775L...7C,2014A&amp;A...565A.114F}. The emergence of \ion{He}{I} $\lambda\lambda$5016, 5876, 6678, and 7065 absorption became clear after $\sim+15$~d. The velocities of the \ion{He}{I} $\lambda\lambda$5876, 6678, 7065 lines along with the velocity of \ion{Fe}{II} $\lambda$5169, based on the respective absorption minima, were measured by \cite{2014A&amp;A...565A.114F}. 
PTF12os shows very similar evolution in the \ion{He}{I} lines. In spectra taken after $+16$~d, the \ion{He}{I} $\lambda\lambda$5016, 5876, 7065 lines can be clearly identified. \ion{He}{I} $\lambda$6678 is difficult to distinguish in early-time spectra owing to strong host contamination, but it can be seen in our spectra taken between +35~d and +53~d. Another minor difference between these SNe can be seen in the \ion{He}{I} $\lambda$5016 line, which is clearly seen even in our earliest spectrum of PTF12os (taken at $+8.5$~d), while it appears somewhat later, after $+11$~d, in iPTF13bvn. 
A similar effect is visible in \ion{Fe}{II} $\lambda$5169; after +10~d this line gradually starts to appear in iPTF13bvn, while it is already there in our earliest spectrum (at +8.5~d) of PTF12os.

\begin{figure*}[ht]
\centering
\includegraphics[width=18cm]{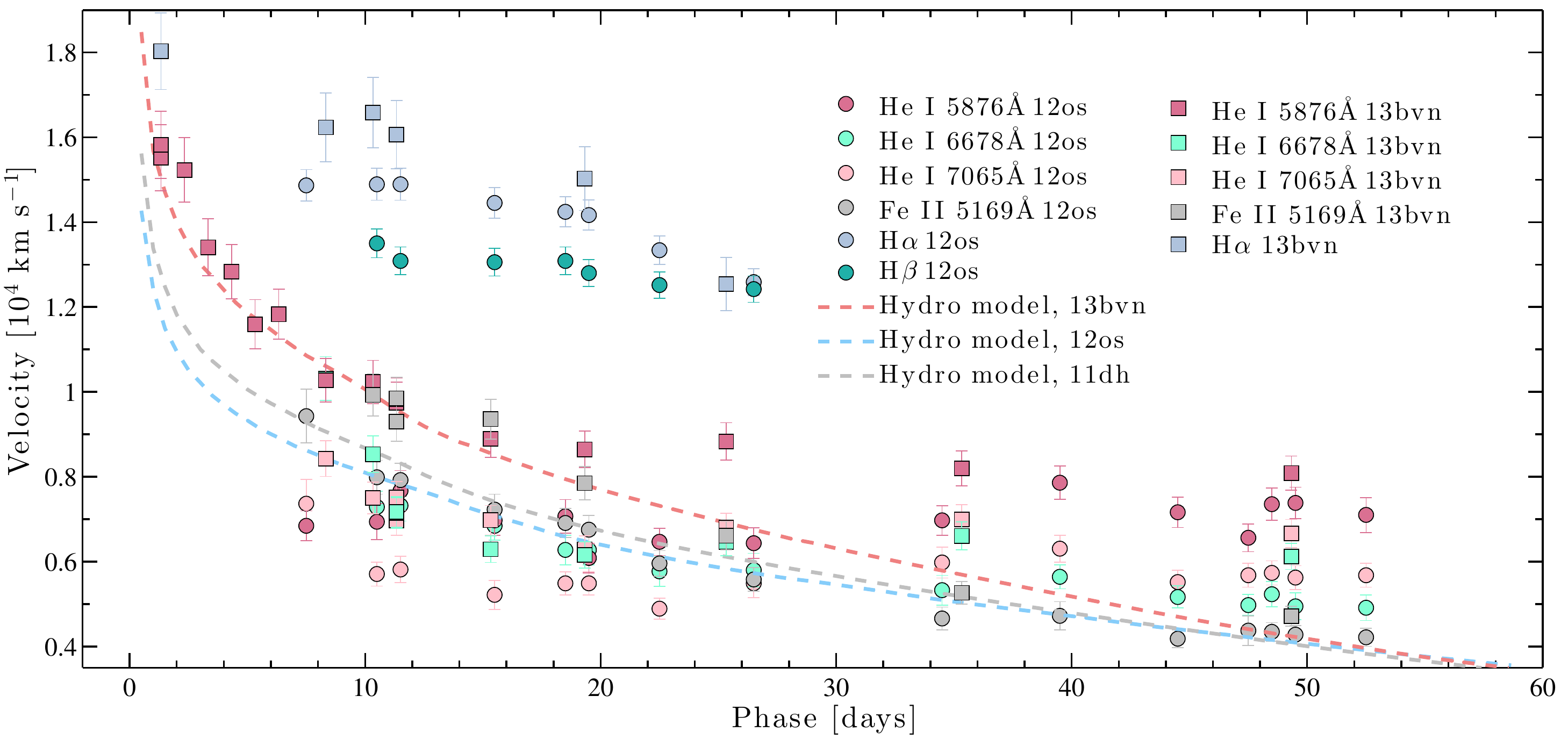}
\caption{Absorption minimum velocity measurements for a selection of spectral lines in iPTF13bvn (squares) and PTF12os (circles). The velocity of \ion{He}{I} $\lambda$5876 is shown in red, \ion{He}{I} $\lambda$6678 in cyan, and \ion{He}{I} $\lambda$7065 in pink. The \ion{Fe}{II} $\lambda$5169 absorption minimum velocities are shown in gray. We also indicate the estimated \ion{H}{$\alpha$} (blue) and \ion{H}{$\beta$} (teal) velocities for both PTF12os and iPTF13bvn. The dashed lines show the photospheric velocity evolution of our best-fitting hydrodynamical models for SN~2011dh (gray), PTF12os (blue), and iPTF13bvn (red).}
\label{fig:vel_evol}
\end{figure*}

We have determined the \ion{He}{I} $\lambda\lambda$5016, 5876, 6678, 7065 and \ion{Fe}{II} $\lambda$5169 velocities in all of the spectra where measurements were possible for both PTF12os and iPTF13bvn. This also included remeasuring velocities in previously published spectra, and it was done by fitting Gaussians in small regions around the absorption minima of the lines and locating the minima in the fits. 
In some cases, where the galaxy contamination is very strong and reasonable Gaussian fits are not possible, the centers of the absorption features were measured by manual inspection of the spectra. While we attempted velocity measurements of the \ion{He}{I} $\lambda$5016 absorption, we found that it is very contaminated in most spectra of both PTF12os and iPTF13bvn; thus, we have chosen to not include this line in our comparisons. We present the absorption-line velocity evolution of PTF12os and iPTF13bvn in Fig.~\ref{fig:vel_evol}.

As an estimate for the photospheric expansion velocity, which is employed in our hydrodynamical modeling in Sect.~\ref{sec:hydro}, we use the \ion{Fe}{II} $\lambda$5169 line \citep[see][]{Dessart:aa}, which evolves from $\sim 10,000$~km~s$^{-1}$ at +10~d to 5800~km~s$^{-1}$ at +50~d in iPTF13bvn, and from $\sim 8000$~km~s$^{-1}$ at +10~d to 4500~km~s$^{-1}$ at +50~d in PTF12os. While PTF12os shows on average slower photospheric velocities, the temporal evolution of the velocities (based on this line) is extremely similar, and the decrease in the expansion velocity is identical within our uncertainties between +10~d and +25~d (see the dashed lines in Fig.~\ref{fig:vel_evol}).

Similar velocity differences as for \ion{Fe}{II} $\lambda$5169 can be seen in the \ion{He}{I} lines, with iPTF13bvn showing on average faster velocities compared to PTF12os. We also note that the velocity evolution of the \ion{He}{I} lines in PTF12os appears relatively flat, with only the \ion{He}{I} $\lambda$6678  showing a small but statistically significant decrease in velocity during the evolution of the SN\footnote{We also find a decline over time in the few velocity measurements that were possible for \ion{He}{I} $\lambda$5016 in PTF12os, but this line is strongly contaminated, as previously discussed.}. Note also that our spectral sequence starts at +8.5~d, and if the same time span were considered for iPTF13bvn, the velocity evolution would look similar. 
Earlier spectra of PTF12os would have been needed to investigate any constant-velocity evolution in the \ion{He}{I} lines, as has been suggested 
for some peculiar SNe~IIb by \cite{2014ApJ...792....7F}. 

Past +50~d, [\ion{O}{I}] $\lambda\lambda$5577, 6300, 6364 emission lines along with the \ion{O}{I} $\lambda$7774 triplet start to emerge in both PTF12os and iPTF13bvn, indicating that their ejecta are gradually becoming optically thin. In our latest spectum of PTF12os, obtained at +215~d, weak \ion{He}{I} signatures are still present, but the spectrum has become dominated by [\ion{O}{I}] $\lambda\lambda$6300, 6364 and [\ion{Ca}{II}] $\lambda\lambda$7291, 7323 emission. The \ion{Mg}{I}] $\lambda$4571 emission line is also present. In our late-time spectra past +250~d of iPTF13bvn, the picture is very similar. The [\ion{O}{I}] $\lambda\lambda$6300, 6364 and [\ion{Ca}{II}] $\lambda\lambda$7291, 7323 doublets dominate. Weak \ion{Mg}{I}] $\lambda$4571 emission can also be identified, although there is significant noise in this region. The [\ion{Ca}{II}] lines started to emerge after +36~d in iPTF13bvn and  after +45~d in PTF12os.

In conclusion, the entire spectral evolution in the optical of PTF12os and iPTF13bvn is very similar, consistent with the comparable photometry (Sect.~\ref{sec:lc}). We will investigate in Sect.~\ref{sec:nebular} the late-time spectra in detail using nebular models.

\subsection{Near-infrared spectroscopy of iPTF13bvn}
\label{sec:13bvnnir}
\begin{figure}
\centering
\includegraphics[width=16cm]{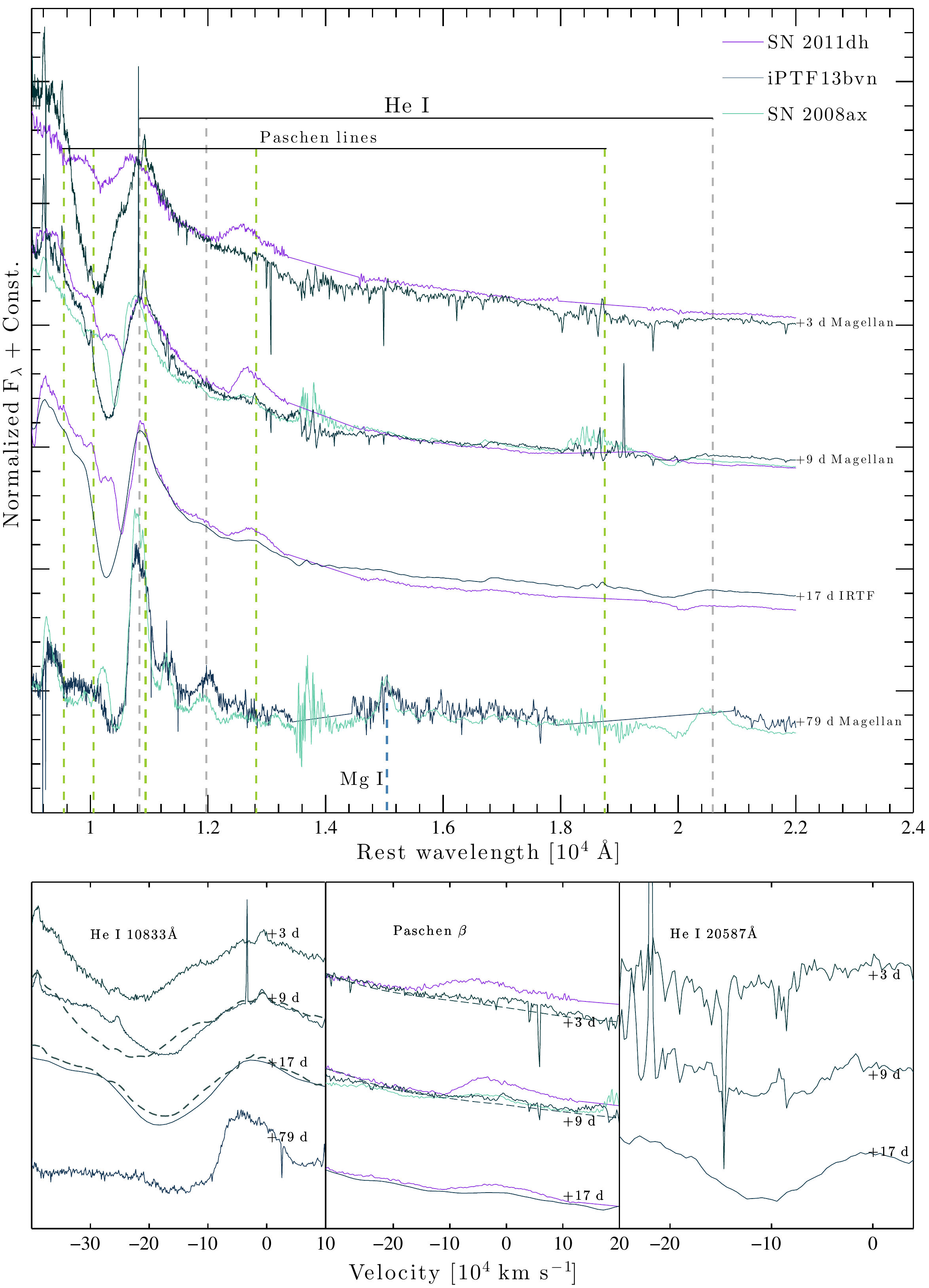}  
 \caption{\label{fig:spec13bvnir}Near-infrared spectral sequence of iPTF13bvn (black lines) compared to SN~2011dh (purple lines) and SN~2008ax (cyan lines) at similar epochs. The spectra have been normalized using the median and shifted by a constant between the epochs. The top panel shows the spectral sequence in wavelength space. In the bottom panels we highlight small sections in velocity space. The bottom-left panel displays the region around \ion{He}{I} $\lambda$10,830; the dashed lines show a smoothed version of the previous spectra in the sequence to highlight the evolution of the line. The bottom-center panel emphasizes the region around Paschen $\beta$; the dashed lines show heavily smoothed versions of the spectra of iPTF13bvn, which should give a rough estimate of the continuum. The bottom-right panel reveals the region around \ion{He}{I} $\lambda$20,590.}
\end{figure}

In addition to the early-time NIR spectra (+3~d, +9~d, and +17~d) published by \cite{2013ApJ...775L...7C}, we have obtained one later NIR spectrum at +79~d with the Magellan telescope in Chile. The NIR spectral sequence, along with comparisons to SN~2011dh and SN~2008ax \citep[Type IIb;][]{2008MNRAS.389..955P,2011MNRAS.413.2140T}, are shown in Fig.~\ref{fig:spec13bvnir}.

In our NIR spectrum of iPTF13bvn obtained at +79~d, in addition to the \ion{He}{I} $\lambda$10,830 and $\lambda$20,590 lines previously identified by \cite{2013ApJ...775L...7C}, emission features around 12,000~\AA\ and 15,000~\AA\ have emerged. These are likely signatures of emission from \ion{He}{I} $\lambda$11,970 or \ion{Si}{I} $\lambda$12,000 and a blend of \ion{Mg}{I} $\lambda\lambda$15,040, 15,048. These identifications are based on the modeling of the NIR spectra of SN~2008ax at +130~d by \cite{2015A&amp;A...573A..12J}. Figure~\ref{fig:spec13bvnir} clearly shows that the spectral evolution in the NIR regime is very similar for iPTF13bvn and SN~2008ax. In the analysis by \cite{2015A&amp;A...573A..12J}, \ion{Si}{I} $\lambda$12,000 from explosively synthesized silicon is preferred over \ion{He}{I} $\lambda$11,970 for the line emerging at 12,000~\AA. We also estimate that the \ion{He}{I} velocities have decreased from $\sim 20,000$~km~s$^{-1}$ in the early-time spectra to $\sim 15,000$~km~s$^{-1}$ at +79~d, measured from the absorption minimum of the \ion{He}{I} $\lambda$10,830 line.

Early-time NIR spectra of SN~2011dh show strong signatures of hydrogen, as seen clearly by the presence of emission features at \ion{Paschen}{$\beta$} and \ion{Paschen}{$\delta$} at 12,820~\AA\ and 10,050~\AA\ (Fig.~\ref{fig:spec13bvnir}). \ion{Paschen}{$\gamma$} is blended with strong emission of \ion{He}{I} $\lambda$10,830. Note also that we do not have data for SN~2011dh in the region of \ion{Paschen}{$\alpha$} owing to telluric absorption.
 
For iPTF13bvn, no clear signatures of hydrogen can be seen. However, compared to an estimate of the continuum, found by heavily smoothing the spectra, there is a possible excess in both the +3~d and +9~d spectra around the \ion{Paschen}{$\beta$} line (Fig.~\ref{fig:spec13bvnir}). Furthermore, the NIR spectrum of the Type IIb SN~2008ax \citep[+10~d;][]{2011MNRAS.413.2140T} is very similar to the spectrum of iPTF13bvn at +9 d, especially in the regions around the Paschen lines. This indicates that the observed NIR spectra of iPTF13bvn do not exclude the possibility of the presence of a small amount of hydrogen at early times. The \ion{Paschen}{$\alpha$} line falls in a region with strong telluric absorption in all our NIR spectra of iPTF13bvn, so we are not able to draw conclusions about the presence of potential broad \ion{Paschen}{$\alpha$} emission. We discuss the presence of hydrogen in the optical spectra of both iPTF13bvn and PTF12os in the following section.

\subsection{Hydrogen in early spectra}
\label{sec:hydrogen}

The similarity of PTF12os to iPTF13bvn would logically lead to a Type Ib classification for PTF12os. However, in our early-time spectra of PTF12os we can identify likely signatures of H$\alpha$, H$\beta$, and H$\gamma$ in absorption at $\sim 14,000$~km~s$^{-1}$. This is illustrated by the dashed green lines in Fig.~\ref{fig:spec12os} and also by the dashed lines in Fig.~\ref{fig:earlyspec}, where we illustrate a comparison of early-time spectra of PTF12os, iPTF13bvn, and SN 2011dh. The spectrum of PTF12os at +19~d is very similar to spectra of the Type IIb SN~2011dh obtained at similar epochs (see also the bottom panel of Fig.~\ref{fig:spec12os}).

\begin{figure}
\centering
\includegraphics[width=4cm]{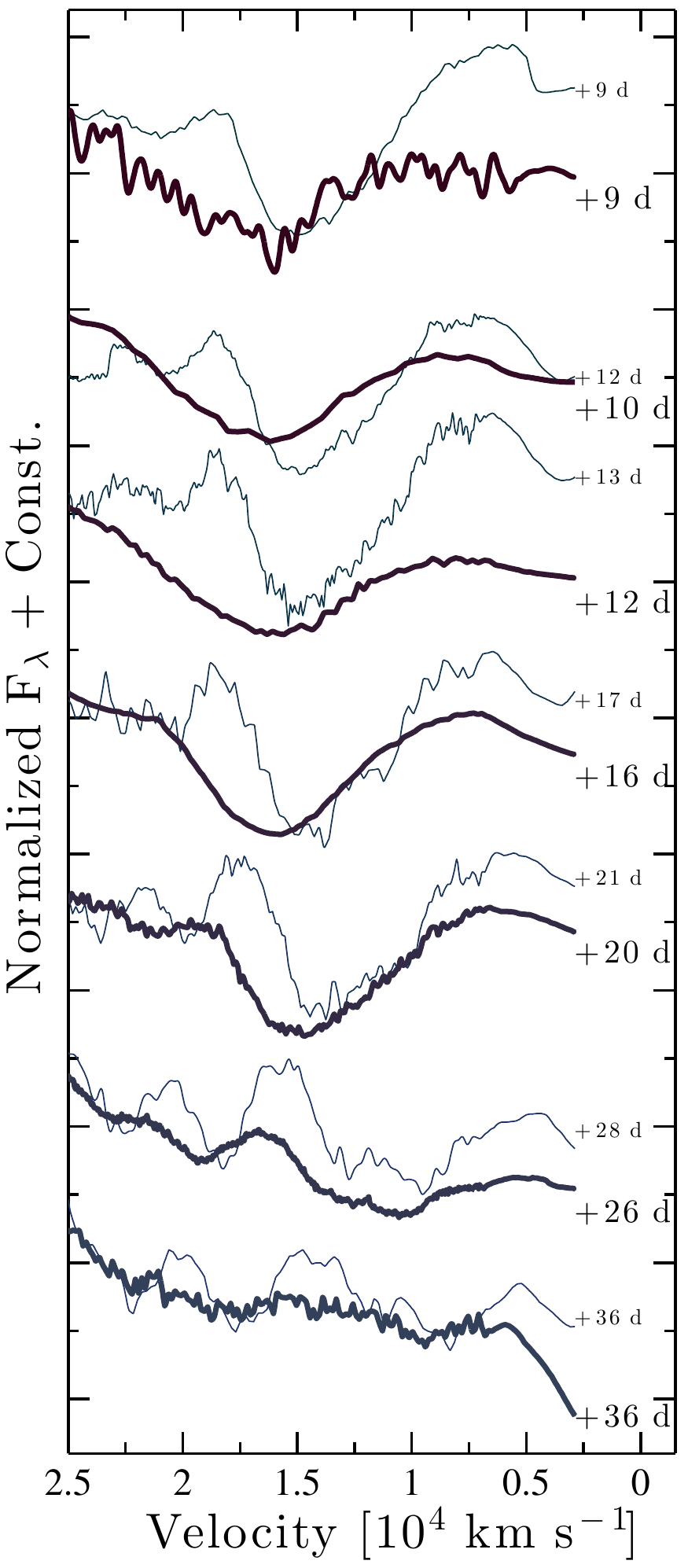}\includegraphics[width=4cm]{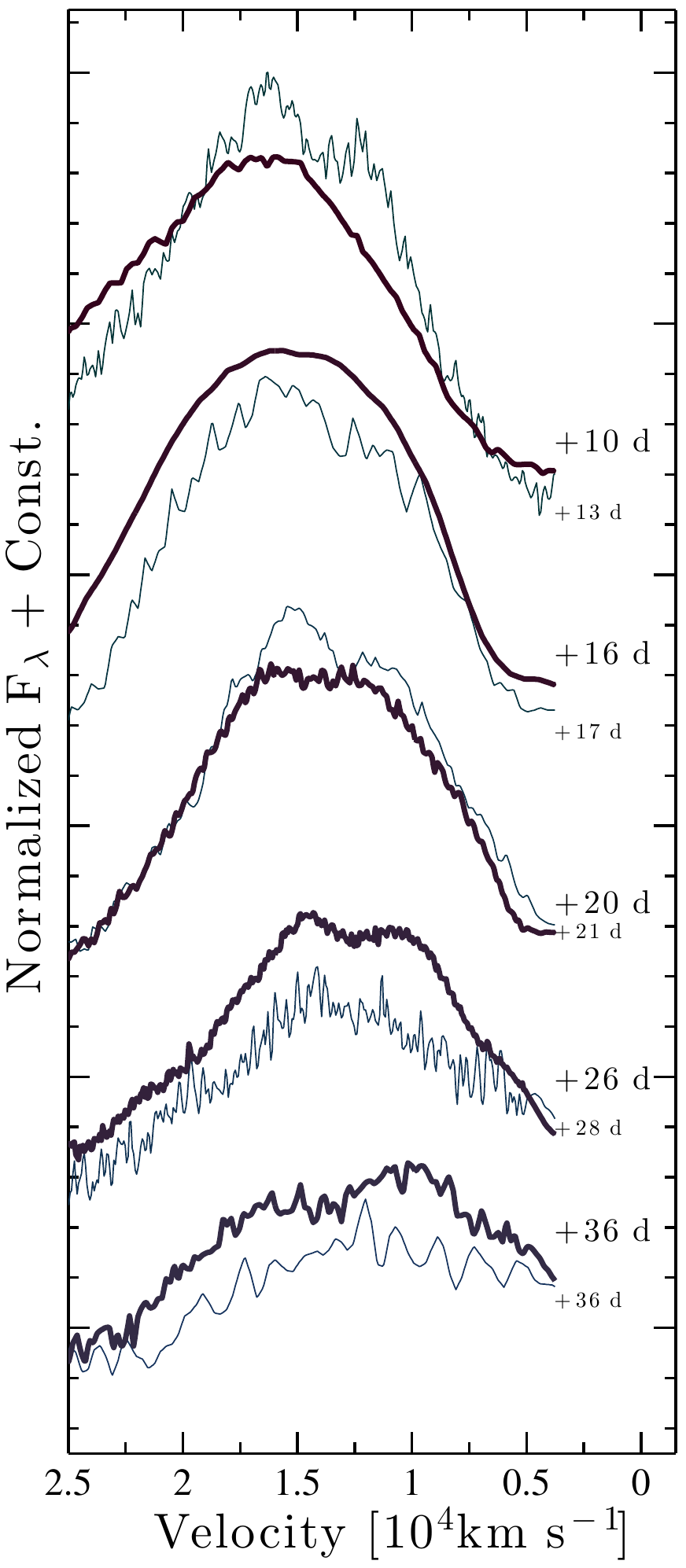}
\caption{Spectral evolution in velocity space around H$\alpha$ (left panel) and H$\beta$ (right panel) of iPTF13bvn (thick black line) and PTF12os (thin dark blue line).}
\label{fig:halphahbeta}
\end{figure}

In iPTF13bvn, no signatures of H$\beta$ or H$\gamma$ can be identified, but the spectra between +10~d and +25~d could still be consistent with the presence of H$\alpha$ since the spectral evolution is very similar to that of PTF12os, especially in this region; the absorption feature around 6300~\AA\ is comparable in strength and it disappears after $\sim$+25~d in both SNe (see Fig.~\ref{fig:halphahbeta}). For PTF12os, based on the comparison to SN~2011dh, it appears very likely that this feature is due to hydrogen, and thus it could be argued that it should be the case also for iPTF13bvn. The early-time NIR spectra of iPTF13bvn (Fig.~\ref{fig:spec13bvnir}) are also very similar to NIR spectra of the Type IIb SN~2008ax around the Paschen lines, so these data do not exclude the presence of a hydrogen envelope, as already discussed in Sect.~\ref{sec:13bvnnir}. The lack of H$\beta$ (and H$\gamma$) signatures in iPTF13bvn may possibly be explained by the higher expansion velocities, making the lines more difficult to identify. 

In spectra of SN~2011dh, a clear P-Cygni signature from H$\alpha$ at $\sim 6300$~\AA\ is seen in spectra obtained after +4~d. The strength of this feature compared to the continuum reached a maximum at around +25~d, whereafter it gradually weakened and disappeared after $\sim$+90~d, leading to the Type~IIb classification. Spectral signatures of H$\beta$ and H$\gamma$ with similar behaviors could also be identified. Monte-Carlo-based spectral modeling was used by \cite{2014A&amp;A...562A..17E} to estimate the mass of the hydrogen envelope of the progenitor of SN~2011dh as 0.01--0.04~\Msun, in agreement with the non-local thermodynamic equilibrium modeling by \cite{2011ApJ...742L..18A}, who estimate the hydrogen mass in the envelope of SN~2011dh to be $0.024$~\msun. 

Based on the sequence shown in Fig.~\ref{fig:earlyspec}, it is clear that much stronger hydrogen features developed in SN~2011dh compared with both PTF12os and iPTF13bvn. The absorption feature at 6300~\AA\ is there at +8.5~d in PTF12os (our earliest spectrum of this SN), and it becomes clear at around +10~d in iPTF13bvn. In both SNe, the absorption disappears after around +25~d (compared to  $\sim$+90~d in SN~2011dh). One interpretation of this spectral behavior is that we are seeing the signature of a hydrogen envelope around the progenitor, just as what is seen in SNe~IIb such as SN~2011dh. If we assume that this absorption is caused by the H$\alpha$ line, we can estimate the expansion velocity of the hydrogen envelope by measuring the position of the absorption minimum using Gaussian fits. We find that H$\alpha$ in iPTF13bvn would be at $\sim 16,000$~km~s$^{-1}$ at +10~d, about 2000~km~s$^{-1}$ faster compared to PTF12os (left panel of Fig.~\ref{fig:halphahbeta} and also Fig.~\ref{fig:vel_evol} for the velocity measurements). We have also investigated the region where we would expect H$\beta$ (right panel of Fig.~\ref{fig:halphahbeta}), and we find that for PTF12os there is absorption at $\sim 13,000$~km~s$^{-1}$ in several of the early spectra before +25~d. The velocity evolutions of the H$\alpha$ and H$\beta$ lines are consistent with what was observed in SN~2011dh, except that PTF12os and especially iPTF13bvn exhibit higher velocities. Faster expansion velocities are expected, since the mass of the hydrogen envelopes appear to be smaller in PTF12os and iPTF13bvn compared with SN~2011dh, given the weaker spectral signatures.

In the modeling effort by \cite{2012MNRAS.422...70H}, it was found that as little as $0.024$~\msun\ of hydrogen in the envelope is enough for H$\beta$ (and a strong signal from H$\alpha$) to show up in the spectra of a SN~IIb/Ib; thus, a hydrogen envelope mass upper limit of $\lesssim 0.02$~\Msun\ is a reasonable estimate for PTF12os (but probably rather high, given the very similar estimate for SN~2011dh, which shows much more long-lasting hydrogen signatures). For iPTF13bvn the hydrogen mass is likely lower, since we see no signs of H$\beta$. However, the higher velocities in combination with the fact that the H$\alpha$ absorption disappears at the same time as for PTF12os could be seen as an argument for at least a comparable hydrogen mass, since the emission from the same amount of hydrogen as in PTF12os should disappear earlier for a higher expansion velocity, assuming that the other physical parameters of the explosion are similar. Detailed physical modeling of the spectra of iPTF13bvn is needed to put a realistic upper limit on the hydrogen mass, especially since the modeling by \cite{2011MNRAS.414.2985D} found that even with an envelope mass as low as $0.001$~\msun, H$\alpha$ will still be present at +15~d --- a very small amount of hydrogen is needed to produce a weak spectral signature at early times.

In the spectral analysis of SE~SNe by \cite{2015arXiv150506645P}, it is also suggested that several other SNe~Ib in the literature besides iPTF13bvn show weak signatures of H$\alpha$ in absorption at early times. This conclusion is also consistent with predictions from binary evolution \citep[e.g.,][]{2015PASA...32...15Y}, where it is found that a small amount of hydrogen can remain in some cases for models otherwise tuned to produce SNe~Ib.

\cite{2013ApJ...767...71M} suggested that the Type IIb vs. Ib classification is time dependent. If the first spectra of PTF12os and iPTF13bvn had been obtained after +25~d, these objects would simply have been classified as SNe~Ib. With knowledge of the early-time spectral evolution they instead appear very similar to the Type IIb SN~2011dh, although likely with a smaller amount of hydrogen. 
In the Type IIb SN~2008ax, the hydrogen signatures disappeared after $\sim$+45~d. There could indeed be a continuum between SNe~IIb and SNe~Ib, with SE~SNe showing varying degrees of stripping of the progenitor. This would be in contrast with the previously suggested bimodal picture in which SNe~IIb and SNe~Ib come from different progenitor channels. 

However, recently \cite{2015arXiv151008049L} analyzed a large sample of SN~IIb and SN~Ib spectra, claiming that the strength of the H$\alpha$ absorption, as measured using the pseudoequivalent width \citep[pEW;][]{2011A&amp;A...526A..81B,2012MNRAS.425.1819S}, shows a bimodal distribution with a clear separation between SNe~IIb and SNe~Ib at all epochs, with SNe~IIb showing stronger absorption. \cite{2015arXiv151008049L} also find that iPTF13bvn seems to be an outlier; in terms of the strength of the H$\alpha$ absorption it is comparable to the SNe~IIb that exhibit the weakest hydrogen signatures. In Fig.~\ref{fig:halphahbeta} it is clear that the strength of H$\alpha$ is also similar for PTF12os at all epochs. Thus, it might indeed be that iPTF13bvn is a special case, and should perhaps be considered a SN~IIb. More early-time spectra of SNe~IIb/Ib, with well-determined explosion dates from early photometry, are needed to further investigate the nature of the hydrogen envelopes of their progenitors.

\begin{figure*}
\centering
\includegraphics[width=18cm]{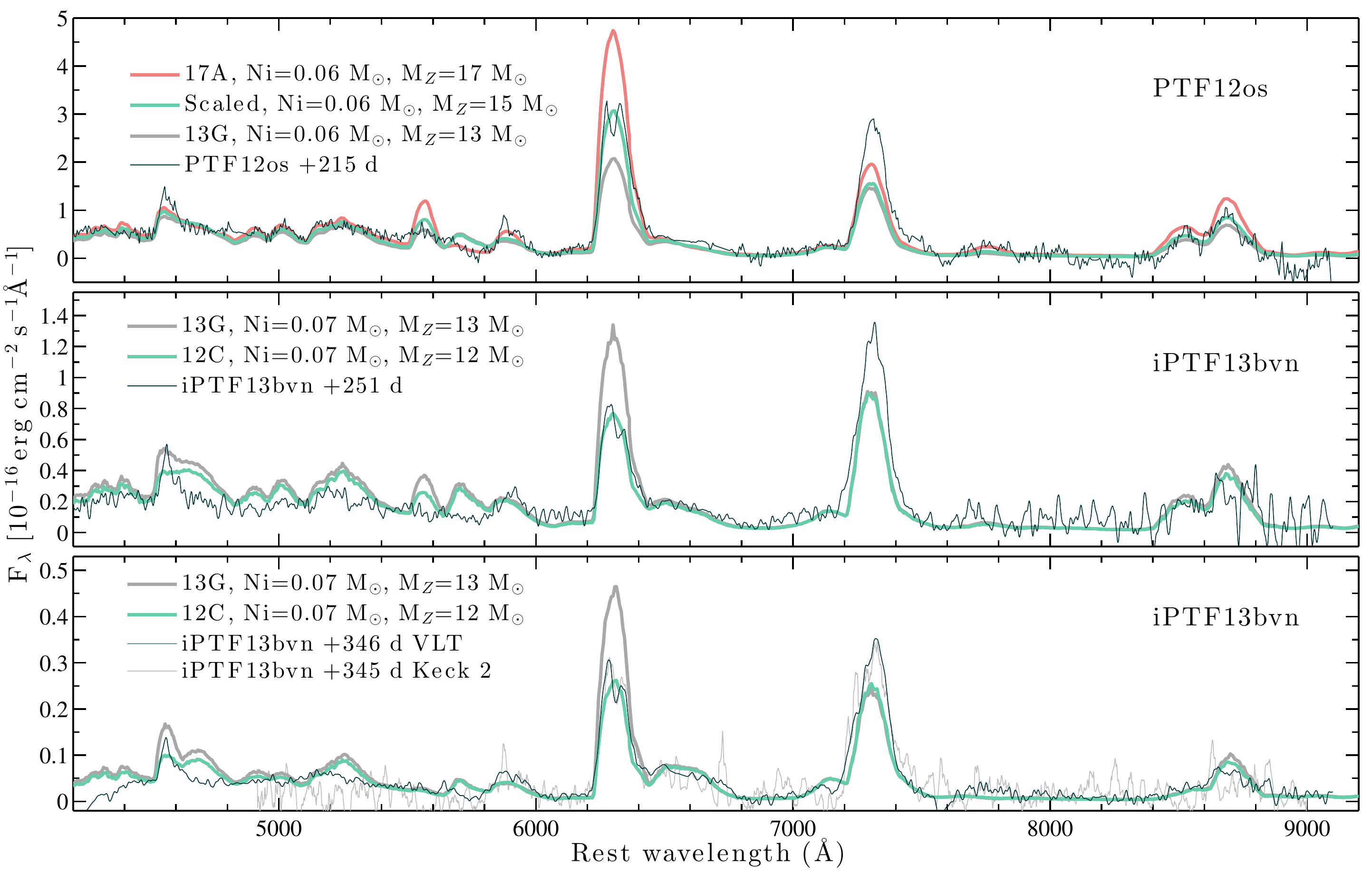}
\caption{Nebular spectra of PTF12os (top panel) and iPTF13bvn (middle and bottom panels) compared with nebular models. For PTF12os we show a flux-calibrated nebular spectrum (black line) obtained at +215~d. Nebular models with $M_{\rm ZAMS}=13$~\msun\ (gray line) and $M_{\rm ZAMS}=17$~\msun\ (red line) are also shown, along with a model scaled in oxygen mass to represent a star with $M_{\rm ZAMS}=15$~\msun\ (green line). The observed spectrum is consistent with the model representing $M_{\rm ZAMS}=15$~\msun. For iPTF13bvn we show three flux-calibrated spectra, one observed at +251~d (black line, middle panel), one at +345~d (gray line, bottom panel) and one at +346~d (black line, bottom panel). These spectra are consistent with a model having $M_{\rm ZAMS}=12$~\msun\ (green lines). All models have been scaled to the epochs of the spectra according to the prescriptions of \cite{2015A&amp;A...573A..12J}.}
\label{fig:neb}
\end{figure*}

\subsection{Nebular spectroscopy and oxygen-mass constraints}
\label{sec:nebular}

Figure~\ref{fig:neb} shows a comparison of late-time spectra of PTF12os (around +215~d) and iPTF13bvn (at +251~d, +345~d, and +346~d), along with fits based on the nebular models by \cite{2015A&amp;A...573A..12J} for SNe~IIb. The predicted luminosity of the [\ion{O}{I}] $\lambda\lambda$6300, 6364 doublet found by \cite{2015A&amp;A...573A..12J} was previously used by \cite{2014A&amp;A...565A.114F} to constrain the ZAMS mass of iPTF13bvn. This was based on nebular $r$-band photometry under the assumption that the $r$-band flux is dominated by the [\ion{O}{I}] $\lambda\lambda$6300, 6364 doublet. An upper limit of $\sim 17$~\Msun\ on the ZAMS mass of iPTF13bvn was derived. Since then, \cite{2015A&amp;A...579A..95K} have arrived at a similar limit on the ZAMS mass using independent nebular spectroscopy obtained at around +300~d. 

We find that our spectra obtained between +251~d and +346~d of iPTF13bvn give even stricter constraints. The same model that was found to be optimal for SN~2011dh \citep[model 12C\footnote{Model 12C has strong mixing, local positron trapping, and no molecular cooling, and it assumes that there is dust formation after +200~d \citep[for further details, see][]{2015A&amp;A...573A..12J}.} in][]{2015A&amp;A...573A..12J} produces good fits to the observed spectra of iPTF13bvn at these epochs, and especially to the [\ion{O}{I}] $\lambda\lambda$6300, 6364 doublet, is very well reproduced (see the middle and bottom panels of Fig.~\ref{fig:neb}). In this model the oxygen mass is $\sim 0.3$~\msun, which translates into a ZAMS mass of 12~\msun. The \ion{Mg}{I}] $\lambda$4571 line is also present in our spectrum at +251~d, and the 12~\msun\ model roughly reproduces this line as well. To produce the model fits for iPTF13bvn we have flux-calibrated our nebular spectra using the extinction-corrected $r$-band flux, scaled the model by the nickel mass derived for iPTF13bvn (0.072~\msun; see Sect.~\ref{sec:hydro}), and scaled the model in time to match the epoch of our spectra according to the prescriptions of \cite{2015A&amp;A...573A..12J}.

We note that in the binary model for iPTF13bvn by \cite{2014AJ....148...68B}, a relatively low oxygen mass is predicted, even though an initial mass of 20~\msun\ for the SN progenitor star is assumed. In a single-star setting, an oxygen mass around 2~\msun\ would be expected for such a star. However, in the model by \cite{2014AJ....148...68B}, the mass transfer to the companion starts during core hydrogen burning, which results in a reduced oxygen mass in the core of the SN progenitor. Thus, our ZAMS estimate of 12~\msun\ should be treated with some caution, since the models by \cite{2015A&amp;A...573A..12J} are based on the single-star nucleosynthesis models of \cite{2007PhR...442..269W}, and binary models can arrive at a \ion{He} core with similar properties to our best-fitting model at the time of explosion for larger initial masses \citep[see, e.g.,][]{2015MNRAS.446.2689E,Yoon:2010aa}. On the other hand, \cite{2016arXiv160405050E} have recently constructed a binary model for the iPTF13bvn system based on both pre-explosion and late-time (+740~d) \textit{HST} data, where the progenitor mass of the SN is constrained to 10--12~\msun, which is in excellent agreement with the result from our nebular modeling.

Following the same procedure as above, we have attempted to fit the same nebular model to PTF12os as for iPTF13bvn, but now scaled to the nickel mass for PTF12os (0.063~\msun; see Sect.~\ref{sec:hydro}). However, this model does not produce sufficiently strong emission lines for this SN (see the gray line in the top panel of Fig.~\ref{fig:neb}). A model with a ZAMS mass of 13~\msun\ \citep[model 13G of][]{2015A&amp;A...573A..12J} and otherwise identical parameters as the model used for iPTF13bvn gives decent matches to both the quasicontinuum and the emission lines of PTF12os (gray line in the top panel of Fig.~\ref{fig:neb}). On the other hand, the [\ion{O}{I}] $\lambda\lambda$6300, 6364 doublet is still too weak.

In Sect.~\ref{sec:bolometric} we found that while the quasibolometric LC peak is lower for PTF12os, the LC is declining slower compared to SN~2011dh past +80~d, and at +105~d the quasibolometric luminosity of both SNe is almost the same. At later times we lack coverage in the $B$ band; however, the fluxes in the $gri$ bands are almost the same in PTF12os at +215~d as in SN~2011dh. The slower decline of PTF12os implies more-efficient $\gamma$-ray trapping compared to SN~2011dh (and iPTF13bvn). If the nebular model were adjusted accordingly, the lines should be stronger for a lower nickel mass at late times.

One way to achieve more-efficient $\gamma$-ray trapping is to simply increase the ZAMS mass of the model, resulting in increased ejecta and oxygen masses. In Fig.~\ref{fig:neb} we also show a model for a star with ZAMS mass of 17~\msun\ \citep[model 17A of][]{2015A&amp;A...573A..12J}. The [\ion{O}{I}] $\lambda\lambda$6300, 6364 lines are significantly too strong in this model, and the ratio between this doublet and the [\ion{Ca}{II}] $\lambda\lambda$7291, 7323 doublet is significantly off compared to the observation. Thus, it appears that a ZAMS mass of 17~\msun\ is too large for PTF12os. If we interpolate the model spectra as a function of the oxygen mass (using models 12C, 13G, and 17A of \citealp{2015A&amp;A...573A..12J}), we find that the [\ion{O}{I}] $\lambda\lambda$6300, 6364 lines are very well reproduced by a model with an oxygen mass of $\sim 0.8$~\msun\ or a ZAMS mass of 15~\msun\ (see the green line in the top panel of Fig.~\ref{fig:neb}). We note that the \ion{He}{} core mass of a $15$~\msun\ model is still roughly consistent with the limits on the ejecta mass derived from our hydrodynamical model of PTF12os (Sect.~\ref{sec:hydro}).

In terms of luminosity, the [\ion{O}{I}] $\lambda\lambda$6300, 6364 doublet luminosity normalized by the energy released by the decaying nickel mass (0.063~\msun) is also significantly lower at both +140~d and +215~d in PTF12os than what is predicted for the model with a ZAMS mass of 17~\msun. We measure $L_{\mathrm{Norm}}\approx0.017$ at +140~d and $L_{\mathrm{Norm}}\approx0.026$ at +215~d, following the description of \cite{2015A&amp;A...573A..12J}. These numbers are roughly comparable to what has been measured for SN~2008ax and SN~1993J at similar epochs. SN~2008ax is well matched by a model with a ZAMS mass of 13\msun.

We have also compared the models presented above to nebular models of iPTF13bvn (for +346d) and PTF12os (for +215d) calculated with the spectral code by \citep{2007ApJ...661..892M,2010MNRAS.408...87M}. The results from these models are also consistent with low-mass ZAMS stars ($<17$~\msun) as the progenitors to these SNe. The best-fitting model of PTF12os has an oxygen mass of $0.73$~\msun, and the model of iPTF13bvn has an oxygen mass of $0.2$~\msun.

In conclusion, we arrive at an upper limit of 0.3~\msun\ for the oxygen mass of iPTF13bvn and 0.8~\msun\ for PTF12os. This translates to an upper limit on the ZAMS mass of 12~\msun\ for iPTF13bvn and 15~\msun\ for PTF12os for single-star ZAMS models.

\section{Hydrodynamical light-curve modeling and progenitor constraints}
\label{sec:hydro}

Following our earlier work on iPTF13bvn \citep{2014A&amp;A...565A.114F}, we have constructed a similar hydrodynamical model for PTF12os. As input data we use the multiband $UVM2$\footnote{We choose to use only the $UVM2$ band to estimate the UV contributions, since this {\it Swift}/UVOT band is least affected by leakage from redder wavelengths \citep[see][for a discussion]{2014A&amp;A...562A..17E}.}, $UB$, and $griz$ LCs of PTF12os, a rough explosion date estimate based on an exponential fit to the early-time $r$-band LC as an initial guess, and the measured photospheric velocities (Sect.~\ref{sec:spectra}), together with a grid of SN models constructed with the hydrodynamical code {\sc hyde} \citep{2015A&amp;A...580A.142E}. This is a one-dimensional code based on flux-limited diffusion, and it follows the framework outlined by \cite{1977ApJS...33..515F}.

To model PTF12os we have generated a new model grid, based on bare \element{He} cores evolved until the verge of core collapse using {\sc MESA} \citep{2010ascl.soft10083P}, and exploded with the latest version of the {\sc hyde} code. To allow direct comparisons we have also redone the fitting of iPTF13bvn and SN~2011dh to this same model grid.

For the IR and UV regions, at epochs where we lack coverage in our datasets on PTF12os and iPTF13bvn, we use bolometric corrections derived from the bolometric LC of SN~2011dh. When we perform the fits to our model grid, we place equal weights on the diffusion phase of the LC ($+1$~d to $+40$~d), the early tail of the LC ($+40$~d to $+100$~d), and the photospheric velocity measurements (which are fitted up to $+40$~d only). From the best-fitting models, shown in the bottom panel of Fig.~\ref{fig:bol_lcs} along with the estimated bolometric LCs, we derive the explosion parameters for PTF12os, iPTF13bvn, and SN~2011dh shown in Table~\ref{tab:hydro}. The results on both iPTF13bvn and SN~2011dh are consistent with previous findings \citep{2014A&amp;A...565A.114F,2014A&amp;A...562A..17E}.

For PTF12os we derive a compact He core of $3.25^{+0.77}_{-0.56}$~\msun, which corresponds to an ejecta mass of $1.9^{+0.77}_{-0.56}$~\msun\ assuming that a 1.4~\msun\ compact remnant remained at the center of the explosion. We also find that $0.063^{+0.020}_{-0.011}$~\msun\ of \element[][56]{Ni} was synthesized in the explosion which had a total kinetic energy of $0.54^{+0.41}_{-0.25}\times10^{51}$~erg, and that the nickel needs to be strongly mixed. 
If we let the explosion time vary by $\pm2.5$~d around the initial guess, we find the explosion epoch of the best-fitting model to be $t_{\rm exp}=2,455,933.0$~days in JD, or 2012 Jan. 6.5 in UT. The errors reported here and in Table~\ref{tab:hydro} are propagated from the uncertainties in the observed quantities, but we do not take the degeneracy between the explosion parameters into account (e.g., ejecta mass vs. explosion energy) when determining the errors.

Our models for all three SNe are qualitatively well fitted to the observed datasets, including both the overall LCs and the photospheric velocities (see Fig.~\ref{fig:vel_evol} for the photospheric-velocity fits for PTF12os and iPTF13bvn). In agreement with the discussion of the quasibolometric LCs in Sect.~\ref{sec:lc}, we find that iPTF13bvn and SN~2011dh has the highest nickel masses (0.072 and 0.075~\msun), followed by PTF12os (0.063~\msun). 

In terms of the ejecta mass, we find that PTF12os has the highest upper limit on the ejecta mass at 2.6~\msun\ and SN~2011dh the lowest at 2.4~\msun\ (see Table.~\ref{tab:hydro}). However, within the uncertainties the differences are not significant. We do not find a clearly lower ejecta mass for iPTF13bvn, although the width of the bolometric LC peak is narrower. This is because the expansion velocities are higher: a higher photospheric expansion velocity leads to a faster evolution of the LC around peak for the same ejecta mass. There is also some indication that iPTF13bvn could have a higher explosion energy compared to SN~2011dh and PTF12os, but within the errors the difference is not significant. Finally, past +70~d, the observed bolometric LC of PTF12os gradually starts to show a somewhat slower decline than what is seen in our best-fitting model. Models with slower late-time declines (higher ejecta mass, or lower energy, or lower velocities) fit less well to the rest of the data we are fitting. Similarly as to what we found when analyzing the quasibolometric LCs in Sect.~\ref{sec:bolometric}, this again indicates that the gamma-ray trapping could be more efficient in PTF12os compared to iPTF13bvn and SN~2011dh.

\begin{deluxetable}{lllcc}
\tabletypesize{}
\tablewidth{0pc}
\tablecaption{Explosion parameters from {\sc hyde}.\,\,\,\,\,\,\,\,\,\,\,\,\,\,\,\,\,\,\,\,\,\,\,\,\,\,\,\,\,\,\,\,\,\,\,\,\,\,\,\,\,\,\,\,\,\,\,\,\,\,\,\,\,\,\,\,\,\,\,\,\,\,\,\,\,\,\,\,\,\,\,\,\,\,\,\,\,\,\,\,\,\,\,\,\,\,\,\,\,\,\,\,\,\,\,\,\,\,\,\,\,\,\,\,\,\,\,\,\,\,\,\,\,\,\,\,\,\,\,\,\,\,\,\,\,\,\,\,\,\,\,\,\,\,\,\,\,\,\,\,\,\,\,\,\,\,\,\,\,\,\,\,\,\,\,\,\,\,\,\,\,\,\label{tab:hydro}}
\tablehead{
\colhead{SN} &
\colhead{$M_{\mathrm{He}}$ (\msun)}&
\colhead{$E$ ($10^{51}$ erg)}&
\colhead{$M_{\mathrm{Ni}}$ (\msun)}&
\colhead{Mix$_{\mathrm{Ni}}$}}
\startdata
SN~2011dh &  $3.31^{+0.53}_{-0.57}$&  $0.64^{+0.38}_{-0.31}$      & $0.075^{+0.028}_{-0.020}$& $1.05^{+0.08}_{-0.00}$    \\
&  &    &  &  \vspace{-0.2cm}     \\
PTF12os&   $3.25^{+0.77}_{-0.56}$&  $0.54^{+0.41}_{-0.25}$       &     $0.063^{+0.020}_{-0.011}$    &     $1.55^{+0.07}_{-0.16}$  \\
&  &    &  &     \vspace{-0.2cm}       \\
iPTF13bvn &   $3.38^{+0.57}_{-0.50}$&  $0.94^{+0.63}_{-0.43}$       &  $0.072^{+0.024}_{-0.016}$    &    $1.28^{+0.46}_{-0.00}$  \\
\enddata
\vspace{-0.7cm}
\tablecomments{\element[]{He} core mass, explosion energy, and mass and distribution of \element[][56]{Ni} derived from the bolometric light curves of PTF12os and iPTF13bvn presented in this work and from the bolometric LC of SN~2011dh by \cite{2014A&amp;A...562A..17E}. The errors include the uncertainties in the distance, extinction, and photospheric velocities, but do not take into account the degeneracy between the explosion parameters in the solutions.}
\end{deluxetable}

\section{Discussion, Summary, and Conclusions}
\label{sec:conclusions}

Our observational data on PTF12os and iPTF13bvn show that these two SNe are SE~SNe, with explosion parameters typical of Type IIb or Ib SNe. Compared to the Type IIb SN~2011dh, the multiband light curves, the bolometric light curves, and the spectral evolution of these three SNe are qualitatively very similar. This is reflected in the output when we fit these SNe to our hydrodynamical model grid (Sect.~\ref{sec:hydro}).

\begin{figure}
\centering
\includegraphics[width=16cm]{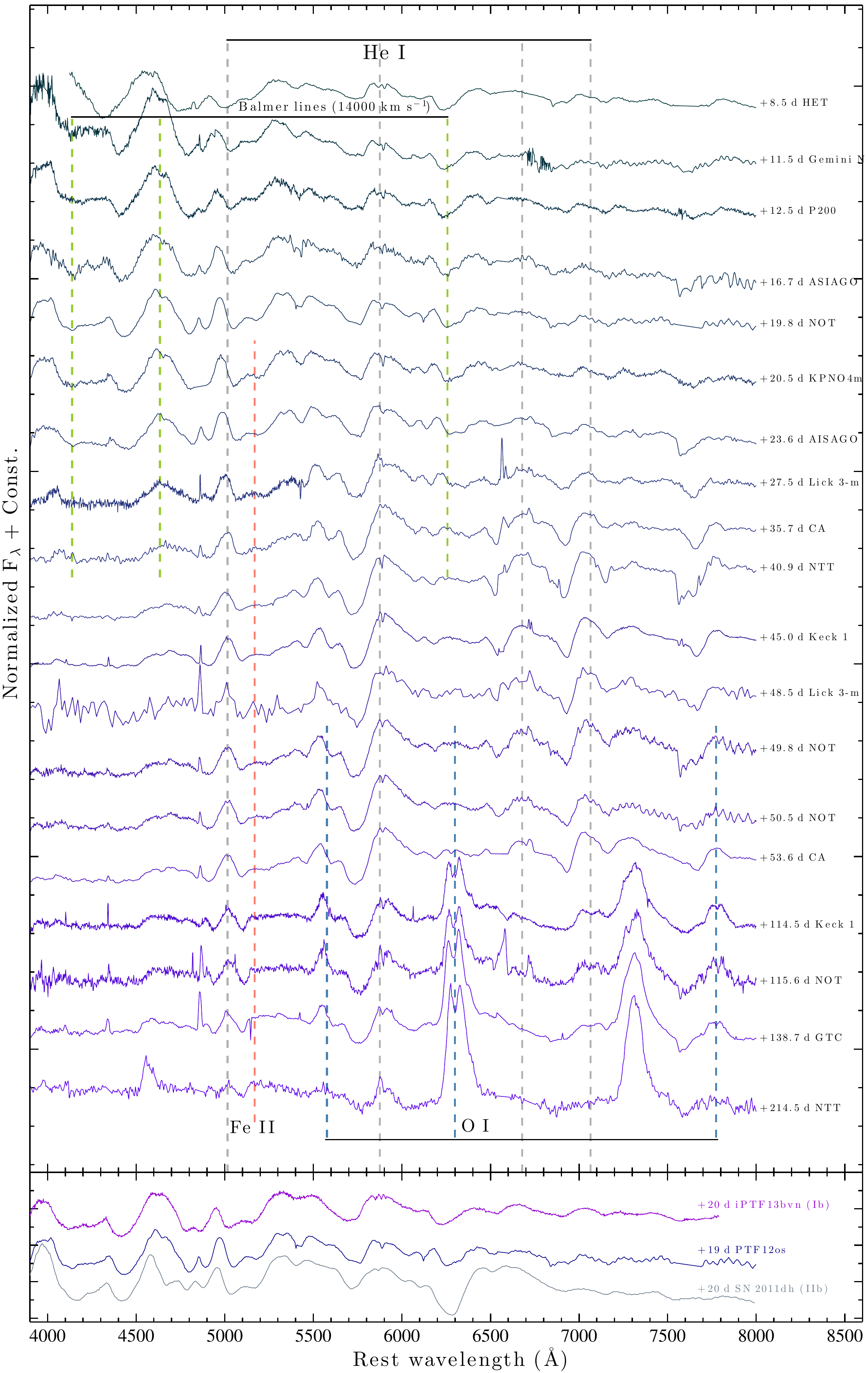}  
\vspace{0.2cm}
 \caption{\label{fig:spec12os}Spectral sequence of PTF12os (SN~2012P) (top panel). Comparison of the visible spectra of PTF12os to iPTF13bvn (Ib) and SN~2011dh (IIb) at approximately +20~d (bottom panel). Thick dashed lines mark the central wavelength of the marked emission lines at rest, except for the Balmer lines, which are marked by green dashed lines at $17, 000$ km~s$^{-1}$. Telluric features near the wavelengths 6860~\AA\, and especially 7600~\AA\, are present in some of the spectra .}
\end{figure}

We classify PTF12os as a SN~IIb, with only a small part ($\lesssim0.02$~\msun) of the hydrogen envelope of the progenitor star remaining prior to the explosion. This conclusion is based on the presence of absorption from several of the Balmer lines in our spectral sequence (Fig.~\ref{fig:spec12os}), and on the temporal evolution of the H$\alpha$ line. Compared to the sample studied by \cite{2015arXiv151008049L}, the velocity of the H$\alpha$ absorption in PTF12os is on the higher end, whereas the pEW of this absorption feature is on the lower end, of the SN~IIb sample. Compared to typical SNe~IIb such as SN~2011dh, the signature from hydrogen is weaker in the spectra of PTF12os (Fig.~\ref{fig:earlyspec}), and it disappeared quite early at around +25~d. Signatures of H$\alpha$ lingered in the spectra of SN~2011dh until approximately +90~d. All of these measurements indicate a lower mass of hydrogen in the envelope surrounding the progenitor of PTF12os compared to the progenitor of SN~2011dh. However, detailed modeling is needed to quantify the hydrogen envelope mass.

For iPTF13bvn the pEW of the H$\alpha$ absorption is close to that of PTF12os and it evolves in a similar way. This means that iPTF13bvn is among the SNe~Ib  with the largest pEW values in the sample studied by \cite{2015arXiv151008049L}, comparable in pEW to the lower end of the SN~IIb sample where PTF12os is also found. It is not completely unreasonable to assume that a similar amount of hydrogen was present in an envelope surrounding the progenitor of iPTF13bvn as for the progenitor of PTF12os. 

A similar mass of hydrogen as in PTF12os for iPTF13bvn would likely result in a radius of the progenitor prior to explosion that is significantly larger than the value of just a few solar radii reported by \cite{2013ApJ...775L...7C} by fitting the equations of \cite{Piro:2012aa} to the early $r$-band LC. When we fitted the temperature, the early bolometric LC, and the early $gri$ LCs of iPTF13bvn simultaneously using the same semianalytical framework, with the explosion parameters from our hydrodynamical model of iPTF13bvn as input, this resulted in a best-fitting radius for the progenitor on the order of 10~\rsun. Hydrodynamical modeling by \cite{2014AJ....148...68B} is also consistent with a significantly larger radius for iPTF13bvn. The binary models by \cite{Yoon:2010aa} show that it is likely that some SE~SNe, with progenitors very similar to those of hydrogen-free SNe~Ib formed in binary systems, will end up with a small part of the hydrogen envelope remaining and that these should spectroscopically be very similar to SNe~IIb. It could be that iPTF13bvn is one of these atypical SNe~Ib, and in that case it should perhaps be considered to be a SN~IIb. It was also pointed out that iPTF13bvn is a clear outlier in terms of the H$\alpha$ absorption strength in the sample studied by \cite{2015arXiv151008049L}. However, we should note that we do not see any sign of the Balmer lines in the spectra of iPTF13bvn other than possibly H$\alpha$.

Regarding our hydrodynamical modeling, there is always a concern that the ejecta masses of SNe~IIb/Ib could be underestimated from the LC-peak widths, owing to significant fallback onto a central black hole. However, in such a picture it becomes difficult to explain the oxygen masses derived from the late-time [\ion{O}{I}] $\lambda\lambda$6300, 6364 emission lines. If there is significant fallback, most of the oxygen should fall back into the black hole, but we find that the late-time oxygen emission of both PTF12os and iPTF13bvn is very consistent with low ZAMS-mass stars. The rest of the explosion parameters we find are also very consistent with those of other SNe~IIb and SNe~Ib in the literature.

Our study of the metallicity in the vicinity of PTF12os and iPTF13bvn shows that both of these SNe exploded in regions of comparable metallicity, with oxygen abundance very close to the solar value. A solar metallicity is consistent with the models by \cite{Yoon:2010aa} that predict a small amount of hydrogen remaining around some SNe~Ib. 

\cite{2016arXiv160405050E} have also constructed a binary model for the iPTF13bvn system based on pre-explosion and late-time (+740~d) \textit{HST} data, where the progenitor of the SN is constrained to a mass range of 10--12~\msun. This is very similar to the mass-constraint from our nebular modeling, and thus it seems that a very consistent picture of this SN is emerging. These \textit{HST} observations were likely still dominated by light from the SN. However, it is possible that in the near future the first direct detection of the light from a SN~Ib binary companion will be possible via another set of \textit{HST} observations.

Finally, we stress again that the overall appearance of PTF12os and iPTF13bvn is very similar indeed. Investigating two such SE~SNe in the same galaxy has allowed a detailed comparison of subtle differences, but generically both the LCs and the spectral evolution are astonishingly similar (and also to the well-studied SN 2011dh). We note that observations of polarization for both SN 2011dh \citep{2015MNRAS.453.4467M} and iPTF13bvn \citep{2015arXiv151002492R} have suggested significant asymmetry and line-of-sight effects. In terms of the overall observables for these SNe, such effects are not dominant.

\begin{acknowledgements}

We  gratefully  acknowledge support  from  the  Knut and  Alice  
Wallenberg  Foundation.  The  Oskar  Klein  Centre  is  funded  by  the
Swedish Research Council.
This research used resources of the National Energy Research Scientific Computing Center, a DOE Office of Science User Facility
supported  by  the  Office  of  Science  of  the  U.S.  Department  of  Energy  under Contract No. DE-AC02-05CH11231. 

We acknowledge help from Rob Fesen, Craig Wheeler, and Shazrene
Mohamed with the SALT data, as well as Jeffrey Silverman for
help with the Keck data.
We thank Alastair Bruce for assistance with the WHT observing and data reduction. 
We especially thank Shri Kulkarni, Mansi Kasliwal, Yi Cao, Anna-Lisa de Cia, Jerod Parrent, Assaf Horesh, Tom Matheson, Melissa Graham, Dan Perley, Eric Bellm, Ofer Yaron, Yen-Chen Pan and Kelsey Clubb for help with observations and/or reductions within the PTF effort.
We also acknowledge observers, organizers, and data reducers who participated in the BLP 
2012P campaign, in particular Andrea Pastorello, Max Stritzinger, Cosimo Inserra, Flora Cellier-Holzem, 
Luca Borsato, Valerio Nascimbeni, Stefano Benetti, and Stefan Taubenberger.
We thank Doug Leonard for discussions regarding complementary MLO data.

This work is partly based on observations obtained with the Nordic Optical Telescope, operated by the Nordic Optical Telescope Scientific Association at the Observatorio del Roque de los Muchachos, La Palma, Spain. We acknowledge the exceptional support we received from the NOT staff throughout this campaign. Also based in part on observations made with the Gran Telescopio Canarias (GTC), installed in the Spanish Observatorio del Roque de los Muchachos of the Instituto de Astrof{\i}sica de Canarias, in the island of La Palma. This work is based in part on observations from the LCOGT network. 
Some of the data presented herein were obtained at  the  W. M. Keck  Observatory,  which  is  operated  as  a  scientific  partnership among the California Institute of Technology, the University of California, and NASA; the observatory was made possible by the generous financial support of the W. M. Keck Foundation. Research at Lick Observatory is partially supported by a generous gift from Google.
CAFOS, AFOSC, and EFOSC2 data were taken within the European supernova
collaboration involved in the ESO-NTT large programme 184.D-1140 led by Stefano
Benetti. Partially based on observations collected at Copernico telescope 
(Asiago, Italy) of the INAF -- Osservatorio Astronomico di Padova, and the 
2.2~m Telescope of the Centro Astronomico Hispano-Aleman, Calar Alto, Spain.
Based in part on observations obtained at the Gemini Observatory, which is operated by the Association of Universities for Research in Astronomy, Inc., under a cooperative agreement with the NSF on behalf of the Gemini partnership: the National Science Foundation (United States), the National Research Council (Canada), CONICYT (Chile), Ministerio de Ciencia, Tecnolog\'{i}a e Innovaci\'{o}n Productiva (Argentina), and Minist\'{e}rio da Ci\^{e}ncia, Tecnologia e Inova\c{c}\~{a}o (Brazil).
The Hobby-Eberly Telescope (HET) is a joint project of the University of Texas at Austin, the Pennsylvania State University, Stanford University, Ludwig-Maximilians-Universit{\"a}t M{\"u}nchen, and Georg-August-Universit{\"a}t G{\"o}ttingen. The HET is named in honor of its principal benefactors, William P. Hobby and Robert E. Eberly. Some of the observations reported in this paper were obtained with the Southern African Large Telescope (SALT). This paper is partly based on observations made with the Italian Telescopio Nazionale Galileo (TNG) operated on the island of La Palma by the Fundaci{\'o}n Galileo Galilei of the INAF (Istituto Nazionale di Astrofisica) at the Spanish Observatorio del Roque de los Muchachos of the Instituto de Astrofisica de Canarias. Partially based on observations obtained with the Apache Point Observatory 3.5-meter telescope, which is owned and operated by the Astrophysical Research Consortium. This paper includes data gathered with the 6.5 meter Magellan Telescopes located at Las Campanas Observatory, Chile. Based in part on observations made with ESO Telescopes at the La Silla Paranal Observatory under programme 093.D-0199(A).
We are grateful for the assistance of the staff members at all observatories 
where we obtained data.

A.G.-Y. is supported by the EU/FP7 via ERC grant No. 307260, the Quantum Universe I-Core program by the Israeli Committee for  Planning  and  Budgeting;  by  Minerva  and  ISF  grants;  by  the Weizmann-UK making connections program; and by Kimmel and ARCHES awards.  
A.V.F.'s research is supported by NASA/{\it HST} grant AR-14295 from
the Space Telescope Science Institute, which is operated by the
Association of Universities for Research in Astronomy, Inc., under
NASA contract NAS5-26555; additional financial assistance was provided
by the Christopher R. Redlich Fund, the TABASGO Foundation, and NSF
grant AST-1211916.
N.E.R. is supported by the PRIN-INAF 2014 with the project: Transient Universe: unveiling new types of stellar explosions with PESSTO.
This work was partly supported by the European Union FP7 program though ERC grant number 320360.

We acknowledge funding from the European Research Council under the European Union's Seventh Framework Programme (FP7/2007-2013)/ERC Grant agreement n$^{\rm o}$ [291222].

\end{acknowledgements}

\bibliography{ngc5806_13bvn12os_accepted}

\begin{thebibliography}{102}
\expandafter\ifx\csname natexlab\endcsname\relax\def\natexlab#1{#1}\fi

\bibitem[{{Ahn} {et~al.}(2014){Ahn}, {Alexandroff}, {Allende Prieto}, {Anders},
  {Anderson}, {Anderton}, {Andrews}, {Aubourg}, {Bailey}, {Bastien}, \&
  et~al.}]{2014ApJS..211...17A}
{Ahn}, C.~P., {Alexandroff}, R., {Allende Prieto}, C., {et~al.} 2014, \apjs,
  211, 17

\bibitem[{{Arcavi} {et~al.}(2012){Arcavi}, {Gal-Yam}, {Ben-Ami}, {Fox},
  {Kasliwal}, {Nugent}, {Howell}, {Parrent}, {Horesh}, {Bellm}, \& {PTF
  Collaboration}}]{2012ATel.3881....1A}
{Arcavi}, I., {Gal-Yam}, A., {Ben-Ami}, S., {et~al.} 2012, The Astronomer's
  Telegram, 3881, 1

\bibitem[{{Arcavi} {et~al.}(2011){Arcavi}, {Gal-Yam}, {Yaron}, {Sternberg},
  {Rabinak}, {Waxman}, {Kasliwal}, {Quimby}, {Ofek}, {Horesh}, {Kulkarni},
  {Filippenko}, {Silverman}, {Cenko}, {Li}, {Bloom}, {Sullivan}, {Nugent},
  {Poznanski}, {Gorbikov}, {Fulton}, {Howell}, {Bersier}, {Riou},
  {Lamotte-Bailey}, {Griga}, {Cohen}, {Hachinger}, {Polishook}, {Xu},
  {Ben-Ami}, {Manulis}, {Walker}, {Maguire}, {Pan}, {Matheson}, {Mazzali},
  {Pian}, {Fox}, {Gehrels}, {Law}, {James}, {Marchant}, {Smith}, {Mottram},
  {Barnsley}, {Kandrashoff}, \& {Clubb}}]{2011ApJ...742L..18A}
{Arcavi}, I., {Gal-Yam}, A., {Yaron}, O., {et~al.} 2011, \apjl, 742, L18

\bibitem[{{Arcavi} {et~al.}(2013{\natexlab{a}}){Arcavi}, {Ofek}, {Cao},
  {Gelino}, {Yaron}, {Vreeswijk}, {Gal-Yam}, {Cenko}, {Kong}, \&
  {Li}}]{2013ATel.5152....1A}
{Arcavi}, I., {Ofek}, E., {Cao}, Y., {et~al.} 2013{\natexlab{a}}, The
  Astronomer's Telegram, 5152, 1

\bibitem[{{Arcavi} {et~al.}(2013{\natexlab{b}}){Arcavi}, {Yaron}, {Gal-Yam}, \&
  {Cao}}]{2013ATel.5140....1A}
{Arcavi}, I., {Yaron}, O., {Gal-Yam}, A., \& {Cao}, Y. 2013{\natexlab{b}}, The
  Astronomer's Telegram, 5140, 1

\bibitem[{{Asplund} {et~al.}(2009){Asplund}, {Grevesse}, {Sauval}, \&
  {Scott}}]{2009ARA&amp;A..47..481A}
{Asplund}, M., {Grevesse}, N., {Sauval}, A.~J., \& {Scott}, P. 2009, \araa, 47,
  481

\bibitem[{{Benvenuto} {et~al.}(2013){Benvenuto}, {Bersten}, \&
  {Nomoto}}]{2013ApJ...762...74B}
{Benvenuto}, O.~G., {Bersten}, M.~C., \& {Nomoto}, K. 2013, \apj, 762, 74

\bibitem[{{Bersten} {et~al.}(2014){Bersten}, {Benvenuto}, {Folatelli},
  {Nomoto}, {Kuncarayakti}, {Srivastav}, {Anupama}, {Quimby}, \&
  {Sahu}}]{2014AJ....148...68B}
{Bersten}, M.~C., {Benvenuto}, O.~G., {Folatelli}, G., {et~al.} 2014, \aj, 148,
  68

\bibitem[{{Bersten} {et~al.}(2012){Bersten}, {Benvenuto}, {Nomoto}, {Ergon},
  {Folatelli}, {Sollerman}, {Benetti}, {Botticella}, {Fraser}, {Kotak},
  {Maeda}, {Ochner}, \& {Tomasella}}]{2012ApJ...757...31B}
{Bersten}, M.~C., {Benvenuto}, O.~G., {Nomoto}, K., {et~al.} 2012, \apj, 757,
  31

\bibitem[{{Bertin}(2013)}]{2013ascl.soft01001B}
{Bertin}, E. 2013, {PSFEx: Point Spread Function Extractor}, Astrophysics
  Source Code Library

\bibitem[{{Bertin} \& {Arnouts}(1996)}]{1996A&amp;AS..117..393B}
{Bertin}, E. \& {Arnouts}, S. 1996, \aaps, 117, 393

\bibitem[{{Blondin} {et~al.}(2011){Blondin}, {Mandel}, \&
  {Kirshner}}]{2011A&amp;A...526A..81B}
{Blondin}, S., {Mandel}, K.~S., \& {Kirshner}, R.~P. 2011, \aap, 526, A81

\bibitem[{{Brown} {et~al.}(2009){Brown}, {Holland}, {Immler}, {Milne},
  {Roming}, {Gehrels}, {Nousek}, {Panagia}, {Still}, \& {Vanden
  Berk}}]{2009AJ....137.4517B}
{Brown}, P.~J., {Holland}, S.~T., {Immler}, S., {et~al.} 2009, \aj, 137, 4517

\bibitem[{{Brown} {et~al.}(2013){Brown}, {Baliber}, {Bianco}, {Bowman},
  {Burleson}, {Conway}, {Crellin}, {Depagne}, {De Vera}, {Dilday}, {Dragomir},
  {Dubberley}, {Eastman}, {Elphick}, {Falarski}, {Foale}, {Ford}, {Fulton},
  {Garza}, {Gomez}, {Graham}, {Greene}, {Haldeman}, {Hawkins}, {Haworth},
  {Haynes}, {Hidas}, {Hjelstrom}, {Howell}, {Hygelund}, {Lister}, {Lobdill},
  {Martinez}, {Mullins}, {Norbury}, {Parrent}, {Paulson}, {Petry}, {Pickles},
  {Posner}, {Rosing}, {Ross}, {Sand}, {Saunders}, {Shobbrook}, {Shporer},
  {Street}, {Thomas}, {Tsapras}, {Tufts}, {Valenti}, {Vander Horst}, {Walker},
  {White}, \& {Willis}}]{2013PASP..125.1031B}
{Brown}, T.~M., {Baliber}, N., {Bianco}, F.~B., {et~al.} 2013, \pasp, 125, 1031

\bibitem[{{Cano}(2013)}]{Cano:2013aa}
{Cano}, Z. 2013, \mnras, 434, 1098

\bibitem[{Cao {et~al.}(2013)Cao, Kasliwal, Arcavi, Horesh, Hancock, Valenti,
  Cenko, Kulkarni, Gal-Yam, Gorbikov, Ofek, Sand, Yaron, Graham, Silverman,
  Wheeler, Marion, Walker, Mazzali, Howell, Bloom, Nugent, Surace, Masci,
  Carpenter, Degenaar, \& Gelino}]{Cao:2013aa}
Cao, Y., Kasliwal, M.~M., Arcavi, I., {et~al.} 2013, ApJL, 775 (2013), L7

\bibitem[{{Cao} {et~al.}(2013){Cao}, {Kasliwal}, {Arcavi}, {Horesh}, {Hancock},
  {Valenti}, {Cenko}, {Kulkarni}, {Gal-Yam}, {Gorbikov}, {Ofek}, {Sand},
  {Yaron}, {Graham}, {Silverman}, {Wheeler}, {Marion}, {Walker}, {Mazzali},
  {Howell}, {Li}, {Kong}, {Bloom}, {Nugent}, {Surace}, {Masci}, {Carpenter},
  {Degenaar}, \& {Gelino}}]{2013ApJ...775L...7C}
{Cao}, Y., {Kasliwal}, M.~M., {Arcavi}, I., {et~al.} 2013, \apjl, 775, L7

\bibitem[{{Cardelli} {et~al.}(1989){Cardelli}, {Clayton}, \&
  {Mathis}}]{1989ApJ...345..245C}
{Cardelli}, J.~A., {Clayton}, G.~C., \& {Mathis}, J.~S. 1989, \apj, 345, 245

\bibitem[{{Cenko} {et~al.}(2006){Cenko}, {Fox}, {Moon}, {Harrison}, {Kulkarni},
  {Henning}, {Guzman}, {Bonati}, {Smith}, {Thicksten}, {Doyle}, {Petrie},
  {Gal-Yam}, {Soderberg}, {Anagnostou}, \& {Laity}}]{2006PASP..118.1396C}
{Cenko}, S.~B., {Fox}, D.~B., {Moon}, D.-S., {et~al.} 2006, \pasp, 118, 1396

\bibitem[{{Claeys} {et~al.}(2011){Claeys}, {de Mink}, {Pols}, {Eldridge}, \&
  {Baes}}]{2011A&amp;A...528A.131C}
{Claeys}, J.~S.~W., {de Mink}, S.~E., {Pols}, O.~R., {Eldridge}, J.~J., \&
  {Baes}, M. 2011, \aap, 528, A131

\bibitem[{{Conti}(1976)}]{conti76}
{Conti}, P. 1976, MSRSL, 9, 193

\bibitem[{{de Vaucouleurs} {et~al.}(1991){de Vaucouleurs}, {de Vaucouleurs},
  {Corwin}, {Buta}, {Paturel}, \& {Fouqu{\'e}}}]{1991rc3..book.....D}
{de Vaucouleurs}, G., {de Vaucouleurs}, A., {Corwin}, Jr., H.~G., {et~al.}
  1991, {Third Reference Catalogue of Bright Galaxies. Volume I: Explanations
  and references. Volume II: Data for galaxies between 0$^{h}$ and 12$^{h}$.
  Volume III: Data for galaxies between 12$^{h}$ and 24$^{h}$.}

\bibitem[{{Dessart} \& {Hillier}(2005)}]{Dessart:aa}
{Dessart}, L. \& {Hillier}, D.~J. 2005, \aap, 439, 671

\bibitem[{{Dessart} {et~al.}(2012){Dessart}, {Hillier}, {Li}, \&
  {Woosley}}]{2012MNRAS.424.2139D}
{Dessart}, L., {Hillier}, D.~J., {Li}, C., \& {Woosley}, S. 2012, \mnras, 424,
  2139

\bibitem[{{Dessart} {et~al.}(2011){Dessart}, {Hillier}, {Livne}, {Yoon},
  {Woosley}, {Waldman}, \& {Langer}}]{2011MNRAS.414.2985D}
{Dessart}, L., {Hillier}, D.~J., {Livne}, E., {et~al.} 2011, \mnras, 414, 2985

\bibitem[{{Dimai} {et~al.}(2012){Dimai}, {Briganti}, \&
  {Brimacombe}}]{2012CBET.2993....1D}
{Dimai}, A., {Briganti}, F., \& {Brimacombe}, J. 2012, Central Bureau
  Electronic Telegrams, 2993, 1

\bibitem[{{Dolphin}(2000)}]{2000PASP..112.1383D}
{Dolphin}, A.~E. 2000, \pasp, 112, 1383

\bibitem[{{Eldridge} {et~al.}(2015){Eldridge}, {Fraser}, {Maund}, \&
  {Smartt}}]{2015MNRAS.446.2689E}
{Eldridge}, J.~J., {Fraser}, M., {Maund}, J.~R., \& {Smartt}, S.~J. 2015,
  \mnras, 446, 2689

\bibitem[{{Eldridge} {et~al.}(2013){Eldridge}, {Fraser}, {Smartt}, {Maund}, \&
  {Crockett}}]{Eldridge:2013aa}
{Eldridge}, J.~J., {Fraser}, M., {Smartt}, S.~J., {Maund}, J.~R., \&
  {Crockett}, R.~M. 2013, \mnras, 436, 774

\bibitem[{{Eldridge} \& {Maund}(2016)}]{2016arXiv160405050E}
{Eldridge}, J.~J. \& {Maund}, J.~R. 2016, ArXiv e-prints

\bibitem[{{Eldridge} \& {Stanway}(2009)}]{2009MNRAS.400.1019E}
{Eldridge}, J.~J. \& {Stanway}, E.~R. 2009, \mnras, 400, 1019

\bibitem[{{Ergon} {et~al.}(2015){Ergon}, {Jerkstrand}, {Sollerman},
  {Elias-Rosa}, {Fransson}, {Fraser}, {Pastorello}, {Kotak}, {Taubenberger},
  {Tomasella}, {Valenti}, {Benetti}, {Helou}, {Kasliwal}, {Maund}, {Smartt}, \&
  {Spyromilio}}]{2015A&amp;A...580A.142E}
{Ergon}, M., {Jerkstrand}, A., {Sollerman}, J., {et~al.} 2015, \aap, 580, A142

\bibitem[{{Ergon} {et~al.}(2014){Ergon}, {Sollerman}, {Fraser}, {Pastorello},
  {Taubenberger}, {Elias-Rosa}, {Bersten}, {Jerkstrand}, {Benetti},
  {Botticella}, {Fransson}, {Harutyunyan}, {Kotak}, {Smartt}, {Valenti},
  {Bufano}, {Cappellaro}, {Fiaschi}, {Howell}, {Kankare}, {Magill}, {Mattila},
  {Maund}, {Naves}, {Ochner}, {Ruiz}, {Smith}, {Tomasella}, \&
  {Turatto}}]{2014A&amp;A...562A..17E}
{Ergon}, M., {Sollerman}, J., {Fraser}, M., {et~al.} 2014, \aap, 562, A17

\bibitem[{{Faber} {et~al.}(2003){Faber}, {Phillips}, {Kibrick}, {Alcott},
  {Allen}, {Burrous}, {Cantrall}, {Clarke}, {Coil}, {Cowley}, {Davis}, {Deich},
  {Dietsch}, {Gilmore}, {Harper}, {Hilyard}, {Lewis}, {McVeigh}, {Newman},
  {Osborne}, {Schiavon}, {Stover}, {Tucker}, {Wallace}, {Wei}, {Wirth}, \&
  {Wright}}]{2003SPIE.4841.1657F}
{Faber}, S.~M., {Phillips}, A.~C., {Kibrick}, R.~I., {et~al.} 2003, in Society
  of Photo-Optical Instrumentation Engineers (SPIE) Conference Series, Vol.
  4841, Instrument Design and Performance for Optical/Infrared Ground-based
  Telescopes, ed. M.~{Iye} \& A.~F.~M. {Moorwood}, 1657--1669

\bibitem[{{Falk} \& {Arnett}(1977)}]{1977ApJS...33..515F}
{Falk}, S.~W. \& {Arnett}, W.~D. 1977, \apjs, 33, 515

\bibitem[{{Filippenko}(1997)}]{1997ARA&amp;A..35..309F}
{Filippenko}, A.~V. 1997, \araa, 35, 309

\bibitem[{{Folatelli} {et~al.}(2014{\natexlab{a}}){Folatelli}, {Bersten},
  {Benvenuto}, {Van Dyk}, {Kuncarayakti}, {Maeda}, {Nozawa}, {Nomoto}, {Hamuy},
  \& {Quimby}}]{2014ApJ...793L..22F}
{Folatelli}, G., {Bersten}, M.~C., {Benvenuto}, O.~G., {et~al.}
  2014{\natexlab{a}}, \apjl, 793, L22

\bibitem[{{Folatelli} {et~al.}(2014{\natexlab{b}}){Folatelli}, {Bersten},
  {Kuncarayakti}, {Olivares Estay}, {Anderson}, {Holmbo}, {Maeda}, {Morrell},
  {Nomoto}, {Pignata}, {Stritzinger}, {Contreras}, {F{\"o}rster}, {Hamuy},
  {Phillips}, {Prieto}, {Valenti}, {Afonso}, {Altenm{\"u}ller}, {Elliott},
  {Greiner}, {Updike}, {Haislip}, {LaCluyze}, {Moore}, \&
  {Reichart}}]{2014ApJ...792....7F}
{Folatelli}, G., {Bersten}, M.~C., {Kuncarayakti}, H., {et~al.}
  2014{\natexlab{b}}, \apj, 792, 7

\bibitem[{{Folatelli} {et~al.}(2016){Folatelli}, {Van Dyk}, {Kuncarayakti},
  {Maeda}, {Bersten}, {Nomoto}, {Pignata}, {Hamuy}, {Quimby}, {Zheng},
  {Filippenko}, {Clubb}, {Smith}, {Elias-Rosa}, {Foley}, \&
  {Miller}}]{2016arXiv160406821F}
{Folatelli}, G., {Van Dyk}, S.~D., {Kuncarayakti}, H., {et~al.} 2016, ArXiv
  e-prints

\bibitem[{{Fox} {et~al.}(2014){Fox}, {Azalee Bostroem}, {Van Dyk},
  {Filippenko}, {Fransson}, {Matheson}, {Cenko}, {Chandra}, {Dwarkadas}, {Li},
  {Parker}, \& {Smith}}]{2014ApJ...790...17F}
{Fox}, O.~D., {Azalee Bostroem}, K., {Van Dyk}, S.~D., {et~al.} 2014, \apj,
  790, 17

\bibitem[{{Fremling} {et~al.}(2014){Fremling}, {Sollerman}, {Taddia}, {Ergon},
  {Valenti}, {Arcavi}, {Ben-Ami}, {Cao}, {Cenko}, {Filippenko}, {Gal-Yam}, \&
  {Howell}}]{2014A&amp;A...565A.114F}
{Fremling}, C., {Sollerman}, J., {Taddia}, F., {et~al.} 2014, \aap, 565, A114

\bibitem[{{Gal-Yam} {et~al.}(2008){Gal-Yam}, {Maoz}, {Guhathakurta}, \&
  {Filippenko}}]{2008ApJ...680..550G}
{Gal-Yam}, A., {Maoz}, D., {Guhathakurta}, P., \& {Filippenko}, A.~V. 2008,
  \apj, 680, 550

\bibitem[{{Gehrels} {et~al.}(2004){Gehrels}, {Chincarini}, {Giommi}, {Mason},
  {Nousek}, {Wells}, {White}, {Barthelmy}, {Burrows}, {Cominsky}, {Hurley},
  {Marshall}, {M{\'e}sz{\'a}ros}, {Roming}, {Angelini}, {Barbier}, {Belloni},
  {Campana}, {Caraveo}, {Chester}, {Citterio}, {Cline}, {Cropper}, {Cummings},
  {Dean}, {Feigelson}, {Fenimore}, {Frail}, {Fruchter}, {Garmire}, {Gendreau},
  {Ghisellini}, {Greiner}, {Hill}, {Hunsberger}, {Krimm}, {Kulkarni}, {Kumar},
  {Lebrun}, {Lloyd-Ronning}, {Markwardt}, {Mattson}, {Mushotzky}, {Norris},
  {Osborne}, {Paczynski}, {Palmer}, {Park}, {Parsons}, {Paul}, {Rees},
  {Reynolds}, {Rhoads}, {Sasseen}, {Schaefer}, {Short}, {Smale}, {Smith},
  {Stella}, {Tagliaferri}, {Takahashi}, {Tashiro}, {Townsley}, {Tueller},
  {Turner}, {Vietri}, {Voges}, {Ward}, {Willingale}, {Zerbi}, \&
  {Zhang}}]{2004ApJ...611.1005G}
{Gehrels}, N., {Chincarini}, G., {Giommi}, P., {et~al.} 2004, \apj, 611, 1005

\bibitem[{{Groh} {et~al.}(2013){Groh}, {Georgy}, \&
  {Ekstr{\"o}m}}]{Groh:2013aa}
{Groh}, J.~H., {Georgy}, C., \& {Ekstr{\"o}m}, S. 2013, \aap, 558, L1

\bibitem[{Groh {et~al.}(2013)Groh, Georgy, \& Ekstrom}]{Groh:2013aa2}
Groh, J.~H., Georgy, C., \& Ekstrom, S. 2013

\bibitem[{Gusev {et~al.}(2012)Gusev, Pilyugin, Sakhibov, Dodonov, Ezhkova, \&
  Khramtsova}]{Gusev:2012aa}
Gusev, A.~S., Pilyugin, L.~S., Sakhibov, F., {et~al.} 2012

\bibitem[{{Hachinger} {et~al.}(2012){Hachinger}, {Mazzali}, {Taubenberger},
  {Hillebrandt}, {Nomoto}, \& {Sauer}}]{2012MNRAS.422...70H}
{Hachinger}, S., {Mazzali}, P.~A., {Taubenberger}, S., {et~al.} 2012, \mnras,
  422, 70

\bibitem[{{Iben} \& {Tutukov}(1985)}]{1985ApJS...58..661I}
{Iben}, Jr., I. \& {Tutukov}, A.~V. 1985, \apjs, 58, 661

\bibitem[{Jerkstrand {et~al.}(2014b)Jerkstrand, Ergon, Smartt, Fransson,
  Sollerman, Taubenberger, \& Spyromilio}]{jerkstrand2014}
Jerkstrand, A., Ergon, M., Smartt, S., {et~al.} 2014b, \aap, submitted

\bibitem[{{Jerkstrand} {et~al.}(2015){Jerkstrand}, {Ergon}, {Smartt},
  {Fransson}, {Sollerman}, {Taubenberger}, {Bersten}, \&
  {Spyromilio}}]{2015A&amp;A...573A..12J}
{Jerkstrand}, A., {Ergon}, M., {Smartt}, S.~J., {et~al.} 2015, \aap, 573, A12

\bibitem[{{Jester} {et~al.}(2005){Jester}, {Schneider}, {Richards}, {Green},
  {Schmidt}, {Hall}, {Strauss}, {Vanden Berk}, {Stoughton}, {Gunn},
  {Brinkmann}, {Kent}, {Smith}, {Tucker}, \& {Yanny}}]{2005AJ....130..873J}
{Jester}, S., {Schneider}, D.~P., {Richards}, G.~T., {et~al.} 2005, \aj, 130,
  873

\bibitem[{{Kankare} {et~al.}(2015){Kankare}, {Kotak}, {Pastorello}, {Fraser},
  {Mattila}, {Smartt}, {Bruce}, {Chambers}, {Elias-Rosa}, {Flewelling},
  {Fremling}, {Harmanen}, {Huber}, {Jerkstrand}, {Kangas}, {Kuncarayakti},
  {Magee}, {Magnier}, {Polshaw}, {Smith}, {Sollerman}, \&
  {Tomasella}}]{2015A&amp;A...581L...4K}
{Kankare}, E., {Kotak}, R., {Pastorello}, A., {et~al.} 2015, \aap, 581, L4

\bibitem[{{Kelly} \& {Kirshner}(2012)}]{2012ApJ...759..107K}
{Kelly}, P.~L. \& {Kirshner}, R.~P. 2012, \apj, 759, 107

\bibitem[{{Kuncarayakti} {et~al.}(2015){Kuncarayakti}, {Maeda}, {Bersten},
  {Folatelli}, {Morrell}, {Hsiao}, {Gonz{\'a}lez-Gait{\'a}n}, {Anderson},
  {Hamuy}, {de Jaeger}, {Guti{\'e}rrez}, \&
  {Kawabata}}]{2015A&amp;A...579A..95K}
{Kuncarayakti}, H., {Maeda}, K., {Bersten}, M.~C., {et~al.} 2015, \aap, 579,
  A95

\bibitem[{{Landolt}(1992)}]{1992AJ....104..340L}
{Landolt}, A.~U. 1992, \aj, 104, 340

\bibitem[{{Law} {et~al.}(2009){Law}, {Kulkarni}, {Dekany}, {Ofek}, {Quimby},
  {Nugent}, {Surace}, {Grillmair}, {Bloom}, {Kasliwal}, {Bildsten}, {Brown},
  {Cenko}, {Ciardi}, {Croner}, {Djorgovski}, {van Eyken}, {Filippenko}, {Fox},
  {Gal-Yam}, {Hale}, {Hamam}, {Helou}, {Henning}, {Howell}, {Jacobsen},
  {Laher}, {Mattingly}, {McKenna}, {Pickles}, {Poznanski}, {Rahmer}, {Rau},
  {Rosing}, {Shara}, {Smith}, {Starr}, {Sullivan}, {Velur}, {Walters}, \&
  {Zolkower}}]{Law:2009aa}
{Law}, N.~M., {Kulkarni}, S.~R., {Dekany}, R.~G., {et~al.} 2009, \pasp, 121,
  1395

\bibitem[{{Leitherer} {et~al.}(1999){Leitherer}, {Schaerer}, {Goldader},
  {Delgado}, {Robert}, {Kune}, {de Mello}, {Devost}, \&
  {Heckman}}]{1999ApJS..123....3L}
{Leitherer}, C., {Schaerer}, D., {Goldader}, J.~D., {et~al.} 1999, \apjs, 123,
  3

\bibitem[{{Li} {et~al.}(2011){Li}, {Leaman}, {Chornock}, {Filippenko},
  {Poznanski}, {Ganeshalingam}, {Wang}, {Modjaz}, {Jha}, {Foley}, \&
  {Smith}}]{2011MNRAS.412.1441L}
{Li}, W., {Leaman}, J., {Chornock}, R., {et~al.} 2011, \mnras, 412, 1441

\bibitem[{{Liu} {et~al.}(2015){Liu}, {Modjaz}, {Bianco}, \&
  {Graur}}]{2015arXiv151008049L}
{Liu}, Y.-Q., {Modjaz}, M., {Bianco}, F.~B., \& {Graur}, O. 2015, ArXiv
  e-prints

\bibitem[{{Lyman} {et~al.}(2014){Lyman}, {Bersier}, {James}, {Mazzali},
  {Eldridge}, {Fraser}, \& {Pian}}]{2014arXiv1406.3667L}
{Lyman}, J., {Bersier}, D., {James}, P., {et~al.} 2014, ArXiv e-prints

\bibitem[{{Ma{\'{\i}}z-Apell{\'a}niz}(2004)}]{2004PASP..116..859M}
{Ma{\'{\i}}z-Apell{\'a}niz}, J. 2004, \pasp, 116, 859

\bibitem[{{Mauerhan} {et~al.}(2015{\natexlab{a}}){Mauerhan}, {Van Dyk},
  {Graham}, {Zheng}, {Clubb}, {Filippenko}, {Valenti}, {Brown}, {Smith},
  {Howell}, \& {Arcavi}}]{2015MNRAS.447.1922M}
{Mauerhan}, J.~C., {Van Dyk}, S.~D., {Graham}, M.~L., {et~al.}
  2015{\natexlab{a}}, \mnras, 447, 1922

\bibitem[{{Mauerhan} {et~al.}(2015{\natexlab{b}}){Mauerhan}, {Williams},
  {Leonard}, {Smith}, {Filippenko}, {Smith}, {Hoffman}, {Huk}, {Clubb},
  {Silverman}, {Cenko}, {Milne}, {Gal-Yam}, \& {Ben-Ami}}]{2015MNRAS.453.4467M}
{Mauerhan}, J.~C., {Williams}, G.~G., {Leonard}, D.~C., {et~al.}
  2015{\natexlab{b}}, \mnras, 453, 4467

\bibitem[{{Maund} {et~al.}(2015){Maund}, {Arcavi}, {Ergon}, {Eldridge},
  {Georgy}, {Cenko}, {Horesh}, {Izzard}, \& {Stancliffe}}]{2015MNRAS.454.2580M}
{Maund}, J.~R., {Arcavi}, I., {Ergon}, M., {et~al.} 2015, \mnras, 454, 2580

\bibitem[{{Maund} \& {Smartt}(2009)}]{Maund:2009}
{Maund}, J.~R. \& {Smartt}, S.~J. 2009, Science, 324, 486

\bibitem[{{Mazzali} {et~al.}(2007){Mazzali}, {Foley}, {Deng}, {Patat}, {Pian},
  {Baade}, {Bloom}, {Filippenko}, {Perley}, {Valenti}, {Wang}, {Kawabata},
  {Maeda}, \& {Nomoto}}]{2007ApJ...661..892M}
{Mazzali}, P.~A., {Foley}, R.~J., {Deng}, J., {et~al.} 2007, \apj, 661, 892

\bibitem[{{Mazzali} {et~al.}(2010){Mazzali}, {Maurer}, {Valenti}, {Kotak}, \&
  {Hunter}}]{2010MNRAS.408...87M}
{Mazzali}, P.~A., {Maurer}, I., {Valenti}, S., {Kotak}, R., \& {Hunter}, D.
  2010, \mnras, 408, 87

\bibitem[{{Milisavljevic} {et~al.}(2013{\natexlab{a}}){Milisavljevic}, {Fesen},
  {Pickering}, {Miszalski}, {Buckley}, {Parrent}, {Marion}, {Silverman},
  {Vinko}, {Wheeler}, {Quimby}, {Jha}, {Mohamed}, {Kasliwal}, \&
  {Soderberg}}]{2013ATel.5142....1M}
{Milisavljevic}, D., {Fesen}, R., {Pickering}, T., {et~al.} 2013{\natexlab{a}},
  The Astronomer's Telegram, 5142, 1

\bibitem[{{Milisavljevic} {et~al.}(2013{\natexlab{b}}){Milisavljevic},
  {Margutti}, {Soderberg}, {Pignata}, {Chomiuk}, {Fesen}, {Bufano}, {Sanders},
  {Parrent}, {Parker}, {Mazzali}, {Pian}, {Pickering}, {Buckley}, {Crawford},
  {Gulbis}, {Hettlage}, {Hooper}, {Nordsieck}, {O'Donoghue}, {Husser},
  {Potter}, {Kniazev}, {Kotze}, {Romero-Colmenero}, {Vaisanen}, {Wolf},
  {Bietenholz}, {Bartel}, {Fransson}, {Walker}, {Brunthaler}, {Chakraborti},
  {Levesque}, {MacFadyen}, {Drescher}, {Bock}, {Marples}, {Anderson},
  {Benetti}, {Reichart}, \& {Ivarsen}}]{2013ApJ...767...71M}
{Milisavljevic}, D., {Margutti}, R., {Soderberg}, A.~M., {et~al.}
  2013{\natexlab{b}}, \apj, 767, 71

\bibitem[{{Ofek}(2014)}]{2014ascl.soft07005O}
{Ofek}, E.~O. 2014, {MATLAB package for astronomy and astrophysics},
  Astrophysics Source Code Library

\bibitem[{{Parrent} {et~al.}(2015){Parrent}, {Milisavljevic}, {Soderberg}, \&
  {Parthasarathy}}]{2015arXiv150506645P}
{Parrent}, J.~T., {Milisavljevic}, D., {Soderberg}, A.~M., \& {Parthasarathy},
  M. 2015, ArXiv e-prints

\bibitem[{{Pastorello} {et~al.}(2008){Pastorello}, {Kasliwal}, {Crockett},
  {Valenti}, {Arbour}, {Itagaki}, {Kaspi}, {Gal-Yam}, {Smartt}, {Griffith},
  {Maguire}, {Ofek}, {Seymour}, {Stern}, \& {Wiethoff}}]{2008MNRAS.389..955P}
{Pastorello}, A., {Kasliwal}, M.~M., {Crockett}, R.~M., {et~al.} 2008, \mnras,
  389, 955

\bibitem[{{Paxton} {et~al.}(2010){Paxton}, {Bildsten}, {Dotter}, {Herwig},
  {Lesaffre}, \& {Timmes}}]{2010ascl.soft10083P}
{Paxton}, B., {Bildsten}, L., {Dotter}, A., {et~al.} 2010, {MESA: Modules for
  Experiments in Stellar Astrophysics}, Astrophysics Source Code Library

\bibitem[{{Pettini} \& {Pagel}(2004)}]{2004MNRAS.348L..59P}
{Pettini}, M. \& {Pagel}, B.~E.~J. 2004, \mnras, 348, L59

\bibitem[{{Pilyugin} {et~al.}(2004){Pilyugin}, {V{\'{\i}}lchez}, \&
  {Contini}}]{2004A&amp;A...425..849P}
{Pilyugin}, L.~S., {V{\'{\i}}lchez}, J.~M., \& {Contini}, T. 2004, \aap, 425,
  849

\bibitem[{{Piro} \& {Nakar}(2013)}]{Piro:2012aa}
{Piro}, A.~L. \& {Nakar}, E. 2013, \apj, 769, 67

\bibitem[{{Poznanski} {et~al.}(2011){Poznanski}, {Ganeshalingam}, {Silverman},
  \& {Filippenko}}]{2011MNRAS.415L..81P}
{Poznanski}, D., {Ganeshalingam}, M., {Silverman}, J.~M., \& {Filippenko},
  A.~V. 2011, \mnras, 415, L81

\bibitem[{{Poznanski} {et~al.}(2012){Poznanski}, {Prochaska}, \&
  {Bloom}}]{2012MNRAS.426.1465P}
{Poznanski}, D., {Prochaska}, J.~X., \& {Bloom}, J.~S. 2012, \mnras, 426, 1465

\bibitem[{{Reilly} {et~al.}(2015){Reilly}, {Maund}, {Baade}, {Wheeler},
  {Silverman}, {Clocchiatti}, {Patat}, {H{\"o}flich}, {Spyromilio}, {Wang}, \&
  {Zelaya}}]{2015arXiv151002492R}
{Reilly}, E., {Maund}, J.~R., {Baade}, D., {et~al.} 2015, ArXiv e-prints

\bibitem[{{Roming} {et~al.}(2004){Roming}, {Hunsberger}, {Mason}, {Nousek},
  {Broos}, {Carter}, {Hancock}, {Huckle}, {Kennedy}, {Killough}, {Koch},
  {McLelland}, {Pryzby}, {Smith}, {Soto}, {Stock}, {Boyd}, \&
  {Still}}]{2004SPIE.5165..262R}
{Roming}, P.~W.~A., {Hunsberger}, S.~D., {Mason}, K.~O., {et~al.} 2004, in
  Society of Photo-Optical Instrumentation Engineers (SPIE) Conference Series,
  Vol. 5165, X-Ray and Gamma-Ray Instrumentation for Astronomy XIII, ed. K.~A.
  {Flanagan} \& O.~H.~W. {Siegmund}, 262--276

\bibitem[{{Roming} {et~al.}(2005){Roming}, {Kennedy}, {Mason}, {Nousek}, {Ahr},
  {Bingham}, {Broos}, {Carter}, {Hancock}, {Huckle}, {Hunsberger}, {Kawakami},
  {Killough}, {Koch}, {McLelland}, {Smith}, {Smith}, {Soto}, {Boyd},
  {Breeveld}, {Holland}, {Ivanushkina}, {Pryzby}, {Still}, \&
  {Stock}}]{2005SSRv..120...95R}
{Roming}, P.~W.~A., {Kennedy}, T.~E., {Mason}, K.~O., {et~al.} 2005, \ssr, 120,
  95

\bibitem[{{Schlafly} \& {Finkbeiner}(2011)}]{2011ApJ...737..103S}
{Schlafly}, E.~F. \& {Finkbeiner}, D.~P. 2011, \apj, 737, 103

\bibitem[{{Silverman} {et~al.}(2012){Silverman}, {Kong}, \&
  {Filippenko}}]{2012MNRAS.425.1819S}
{Silverman}, J.~M., {Kong}, J.~J., \& {Filippenko}, A.~V. 2012, \mnras, 425,
  1819

\bibitem[{{Simcoe} {et~al.}(2013){Simcoe}, {Burgasser}, {Schechter}, {Fishner},
  {Bernstein}, {Bigelow}, {Pipher}, {Forrest}, {McMurtry}, {Smith}, \&
  {Bochanski}}]{2013PASP..125..270S}
{Simcoe}, R.~A., {Burgasser}, A.~J., {Schechter}, P.~L., {et~al.} 2013, \pasp,
  125, 270

\bibitem[{{Smartt} {et~al.}(2009){Smartt}, {Eldridge}, {Crockett}, \&
  {Maund}}]{Smartt:2009}
{Smartt}, S.~J., {Eldridge}, J.~J., {Crockett}, R.~M., \& {Maund}, J.~R. 2009,
  \mnras, 395, 1409

\bibitem[{{Smith}(2014)}]{2014ARA&amp;A..52..487S}
{Smith}, N. 2014, \araa, 52, 487

\bibitem[{{Srivastav} {et~al.}(2014){Srivastav}, {Anupama}, \&
  {Sahu}}]{2014MNRAS.445.1932S}
{Srivastav}, S., {Anupama}, G.~C., \& {Sahu}, D.~K. 2014, \mnras, 445, 1932

\bibitem[{{Strotjohann} {et~al.}(2015){Strotjohann}, {Ofek}, {Gal-Yam},
  {Sullivan}, {Kulkarni}, {Shaviv}, {Fremling}, {Kasliwal}, {Nugent}, {Cao},
  {Arcavi}, {Sollerman}, {Filippenko}, {Yaron}, {Laher}, \&
  {Surace}}]{2015ApJ...811..117S}
{Strotjohann}, N.~L., {Ofek}, E.~O., {Gal-Yam}, A., {et~al.} 2015, \apj, 811,
  117

\bibitem[{{Taddia} {et~al.}(2015){Taddia}, {Sollerman}, {Leloudas},
  {Stritzinger}, {Valenti}, {Galbany}, {Kessler}, {Schneider}, \&
  {Wheeler}}]{2015A&amp;A...574A..60T}
{Taddia}, F., {Sollerman}, J., {Leloudas}, G., {et~al.} 2015, \aap, 574, A60

\bibitem[{{Taddia} {et~al.}(2013){Taddia}, {Sollerman}, {Razza}, {Gafton},
  {Pastorello}, {Fransson}, {Stritzinger}, {Leloudas}, \&
  {Ergon}}]{2013A&amp;A...558A.143T}
{Taddia}, F., {Sollerman}, J., {Razza}, A., {et~al.} 2013, \aap, 558, A143

\bibitem[{{Taubenberger} {et~al.}(2011){Taubenberger}, {Navasardyan}, {Maurer},
  {Zampieri}, {Chugai}, {Benetti}, {Agnoletto}, {Bufano}, {Elias-Rosa},
  {Turatto}, {Patat}, {Cappellaro}, {Mazzali}, {Iijima}, {Valenti},
  {Harutyunyan}, {Claudi}, \& {Dolci}}]{2011MNRAS.413.2140T}
{Taubenberger}, S., {Navasardyan}, H., {Maurer}, J.~I., {et~al.} 2011, \mnras,
  413, 2140

\bibitem[{{Tully} {et~al.}(2013){Tully}, {Courtois}, {Dolphin}, {Fisher},
  {H{\'e}raudeau}, {Jacobs}, {Karachentsev}, {Makarov}, {Makarova},
  {Mitronova}, {Rizzi}, {Shaya}, {Sorce}, \& {Wu}}]{2013AJ....146...86T}
{Tully}, R.~B., {Courtois}, H.~M., {Dolphin}, A.~E., {et~al.} 2013, \aj, 146,
  86

\bibitem[{{Tully} {et~al.}(2009){Tully}, {Rizzi}, {Shaya}, {Courtois},
  {Makarov}, \& {Jacobs}}]{Tully:2009}
{Tully}, R.~B., {Rizzi}, L., {Shaya}, E.~J., {et~al.} 2009, \aj, 138, 323

\bibitem[{{Turatto} {et~al.}(2003){Turatto}, {Benetti}, \&
  {Cappellaro}}]{2003fthp.conf..200T}
{Turatto}, M., {Benetti}, S., \& {Cappellaro}, E. 2003, in From Twilight to
  Highlight: The Physics of Supernovae, ed. W.~{Hillebrandt} \&
  B.~{Leibundgut}, 200

\bibitem[{{Van Dyk} {et~al.}(2012){Van Dyk}, {Gal-Yam}, {Arcavi}, {Kasliwal},
  {Horesh}, \& {PTF Collaboration}}]{2012ATel.3884....1V}
{Van Dyk}, S.~D., {Gal-Yam}, A., {Arcavi}, I., {et~al.} 2012, The Astronomer's
  Telegram, 3884, 1

\bibitem[{{Van Dyk} {et~al.}(2011){Van Dyk}, {Li}, {Cenko}, {Kasliwal},
  {Horesh}, {Ofek}, {Kraus}, {Silverman}, {Arcavi}, {Filippenko}, {Gal-Yam},
  {Quimby}, {Kulkarni}, {Yaron}, \& {Polishook}}]{2011ApJ...741L..28V}
{Van Dyk}, S.~D., {Li}, W., {Cenko}, S.~B., {et~al.} 2011, \apjl, 741, L28

\bibitem[{{Vink{\'o}} {et~al.}(2009){Vink{\'o}}, {S{\'a}rneczky}, {Balog},
  {Immler}, {Sugerman}, {Brown}, {Misselt}, {Szab{\'o}}, {Csizmadia}, {Kun},
  {Klagyivik}, {Foley}, {Filippenko}, {Cs{\'a}k}, \&
  {Kiss}}]{2009ApJ...695..619V}
{Vink{\'o}}, J., {S{\'a}rneczky}, K., {Balog}, Z., {et~al.} 2009, \apj, 695,
  619

\bibitem[{{Wheeler} {et~al.}(2015){Wheeler}, {Johnson}, \&
  {Clocchiatti}}]{2015MNRAS.450.1295W}
{Wheeler}, J.~C., {Johnson}, V., \& {Clocchiatti}, A. 2015, \mnras, 450, 1295

\bibitem[{{Woosley} \& {Heger}(2007)}]{2007PhR...442..269W}
{Woosley}, S.~E. \& {Heger}, A. 2007, \physrep, 442, 269

\bibitem[{{Yaron} \& {Gal-Yam}(2012)}]{Yaron:2012aa}
{Yaron}, O. \& {Gal-Yam}, A. 2012, \pasp, 124, 668

\bibitem[{{Yoon}(2015)}]{2015PASA...32...15Y}
{Yoon}, S.-C. 2015, \pasa, 32, 15

\bibitem[{{Yoon} {et~al.}(2010){Yoon}, {Woosley}, \& {Langer}}]{Yoon:2010aa}
{Yoon}, S.-C., {Woosley}, S.~E., \& {Langer}, N. 2010, \apj, 725, 940

\end{thebibliography}

\onecolumn

\clearpage

\clearpage

\clearpage

\clearpage

\begin{deluxetable}{lrlrlrlrlrlrll} 
\tabletypesize{\scriptsize} 
\tablecolumns{14} 
\tablewidth{0pt} 
\tablecaption{Photometry table for iPTF13bvn ($UBgriz$)\label{tab:13bvnphot}}
\tablehead{ 
\colhead{Phase} & \colhead{$U$} & \colhead{$\sigma$} & \colhead{$B$} & \colhead{$\sigma$} & \colhead{$g$} & \colhead{$\sigma$} & \colhead{$r$} & \colhead{$\sigma$} & \colhead{$i$} & \colhead{$\sigma$} & \colhead{$z$} & \colhead{$\sigma$} & \colhead{$tel.$} \\ 
\colhead{JD $-$ 2,456,459.17} & \colhead{[mag]} & \colhead{[mag]} & \colhead{[mag]} & \colhead{[mag]} & \colhead{[mag]} & \colhead{[mag]} & \colhead{[mag]} & \colhead{[mag]} & \colhead{[mag]} & \colhead{[mag]} & \colhead{[mag]} & \colhead{[mag]} & \colhead{} \\ 
}
\startdata 
$   0.57$ & $...$ & $...$ & $...$ & $...$ & $...$ & $...$ & $18.61$ & $0.05$ & $...$ & $...$ & $...$ & $...$ & $  P48$ \\ 
$   0.62$ & $...$ & $...$ & $...$ & $...$ & $...$ & $...$ & $18.62$ & $0.05$ & $...$ & $...$ & $...$ & $...$ & $  P48$ \\ 
$   1.57$ & $...$ & $...$ & $...$ & $...$ & $...$ & $...$ & $17.61$ & $0.04$ & $...$ & $...$ & $...$ & $...$ & $  P48$ \\ 
$   1.62$ & $...$ & $...$ & $...$ & $...$ & $...$ & $...$ & $17.55$ & $0.04$ & $...$ & $...$ & $...$ & $...$ & $  P48$ \\ 
$   1.71$ & $...$ & $...$ & $...$ & $...$ & $17.83$ & $0.01$ & $17.50$ & $0.01$ & $17.67$ & $0.01$ & $...$ & $...$ & $  P60$ \\ 
$   2.08$ & $...$ & $...$ & $17.91$ & $0.02$ & $...$ & $...$ & $...$ & $...$ & $...$ & $...$ & $...$ & $...$ & $LCOGT$ \\ 
$   2.16$ & $...$ & $...$ & $...$ & $...$ & $17.63$ & $0.01$ & $17.21$ & $0.01$ & $...$ & $...$ & $...$ & $...$ & $LCOGT$ \\ 
$   2.18$ & $...$ & $...$ & $...$ & $...$ & $...$ & $...$ & $...$ & $...$ & $17.32$ & $0.01$ & $17.11$ & $0.04$ & $LCOGT$ \\ 
$   2.33$ & $...$ & $...$ & $17.84$ & $0.02$ & $...$ & $...$ & $...$ & $...$ & $...$ & $...$ & $...$ & $...$ & $LCOGT$ \\ 
$   2.34$ & $...$ & $...$ & $...$ & $...$ & $17.57$ & $0.02$ & $17.12$ & $0.02$ & $...$ & $...$ & $...$ & $...$ & $LCOGT$ \\ 
$   2.36$ & $...$ & $...$ & $...$ & $...$ & $...$ & $...$ & $...$ & $...$ & $17.37$ & $0.02$ & $...$ & $...$ & $LCOGT$ \\ 
$   2.39$ & $17.68$ & $0.06$ & $...$ & $...$ & $...$ & $...$ & $...$ & $...$ & $...$ & $...$ & $...$ & $...$ & $LCOGT$ \\ 
$   2.42$ & $...$ & $...$ & $...$ & $...$ & $...$ & $...$ & $...$ & $...$ & $...$ & $...$ & $17.19$ & $0.04$ & $LCOGT$ \\ 
$   2.50$ & $...$ & $...$ & $...$ & $...$ & $...$ & $...$ & $17.06$ & $0.01$ & $17.23$ & $0.01$ & $...$ & $...$ & $  P60$ \\ 
$   2.56$ & $...$ & $...$ & $...$ & $...$ & $...$ & $...$ & $17.08$ & $0.04$ & $...$ & $...$ & $...$ & $...$ & $  P48$ \\ 
$   2.60$ & $...$ & $...$ & $...$ & $...$ & $...$ & $...$ & $17.06$ & $0.04$ & $...$ & $...$ & $...$ & $...$ & $  P48$ \\ 
$   2.63$ & $...$ & $...$ & $...$ & $...$ & $...$ & $...$ & $17.04$ & $0.01$ & $17.20$ & $0.01$ & $...$ & $...$ & $  P60$ \\ 
$   2.67$ & $...$ & $...$ & $...$ & $...$ & $17.49$ & $0.01$ & $...$ & $...$ & $17.18$ & $0.01$ & $...$ & $...$ & $  P60$ \\ 
$   3.03$ & $...$ & $...$ & $...$ & $...$ & $...$ & $...$ & $...$ & $...$ & $...$ & $...$ & $16.86$ & $0.06$ & $LCOGT$ \\ 
$   3.36$ & $...$ & $...$ & $17.48$ & $0.02$ & $...$ & $...$ & $...$ & $...$ & $...$ & $...$ & $...$ & $...$ & $LCOGT$ \\ 
$   3.39$ & $17.41$ & $0.07$ & $...$ & $...$ & $...$ & $...$ & $...$ & $...$ & $...$ & $...$ & $...$ & $...$ & $LCOGT$ \\ 
$   3.41$ & $...$ & $...$ & $...$ & $...$ & $...$ & $...$ & $...$ & $...$ & $...$ & $...$ & $16.75$ & $0.03$ & $LCOGT$ \\ 
$   3.53$ & $...$ & $...$ & $17.45$ & $0.02$ & $...$ & $...$ & $...$ & $...$ & $...$ & $...$ & $...$ & $...$ & $LCOGT$ \\ 
$   3.54$ & $...$ & $...$ & $...$ & $...$ & $17.26$ & $0.01$ & $16.70$ & $0.01$ & $...$ & $...$ & $...$ & $...$ & $LCOGT$ \\ 
$   3.56$ & $...$ & $...$ & $...$ & $...$ & $...$ & $...$ & $...$ & $...$ & $16.90$ & $0.02$ & $...$ & $...$ & $LCOGT$ \\ 
$   3.66$ & $...$ & $...$ & $...$ & $...$ & $...$ & $...$ & $...$ & $...$ & $...$ & $...$ & $16.70$ & $0.04$ & $LCOGT$ \\ 
$   4.33$ & $...$ & $...$ & $17.14$ & $0.02$ & $16.92$ & $0.01$ & $16.42$ & $0.02$ & $16.56$ & $0.02$ & $...$ & $...$ & $LCOGT$ \\ 
$   4.39$ & $...$ & $...$ & $...$ & $...$ & $...$ & $...$ & $...$ & $...$ & $...$ & $...$ & $16.45$ & $0.05$ & $LCOGT$ \\ 
$   6.30$ & $16.43$ & $0.05$ & $...$ & $...$ & $...$ & $...$ & $...$ & $...$ & $...$ & $...$ & $...$ & $...$ & $LCOGT$ \\ 
$   6.40$ & $...$ & $...$ & $16.57$ & $0.02$ & $...$ & $...$ & $...$ & $...$ & $...$ & $...$ & $...$ & $...$ & $LCOGT$ \\ 
$   6.41$ & $...$ & $...$ & $...$ & $...$ & $16.35$ & $0.01$ & $15.96$ & $0.02$ & $16.12$ & $0.02$ & $16.10$ & $0.04$ & $LCOGT$ \\ 
$   8.40$ & $...$ & $...$ & $16.15$ & $0.01$ & $...$ & $...$ & $...$ & $...$ & $...$ & $...$ & $...$ & $...$ & $LCOGT$ \\ 
$   8.41$ & $...$ & $...$ & $...$ & $...$ & $15.99$ & $0.01$ & $15.66$ & $0.01$ & $15.79$ & $0.01$ & $15.74$ & $0.04$ & $LCOGT$ \\ 
$   8.61$ & $...$ & $...$ & $...$ & $...$ & $15.96$ & $0.01$ & $15.68$ & $0.01$ & $15.81$ & $0.01$ & $...$ & $...$ & $  P60$ \\ 
$   9.42$ & $15.67$ & $0.05$ & $16.03$ & $0.01$ & $15.89$ & $0.01$ & $15.61$ & $0.01$ & $15.74$ & $0.01$ & $...$ & $...$ & $LCOGT$ \\ 
$  10.40$ & $...$ & $...$ & $15.99$ & $0.01$ & $...$ & $...$ & $...$ & $...$ & $...$ & $...$ & $...$ & $...$ & $LCOGT$ \\ 
$  10.41$ & $...$ & $...$ & $...$ & $...$ & $15.83$ & $0.01$ & $15.53$ & $0.01$ & $15.64$ & $0.01$ & $15.62$ & $0.02$ & $LCOGT$ \\ 
$  10.59$ & $...$ & $...$ & $...$ & $...$ & $15.75$ & $0.01$ & $15.48$ & $0.01$ & $15.61$ & $0.01$ & $...$ & $...$ & $  P60$ \\ 
$  11.55$ & $...$ & $...$ & $...$ & $...$ & $15.68$ & $0.01$ & $15.41$ & $0.01$ & $15.53$ & $0.01$ & $...$ & $...$ & $  P60$ \\ 
$  12.24$ & $...$ & $...$ & $15.80$ & $0.01$ & $...$ & $...$ & $...$ & $...$ & $...$ & $...$ & $...$ & $...$ & $LCOGT$ \\ 
$  12.26$ & $...$ & $...$ & $...$ & $...$ & $...$ & $...$ & $15.33$ & $0.02$ & $15.47$ & $0.02$ & $...$ & $...$ & $LCOGT$ \\ 
$  12.40$ & $...$ & $...$ & $15.89$ & $0.01$ & $...$ & $...$ & $...$ & $...$ & $...$ & $...$ & $...$ & $...$ & $LCOGT$ \\ 
$  12.41$ & $...$ & $...$ & $...$ & $...$ & $...$ & $...$ & $15.36$ & $0.01$ & $15.46$ & $0.01$ & $15.38$ & $0.03$ & $LCOGT$ \\ 
$  12.52$ & $...$ & $...$ & $...$ & $...$ & $15.63$ & $0.01$ & $15.34$ & $0.01$ & $15.46$ & $0.01$ & $...$ & $...$ & $  P60$ \\ 
$  13.50$ & $...$ & $...$ & $15.79$ & $0.01$ & $...$ & $...$ & $...$ & $...$ & $...$ & $...$ & $...$ & $...$ & $LCOGT$ \\ 
$  13.51$ & $...$ & $...$ & $...$ & $...$ & $15.63$ & $0.00$ & $15.29$ & $0.01$ & $...$ & $...$ & $...$ & $...$ & $LCOGT$ \\ 
$  13.52$ & $...$ & $...$ & $...$ & $...$ & $...$ & $...$ & $...$ & $...$ & $15.41$ & $0.01$ & $...$ & $...$ & $LCOGT$ \\ 
$  13.52$ & $...$ & $...$ & $...$ & $...$ & $...$ & $...$ & $15.28$ & $0.01$ & $15.42$ & $0.01$ & $...$ & $...$ & $  P60$ \\ 
$  13.53$ & $...$ & $...$ & $...$ & $...$ & $15.59$ & $0.01$ & $...$ & $...$ & $...$ & $...$ & $...$ & $...$ & $  P60$ \\ 
$  14.61$ & $...$ & $...$ & $...$ & $...$ & $15.56$ & $0.01$ & $15.24$ & $0.01$ & $15.36$ & $0.01$ & $...$ & $...$ & $  P60$ \\ 
$  15.54$ & $...$ & $...$ & $...$ & $...$ & $15.54$ & $0.01$ & $15.19$ & $0.01$ & $15.32$ & $0.01$ & $...$ & $...$ & $  P60$ \\ 
$  16.28$ & $...$ & $...$ & $15.79$ & $0.01$ & $15.55$ & $0.01$ & $15.16$ & $0.01$ & $15.27$ & $0.01$ & $...$ & $...$ & $LCOGT$ \\ 
$  16.38$ & $...$ & $...$ & $15.81$ & $0.01$ & $...$ & $...$ & $...$ & $...$ & $...$ & $...$ & $...$ & $...$ & $LCOGT$ \\ 
$  16.39$ & $...$ & $...$ & $...$ & $...$ & $15.54$ & $0.01$ & $15.15$ & $0.01$ & $15.27$ & $0.01$ & $15.22$ & $0.02$ & $LCOGT$ \\ 
$  16.52$ & $...$ & $...$ & $...$ & $...$ & $15.53$ & $0.01$ & $15.16$ & $0.01$ & $15.26$ & $0.01$ & $...$ & $...$ & $  P60$ \\ 
$  17.38$ & $...$ & $...$ & $15.84$ & $0.01$ & $...$ & $...$ & $...$ & $...$ & $...$ & $...$ & $...$ & $...$ & $LCOGT$ \\ 
$  17.39$ & $...$ & $...$ & $...$ & $...$ & $15.61$ & $0.01$ & $15.17$ & $0.01$ & $15.27$ & $0.01$ & $15.14$ & $0.02$ & $LCOGT$ \\ 
$  17.55$ & $...$ & $...$ & $...$ & $...$ & $15.55$ & $0.01$ & $...$ & $...$ & $...$ & $...$ & $...$ & $...$ & $  P60$ \\ 
$  18.55$ & $...$ & $...$ & $...$ & $...$ & $...$ & $...$ & $15.12$ & $0.01$ & $15.20$ & $0.01$ & $...$ & $...$ & $  P60$ \\ 
$  18.57$ & $...$ & $...$ & $...$ & $...$ & $15.61$ & $0.01$ & $15.10$ & $0.02$ & $...$ & $...$ & $...$ & $...$ & $LCOGT$ \\ 
$  18.58$ & $...$ & $...$ & $...$ & $...$ & $...$ & $...$ & $...$ & $...$ & $15.21$ & $0.04$ & $...$ & $...$ & $LCOGT$ \\ 
$  19.57$ & $...$ & $...$ & $...$ & $...$ & $15.68$ & $0.00$ & $15.14$ & $0.01$ & $15.20$ & $0.01$ & $15.08$ & $0.03$ & $LCOGT$ \\ 
$  20.56$ & $...$ & $...$ & $16.12$ & $0.01$ & $...$ & $...$ & $...$ & $...$ & $...$ & $...$ & $...$ & $...$ & $LCOGT$ \\ 
$  20.57$ & $...$ & $...$ & $...$ & $...$ & $15.73$ & $0.00$ & $...$ & $...$ & $...$ & $...$ & $...$ & $...$ & $LCOGT$ \\ 
$  20.58$ & $...$ & $...$ & $...$ & $...$ & $...$ & $...$ & $...$ & $...$ & $15.19$ & $0.01$ & $...$ & $...$ & $  P60$ \\ 
$  20.58$ & $...$ & $...$ & $...$ & $...$ & $...$ & $...$ & $15.14$ & $0.01$ & $...$ & $...$ & $...$ & $...$ & $LCOGT$ \\ 
$  20.58$ & $...$ & $...$ & $...$ & $...$ & $15.72$ & $0.01$ & $15.15$ & $0.01$ & $...$ & $...$ & $...$ & $...$ & $  P60$ \\ 
$  20.58$ & $...$ & $...$ & $...$ & $...$ & $...$ & $...$ & $...$ & $...$ & $15.22$ & $0.01$ & $15.01$ & $0.02$ & $LCOGT$ \\ 
$  21.09$ & $16.37$ & $0.02$ & $...$ & $...$ & $...$ & $...$ & $...$ & $...$ & $...$ & $...$ & $...$ & $...$ & $LCOGT$ \\ 
$  21.19$ & $...$ & $...$ & $16.12$ & $0.02$ & $15.78$ & $0.02$ & $15.17$ & $0.03$ & $15.20$ & $0.02$ & $...$ & $...$ & $LCOGT$ \\ 
$  21.38$ & $...$ & $...$ & $16.18$ & $0.01$ & $...$ & $...$ & $...$ & $...$ & $...$ & $...$ & $...$ & $...$ & $LCOGT$ \\ 
$  21.39$ & $...$ & $...$ & $...$ & $...$ & $15.81$ & $0.01$ & $15.17$ & $0.01$ & $15.22$ & $0.01$ & $15.16$ & $0.02$ & $LCOGT$ \\ 
$  21.53$ & $...$ & $...$ & $...$ & $...$ & $15.80$ & $0.01$ & $15.19$ & $0.01$ & $15.20$ & $0.01$ & $...$ & $...$ & $  P60$ \\ 
$  21.56$ & $...$ & $...$ & $16.25$ & $0.01$ & $...$ & $...$ & $...$ & $...$ & $...$ & $...$ & $...$ & $...$ & $LCOGT$ \\ 
$  21.57$ & $...$ & $...$ & $...$ & $...$ & $15.83$ & $0.01$ & $15.18$ & $0.01$ & $15.23$ & $0.01$ & $15.11$ & $0.03$ & $LCOGT$ \\ 
$  22.52$ & $...$ & $...$ & $...$ & $...$ & $...$ & $...$ & $15.24$ & $0.01$ & $15.22$ & $0.01$ & $...$ & $...$ & $  P60$ \\ 
$  22.52$ & $...$ & $...$ & $16.34$ & $0.01$ & $...$ & $...$ & $...$ & $...$ & $...$ & $...$ & $...$ & $...$ & $LCOGT$ \\ 
$  22.54$ & $...$ & $...$ & $...$ & $...$ & $15.89$ & $0.01$ & $15.25$ & $0.01$ & $15.24$ & $0.01$ & $15.13$ & $0.03$ & $LCOGT$ \\ 
$  23.56$ & $...$ & $...$ & $...$ & $...$ & $16.02$ & $0.01$ & $...$ & $...$ & $...$ & $...$ & $...$ & $...$ & $  P60$ \\ 
$  24.03$ & $...$ & $...$ & $16.63$ & $0.02$ & $16.09$ & $0.01$ & $15.34$ & $0.01$ & $15.35$ & $0.01$ & $...$ & $...$ & $LCOGT$ \\ 
$  24.54$ & $...$ & $...$ & $...$ & $...$ & $16.18$ & $0.01$ & $15.41$ & $0.01$ & $...$ & $...$ & $...$ & $...$ & $LCOGT$ \\ 
$  26.52$ & $...$ & $...$ & $...$ & $...$ & $16.45$ & $0.01$ & $15.60$ & $0.01$ & $15.49$ & $0.01$ & $...$ & $...$ & $  P60$ \\ 
$  26.52$ & $...$ & $...$ & $17.06$ & $0.02$ & $...$ & $...$ & $...$ & $...$ & $...$ & $...$ & $...$ & $...$ & $LCOGT$ \\ 
$  26.54$ & $...$ & $...$ & $...$ & $...$ & $16.44$ & $0.01$ & $15.60$ & $0.01$ & $15.48$ & $0.01$ & $15.35$ & $0.03$ & $LCOGT$ \\ 
$  27.18$ & $18.01$ & $0.10$ & $...$ & $...$ & $...$ & $...$ & $...$ & $...$ & $...$ & $...$ & $...$ & $...$ & $LCOGT$ \\ 
$  27.52$ & $...$ & $...$ & $...$ & $...$ & $16.56$ & $0.01$ & $15.69$ & $0.01$ & $15.57$ & $0.01$ & $...$ & $...$ & $  P60$ \\ 
$  28.41$ & $...$ & $...$ & $17.26$ & $0.02$ & $16.67$ & $0.01$ & $15.77$ & $0.02$ & $15.62$ & $0.01$ & $...$ & $...$ & $LCOGT$ \\ 
$  28.52$ & $...$ & $...$ & $17.33$ & $0.02$ & $...$ & $...$ & $...$ & $...$ & $...$ & $...$ & $...$ & $...$ & $LCOGT$ \\ 
$  28.54$ & $...$ & $...$ & $...$ & $...$ & $16.69$ & $0.01$ & $15.79$ & $0.01$ & $15.64$ & $0.01$ & $15.43$ & $0.02$ & $LCOGT$ \\ 
$  28.55$ & $...$ & $...$ & $...$ & $...$ & $16.71$ & $0.01$ & $15.79$ & $0.01$ & $15.65$ & $0.01$ & $...$ & $...$ & $  P60$ \\ 
$  29.51$ & $...$ & $...$ & $...$ & $...$ & $16.80$ & $0.01$ & $15.87$ & $0.01$ & $15.72$ & $0.01$ & $...$ & $...$ & $  P60$ \\ 
$  29.55$ & $...$ & $...$ & $...$ & $...$ & $16.82$ & $0.01$ & $15.90$ & $0.02$ & $...$ & $...$ & $...$ & $...$ & $LCOGT$ \\ 
$  29.56$ & $...$ & $...$ & $...$ & $...$ & $...$ & $...$ & $...$ & $...$ & $...$ & $...$ & $15.40$ & $0.05$ & $LCOGT$ \\ 
$  35.53$ & $18.48$ & $0.51$ & $18.05$ & $0.08$ & $...$ & $...$ & $...$ & $...$ & $...$ & $...$ & $...$ & $...$ & $LCOGT$ \\ 
$  36.53$ & $...$ & $...$ & $18.00$ & $0.05$ & $...$ & $...$ & $...$ & $...$ & $...$ & $...$ & $...$ & $...$ & $LCOGT$ \\ 
$  40.54$ & $...$ & $...$ & $18.17$ & $0.05$ & $...$ & $...$ & $...$ & $...$ & $...$ & $...$ & $...$ & $...$ & $LCOGT$ \\ 
$  40.56$ & $...$ & $...$ & $...$ & $...$ & $17.54$ & $0.02$ & $16.61$ & $0.02$ & $16.38$ & $0.02$ & $15.84$ & $0.04$ & $LCOGT$ \\ 
$  45.07$ & $...$ & $...$ & $18.24$ & $0.04$ & $17.66$ & $0.01$ & $...$ & $...$ & $...$ & $...$ & $...$ & $...$ & $LCOGT$ \\ 
$  45.13$ & $...$ & $...$ & $...$ & $...$ & $...$ & $...$ & $16.67$ & $0.02$ & $16.47$ & $0.02$ & $...$ & $...$ & $LCOGT$ \\ 
$  45.52$ & $...$ & $...$ & $...$ & $...$ & $17.69$ & $0.01$ & $16.73$ & $0.01$ & $16.50$ & $0.01$ & $...$ & $...$ & $  P60$ \\ 
$  45.78$ & $...$ & $...$ & $...$ & $...$ & $...$ & $...$ & $16.73$ & $0.03$ & $16.50$ & $0.03$ & $...$ & $...$ & $LCOGT$ \\ 
$  46.51$ & $...$ & $...$ & $...$ & $...$ & $17.74$ & $0.01$ & $16.76$ & $0.01$ & $16.52$ & $0.01$ & $...$ & $...$ & $  P60$ \\ 
$  47.51$ & $...$ & $...$ & $...$ & $...$ & $17.74$ & $0.01$ & $16.79$ & $0.01$ & $16.52$ & $0.01$ & $...$ & $...$ & $  P60$ \\ 
$  47.75$ & $...$ & $...$ & $...$ & $...$ & $17.74$ & $0.02$ & $...$ & $...$ & $16.62$ & $0.03$ & $...$ & $...$ & $LCOGT$ \\ 
$  48.49$ & $...$ & $...$ & $...$ & $...$ & $17.78$ & $0.01$ & $16.82$ & $0.01$ & $16.57$ & $0.01$ & $...$ & $...$ & $  P60$ \\ 
$  49.35$ & $...$ & $...$ & $18.38$ & $0.03$ & $17.73$ & $0.01$ & $...$ & $...$ & $...$ & $...$ & $...$ & $...$ & $LCOGT$ \\ 
$  49.37$ & $...$ & $...$ & $...$ & $...$ & $...$ & $...$ & $16.86$ & $0.01$ & $16.58$ & $0.01$ & $...$ & $...$ & $LCOGT$ \\ 
$  49.48$ & $...$ & $...$ & $...$ & $...$ & $...$ & $...$ & $...$ & $...$ & $...$ & $...$ & $15.98$ & $0.05$ & $LCOGT$ \\ 
$  49.52$ & $...$ & $...$ & $...$ & $...$ & $17.79$ & $0.01$ & $16.85$ & $0.01$ & $16.60$ & $0.01$ & $...$ & $...$ & $  P60$ \\ 
$  50.49$ & $...$ & $...$ & $...$ & $...$ & $17.81$ & $0.01$ & $16.88$ & $0.01$ & $16.65$ & $0.01$ & $...$ & $...$ & $  P60$ \\ 
$  51.30$ & $...$ & $...$ & $18.38$ & $0.03$ & $17.77$ & $0.01$ & $...$ & $...$ & $...$ & $...$ & $...$ & $...$ & $LCOGT$ \\ 
$  51.31$ & $...$ & $...$ & $...$ & $...$ & $...$ & $...$ & $16.91$ & $0.01$ & $16.68$ & $0.02$ & $...$ & $...$ & $LCOGT$ \\ 
$  51.47$ & $19.03$ & $0.49$ & $...$ & $...$ & $...$ & $...$ & $...$ & $...$ & $...$ & $...$ & $...$ & $...$ & $LCOGT$ \\ 
$  51.48$ & $...$ & $...$ & $...$ & $...$ & $...$ & $...$ & $...$ & $...$ & $...$ & $...$ & $16.04$ & $0.07$ & $LCOGT$ \\ 
$  51.49$ & $...$ & $...$ & $18.45$ & $0.06$ & $17.84$ & $0.02$ & $16.93$ & $0.02$ & $16.69$ & $0.03$ & $...$ & $...$ & $LCOGT$ \\ 
$  52.29$ & $...$ & $...$ & $...$ & $...$ & $...$ & $...$ & $...$ & $...$ & $...$ & $...$ & $16.08$ & $0.05$ & $LCOGT$ \\ 
$  52.79$ & $...$ & $...$ & $18.38$ & $0.05$ & $17.74$ & $0.02$ & $...$ & $...$ & $...$ & $...$ & $...$ & $...$ & $LCOGT$ \\ 
$  53.48$ & $19.10$ & $0.34$ & $...$ & $...$ & $...$ & $...$ & $...$ & $...$ & $...$ & $...$ & $...$ & $...$ & $LCOGT$ \\ 
$  53.49$ & $...$ & $...$ & $...$ & $...$ & $...$ & $...$ & $...$ & $...$ & $...$ & $...$ & $16.04$ & $0.03$ & $LCOGT$ \\ 
$  53.51$ & $...$ & $...$ & $18.54$ & $0.05$ & $17.89$ & $0.01$ & $17.00$ & $0.02$ & $16.67$ & $0.03$ & $...$ & $...$ & $LCOGT$ \\ 
$  53.75$ & $...$ & $...$ & $18.37$ & $0.05$ & $...$ & $...$ & $...$ & $...$ & $...$ & $...$ & $...$ & $...$ & $LCOGT$ \\ 
$  53.79$ & $...$ & $...$ & $...$ & $...$ & $17.79$ & $0.01$ & $17.00$ & $0.02$ & $16.72$ & $0.03$ & $...$ & $...$ & $LCOGT$ \\ 
$  54.39$ & $...$ & $...$ & $...$ & $...$ & $...$ & $...$ & $...$ & $...$ & $...$ & $...$ & $16.00$ & $0.05$ & $LCOGT$ \\ 
$  55.78$ & $...$ & $...$ & $18.33$ & $0.04$ & $...$ & $...$ & $...$ & $...$ & $...$ & $...$ & $...$ & $...$ & $LCOGT$ \\ 
$  56.40$ & $...$ & $...$ & $18.51$ & $0.07$ & $17.90$ & $0.02$ & $17.04$ & $0.02$ & $16.69$ & $0.03$ & $...$ & $...$ & $LCOGT$ \\ 
$  57.74$ & $...$ & $...$ & $18.44$ & $0.04$ & $17.84$ & $0.01$ & $17.06$ & $0.02$ & $16.80$ & $0.02$ & $...$ & $...$ & $LCOGT$ \\ 
$  59.30$ & $19.37$ & $0.31$ & $18.42$ & $0.02$ & $...$ & $...$ & $...$ & $...$ & $...$ & $...$ & $...$ & $...$ & $LCOGT$ \\ 
$  59.37$ & $...$ & $...$ & $...$ & $...$ & $17.93$ & $0.01$ & $17.12$ & $0.02$ & $16.83$ & $0.02$ & $16.08$ & $0.03$ & $LCOGT$ \\ 
$  59.74$ & $19.12$ & $0.35$ & $18.47$ & $0.04$ & $...$ & $...$ & $...$ & $...$ & $...$ & $...$ & $...$ & $...$ & $LCOGT$ \\ 
$  60.34$ & $...$ & $...$ & $18.54$ & $0.05$ & $17.90$ & $0.02$ & $17.12$ & $0.02$ & $16.88$ & $0.02$ & $...$ & $...$ & $LCOGT$ \\ 
$  63.44$ & $...$ & $...$ & $...$ & $...$ & $18.03$ & $0.04$ & $17.26$ & $0.03$ & $16.96$ & $0.03$ & $16.18$ & $0.03$ & $LCOGT$ \\ 
$  65.36$ & $...$ & $...$ & $18.48$ & $0.06$ & $...$ & $...$ & $...$ & $...$ & $...$ & $...$ & $...$ & $...$ & $LCOGT$ \\ 
$  66.45$ & $...$ & $...$ & $...$ & $...$ & $18.07$ & $0.05$ & $17.24$ & $0.03$ & $17.05$ & $0.03$ & $16.16$ & $0.03$ & $LCOGT$ \\ 
$  67.73$ & $...$ & $...$ & $...$ & $...$ & $18.08$ & $0.07$ & $...$ & $...$ & $...$ & $...$ & $...$ & $...$ & $LCOGT$ \\ 
$  67.74$ & $...$ & $...$ & $...$ & $...$ & $...$ & $...$ & $...$ & $...$ & $17.09$ & $0.10$ & $...$ & $...$ & $LCOGT$ \\ 
$  67.75$ & $...$ & $...$ & $...$ & $...$ & $...$ & $...$ & $...$ & $...$ & $...$ & $...$ & $16.16$ & $0.07$ & $LCOGT$ \\ 
$  69.49$ & $...$ & $...$ & $...$ & $...$ & $18.16$ & $0.01$ & $17.30$ & $0.01$ & $17.05$ & $0.01$ & $...$ & $...$ & $  P60$ \\ 
$  69.71$ & $...$ & $...$ & $18.62$ & $0.05$ & $...$ & $...$ & $...$ & $...$ & $...$ & $...$ & $...$ & $...$ & $LCOGT$ \\ 
$  69.74$ & $...$ & $...$ & $...$ & $...$ & $18.05$ & $0.01$ & $17.32$ & $0.01$ & $17.05$ & $0.01$ & $16.25$ & $0.03$ & $LCOGT$ \\ 
$  70.30$ & $...$ & $...$ & $18.60$ & $0.05$ & $...$ & $...$ & $17.32$ & $0.01$ & $17.13$ & $0.02$ & $...$ & $...$ & $LCOGT$ \\ 
$  70.34$ & $...$ & $...$ & $...$ & $...$ & $18.10$ & $0.05$ & $...$ & $...$ & $...$ & $...$ & $...$ & $...$ & $LCOGT$ \\ 
$  70.70$ & $...$ & $...$ & $...$ & $...$ & $18.10$ & $0.01$ & $17.39$ & $0.01$ & $17.09$ & $0.02$ & $16.30$ & $0.03$ & $LCOGT$ \\ 
$  71.33$ & $...$ & $...$ & $18.51$ & $0.04$ & $...$ & $...$ & $...$ & $...$ & $17.08$ & $0.03$ & $16.25$ & $0.04$ & $LCOGT$ \\ 
$  72.43$ & $...$ & $...$ & $...$ & $...$ & $18.06$ & $0.03$ & $17.39$ & $0.03$ & $17.16$ & $0.03$ & $...$ & $...$ & $LCOGT$ \\ 
$  74.47$ & $...$ & $...$ & $...$ & $...$ & $18.20$ & $0.02$ & $17.43$ & $0.02$ & $17.13$ & $0.02$ & $...$ & $...$ & $  P60$ \\ 
$  77.32$ & $...$ & $...$ & $...$ & $...$ & $18.16$ & $0.01$ & $17.48$ & $0.02$ & $...$ & $...$ & $...$ & $...$ & $LCOGT$ \\ 
$  77.33$ & $...$ & $...$ & $...$ & $...$ & $...$ & $...$ & $...$ & $...$ & $17.13$ & $0.05$ & $...$ & $...$ & $LCOGT$ \\ 
$  78.49$ & $...$ & $...$ & $...$ & $...$ & $...$ & $...$ & $17.53$ & $0.01$ & $17.25$ & $0.02$ & $...$ & $...$ & $  P60$ \\ 
$  78.50$ & $...$ & $...$ & $...$ & $...$ & $18.23$ & $0.01$ & $...$ & $...$ & $...$ & $...$ & $...$ & $...$ & $  P60$ \\ 
$  81.05$ & $...$ & $...$ & $...$ & $...$ & $18.28$ & $0.02$ & $17.61$ & $0.02$ & $17.31$ & $0.01$ & $...$ & $...$ & $LCOGT$ \\ 
$  82.05$ & $...$ & $...$ & $...$ & $...$ & $18.19$ & $0.02$ & $17.56$ & $0.02$ & $17.26$ & $0.02$ & $...$ & $...$ & $LCOGT$ \\ 
$  82.49$ & $...$ & $...$ & $...$ & $...$ & $...$ & $...$ & $17.60$ & $0.01$ & $17.29$ & $0.02$ & $...$ & $...$ & $  P60$ \\ 
$  83.42$ & $...$ & $...$ & $...$ & $...$ & $18.29$ & $0.03$ & $17.67$ & $0.04$ & $17.32$ & $0.03$ & $...$ & $...$ & $LCOGT$ \\ 
$  83.48$ & $...$ & $...$ & $...$ & $...$ & $18.37$ & $0.02$ & $17.63$ & $0.02$ & $17.33$ & $0.02$ & $...$ & $...$ & $  P60$ \\ 
$  86.05$ & $...$ & $...$ & $...$ & $...$ & $18.27$ & $0.03$ & $17.70$ & $0.02$ & $...$ & $...$ & $...$ & $...$ & $LCOGT$ \\ 
$  86.06$ & $...$ & $...$ & $...$ & $...$ & $...$ & $...$ & $...$ & $...$ & $17.39$ & $0.02$ & $...$ & $...$ & $LCOGT$ \\ 
$ 226.56$ & $...$ & $...$ & $...$ & $...$ & $21.09$ & $0.04$ & $...$ & $...$ & $...$ & $...$ & $...$ & $...$ & $  NOT$ \\ 
$ 226.57$ & $...$ & $...$ & $...$ & $...$ & $...$ & $...$ & $20.35$ & $0.08$ & $...$ & $...$ & $...$ & $...$ & $  NOT$ \\ 
$ 239.55$ & $...$ & $...$ & $...$ & $...$ & $21.38$ & $0.08$ & $...$ & $...$ & $...$ & $...$ & $...$ & $...$ & $  NOT$ \\ 
$ 239.57$ & $...$ & $...$ & $...$ & $...$ & $...$ & $...$ & $20.58$ & $0.09$ & $...$ & $...$ & $...$ & $...$ & $  NOT$ \\ 
$ 262.48$ & $...$ & $...$ & $...$ & $...$ & $21.81$ & $0.10$ & $...$ & $...$ & $...$ & $...$ & $...$ & $...$ & $  P200$ \\ 
$ 262.49$ & $...$ & $...$ & $...$ & $...$ & $...$ & $...$ & $20.99$ & $0.20$ & $...$ & $...$ & $...$ & $...$ & $  P200$ \\ 
$ 262.50$ & $...$ & $...$ & $...$ & $...$ & $...$ & $...$ & $...$ & $...$ & $20.07$ & $0.15$ & $...$ & $...$ & $  NOT$ \\ 
$ 293.79$ & $...$ & $...$ & $...$ & $...$ & $22.23$ & $0.31$ & $21.42$ & $0.21$ & $...$ & $...$ & $...$ & $...$ & $  P60$ \\ 
$ 320.34$ & $...$ & $...$ & $...$ & $...$ & $22.94$ & $0.29$ & $...$ & $...$ & $...$ & $...$ & $...$ & $...$ & $  NOT$ \\ 
$ 320.35$ & $...$ & $...$ & $...$ & $...$ & $...$ & $...$ & $21.71$ & $0.18$ & $...$ & $...$ & $...$ & $...$ & $  NOT$ \\ 
$ 354.28$ & $...$ & $...$ & $...$ & $...$ & $...$ & $...$ & $21.99$ & $0.28$ & $...$ & $...$ & $...$ & $...$ & $  NOT$ \\ 
\enddata 
 \tablecomments{$U$ and $B$ magnitudes are given in the Vega system. The $griz$ magnitudes are given in the AB system.}

\end{deluxetable}

\begin{deluxetable}{lrlrlrll} 
\tabletypesize{\scriptsize} 
\tablecolumns{8} 
\tablewidth{0pt} 
\tablecaption{Photometry table for iPTF13bvn ($VRI$)\label{tab:13bvnphot2}}
\tablehead{ 
\colhead{Phase} & \colhead{$V$} & \colhead{$\sigma$} & \colhead{$R$} & \colhead{$\sigma$} & \colhead{$I$} & \colhead{$\sigma$} & \colhead{$tel.$} \\ 
\colhead{JD $-$ 2,456,459.17} & \colhead{[mag]} & \colhead{[mag]} & \colhead{[mag]} & \colhead{[mag]} & \colhead{[mag]} & \colhead{[mag]} & \colhead{} \\ 
}
\startdata 
$   2.08$ & $17.33$ & $0.02$ & $...$ & $...$ & $...$ & $...$ & $LCOGT$ \\ 
$   2.14$ & $...$ & $...$ & $...$ & $...$ & $...$ & $...$ & $LCOGT$ \\ 
$   2.33$ & $17.21$ & $0.02$ & $...$ & $...$ & $...$ & $...$ & $LCOGT$ \\ 
$   2.39$ & $...$ & $...$ & $...$ & $...$ & $...$ & $...$ & $LCOGT$ \\ 
$   2.41$ & $...$ & $...$ & $16.93$ & $0.02$ & $16.86$ & $0.02$ & $LCOGT$ \\ 
$   3.03$ & $...$ & $...$ & $...$ & $...$ & $16.59$ & $0.05$ & $LCOGT$ \\ 
$   3.36$ & $16.90$ & $0.01$ & $...$ & $...$ & $...$ & $...$ & $LCOGT$ \\ 
$   3.39$ & $...$ & $...$ & $...$ & $...$ & $...$ & $...$ & $LCOGT$ \\ 
$   3.41$ & $...$ & $...$ & $16.55$ & $0.01$ & $16.43$ & $0.01$ & $LCOGT$ \\ 
$   3.53$ & $16.87$ & $0.02$ & $...$ & $...$ & $...$ & $...$ & $LCOGT$ \\ 
$   3.58$ & $...$ & $...$ & $...$ & $...$ & $...$ & $...$ & $LCOGT$ \\ 
$   3.65$ & $...$ & $...$ & $16.50$ & $0.02$ & $16.48$ & $0.04$ & $LCOGT$ \\ 
$   4.33$ & $16.62$ & $0.02$ & $...$ & $...$ & $...$ & $...$ & $LCOGT$ \\ 
$   4.39$ & $...$ & $...$ & $16.21$ & $0.01$ & $16.16$ & $0.03$ & $LCOGT$ \\ 
$   6.30$ & $...$ & $...$ & $...$ & $...$ & $...$ & $...$ & $LCOGT$ \\ 
$   6.40$ & $16.08$ & $0.01$ & $15.78$ & $0.02$ & $15.72$ & $0.02$ & $LCOGT$ \\ 
$   8.40$ & $15.75$ & $0.01$ & $15.48$ & $0.01$ & $15.40$ & $0.02$ & $LCOGT$ \\ 
$   9.42$ & $15.59$ & $0.01$ & $...$ & $...$ & $...$ & $...$ & $LCOGT$ \\ 
$   9.44$ & $...$ & $...$ & $...$ & $...$ & $...$ & $...$ & $LCOGT$ \\ 
$  10.40$ & $15.57$ & $0.01$ & $15.32$ & $0.01$ & $15.24$ & $0.01$ & $LCOGT$ \\ 
$  12.24$ & $15.36$ & $0.01$ & $...$ & $...$ & $...$ & $...$ & $LCOGT$ \\ 
$  12.40$ & $15.38$ & $0.01$ & $...$ & $...$ & $...$ & $...$ & $LCOGT$ \\ 
$  13.50$ & $15.32$ & $0.01$ & $...$ & $...$ & $...$ & $...$ & $LCOGT$ \\ 
$  16.28$ & $15.24$ & $0.01$ & $...$ & $...$ & $...$ & $...$ & $LCOGT$ \\ 
$  16.38$ & $15.26$ & $0.01$ & $14.96$ & $0.01$ & $14.85$ & $0.01$ & $LCOGT$ \\ 
$  17.38$ & $15.24$ & $0.01$ & $14.93$ & $0.01$ & $14.78$ & $0.01$ & $LCOGT$ \\ 
$  18.57$ & $15.36$ & $0.04$ & $14.95$ & $0.02$ & $14.75$ & $0.04$ & $LCOGT$ \\ 
$  20.56$ & $15.31$ & $0.01$ & $14.93$ & $0.01$ & $14.75$ & $0.01$ & $LCOGT$ \\ 
$  21.09$ & $...$ & $...$ & $...$ & $...$ & $...$ & $...$ & $LCOGT$ \\ 
$  21.19$ & $15.32$ & $0.02$ & $...$ & $...$ & $...$ & $...$ & $LCOGT$ \\ 
$  21.38$ & $15.33$ & $0.01$ & $14.92$ & $0.01$ & $14.77$ & $0.01$ & $LCOGT$ \\ 
$  21.56$ & $15.38$ & $0.01$ & $14.96$ & $0.01$ & $14.75$ & $0.01$ & $LCOGT$ \\ 
$  22.52$ & $15.44$ & $0.01$ & $15.02$ & $0.01$ & $14.78$ & $0.01$ & $LCOGT$ \\ 
$  24.03$ & $15.61$ & $0.01$ & $...$ & $...$ & $...$ & $...$ & $LCOGT$ \\ 
$  26.52$ & $15.87$ & $0.01$ & $15.34$ & $0.01$ & $15.04$ & $0.02$ & $LCOGT$ \\ 
$  27.18$ & $...$ & $...$ & $...$ & $...$ & $...$ & $...$ & $LCOGT$ \\ 
$  28.41$ & $16.11$ & $0.01$ & $...$ & $...$ & $...$ & $...$ & $LCOGT$ \\ 
$  28.52$ & $16.12$ & $0.01$ & $15.52$ & $0.01$ & $15.20$ & $0.02$ & $LCOGT$ \\ 
$  29.55$ & $...$ & $...$ & $...$ & $...$ & $15.30$ & $0.05$ & $LCOGT$ \\ 
$  35.53$ & $...$ & $...$ & $...$ & $...$ & $...$ & $...$ & $LCOGT$ \\ 
$  36.53$ & $...$ & $...$ & $...$ & $...$ & $...$ & $...$ & $LCOGT$ \\ 
$  40.54$ & $16.95$ & $0.02$ & $16.27$ & $0.02$ & $15.82$ & $0.02$ & $LCOGT$ \\ 
$  45.07$ & $17.06$ & $0.02$ & $...$ & $...$ & $...$ & $...$ & $LCOGT$ \\ 
$  49.35$ & $17.13$ & $0.01$ & $...$ & $...$ & $...$ & $...$ & $LCOGT$ \\ 
$  49.48$ & $...$ & $...$ & $...$ & $...$ & $15.98$ & $0.05$ & $LCOGT$ \\ 
$  51.30$ & $17.25$ & $0.01$ & $...$ & $...$ & $...$ & $...$ & $LCOGT$ \\ 
$  51.47$ & $...$ & $...$ & $16.56$ & $0.05$ & $15.92$ & $0.13$ & $LCOGT$ \\ 
$  51.49$ & $17.30$ & $0.02$ & $...$ & $...$ & $...$ & $...$ & $LCOGT$ \\ 
$  52.29$ & $...$ & $...$ & $16.62$ & $0.02$ & $16.07$ & $0.02$ & $LCOGT$ \\ 
$  52.79$ & $17.22$ & $0.02$ & $...$ & $...$ & $...$ & $...$ & $LCOGT$ \\ 
$  53.48$ & $...$ & $...$ & $16.69$ & $0.03$ & $16.05$ & $0.02$ & $LCOGT$ \\ 
$  53.51$ & $17.26$ & $0.02$ & $...$ & $...$ & $...$ & $...$ & $LCOGT$ \\ 
$  53.75$ & $...$ & $...$ & $...$ & $...$ & $...$ & $...$ & $LCOGT$ \\ 
$  53.79$ & $17.25$ & $0.03$ & $...$ & $...$ & $...$ & $...$ & $LCOGT$ \\ 
$  54.39$ & $...$ & $...$ & $16.62$ & $0.02$ & $16.06$ & $0.03$ & $LCOGT$ \\ 
$  55.78$ & $17.25$ & $0.02$ & $...$ & $...$ & $...$ & $...$ & $LCOGT$ \\ 
$  56.40$ & $...$ & $...$ & $...$ & $...$ & $...$ & $...$ & $LCOGT$ \\ 
$  57.74$ & $17.30$ & $0.02$ & $...$ & $...$ & $...$ & $...$ & $LCOGT$ \\ 
$  59.30$ & $17.38$ & $0.02$ & $...$ & $...$ & $...$ & $...$ & $LCOGT$ \\ 
$  59.31$ & $...$ & $...$ & $16.74$ & $0.02$ & $16.23$ & $0.02$ & $LCOGT$ \\ 
$  59.74$ & $17.35$ & $0.02$ & $...$ & $...$ & $...$ & $...$ & $LCOGT$ \\ 
$  59.75$ & $...$ & $...$ & $16.71$ & $0.02$ & $16.18$ & $0.02$ & $LCOGT$ \\ 
$  60.34$ & $17.48$ & $0.03$ & $16.84$ & $0.01$ & $16.23$ & $0.02$ & $LCOGT$ \\ 
$  65.36$ & $17.51$ & $0.03$ & $16.85$ & $0.02$ & $16.27$ & $0.06$ & $LCOGT$ \\ 
$  67.73$ & $17.52$ & $0.02$ & $16.91$ & $0.02$ & $16.30$ & $0.04$ & $LCOGT$ \\ 
$  69.71$ & $17.59$ & $0.02$ & $17.04$ & $0.02$ & $16.40$ & $0.03$ & $LCOGT$ \\ 
$  70.30$ & $17.63$ & $0.02$ & $17.03$ & $0.02$ & $...$ & $...$ & $LCOGT$ \\ 
$  70.35$ & $17.60$ & $0.02$ & $17.06$ & $0.01$ & $16.29$ & $0.03$ & $LCOGT$ \\ 
$  71.33$ & $17.54$ & $0.02$ & $16.98$ & $0.03$ & $16.35$ & $0.04$ & $LCOGT$ \\ 
\enddata 
 \tablecomments{Magnitudes are given in the Vega system.}
\end{deluxetable}

\begin{deluxetable}{lrlrlrlrlrlrll} 
\tabletypesize{\scriptsize} 
\tablecolumns{14} 
\tablewidth{0pt} 
\tablecaption{Photometry table for PTF12os/SN2012P\label{tab:12osphot}}
\tablehead{ 
\colhead{Phase} & \colhead{$U$} & \colhead{$\sigma$} & \colhead{$B$} & \colhead{$\sigma$} & \colhead{$g$} & \colhead{$\sigma$} & \colhead{$r$} & \colhead{$\sigma$} & \colhead{$i$} & \colhead{$\sigma$} & \colhead{$z$} & \colhead{$\sigma$} & \colhead{$tel.$} \\ 
\colhead{JD $-$ 2,455,933.00} & \colhead{[mag]} & \colhead{[mag]} & \colhead{[mag]} & \colhead{[mag]} & \colhead{[mag]} & \colhead{[mag]} & \colhead{[mag]} & \colhead{[mag]} & \colhead{[mag]} & \colhead{[mag]} & \colhead{[mag]} & \colhead{[mag]} & \colhead{} \\ 
}
\startdata 
$   3.98$ & $...$ & $...$ & $...$ & $...$ & $...$ & $...$ & $17.25$ & $0.03$ & $...$ & $...$ & $...$ & $...$ & $  P48$ \\ 
$   6.98$ & $...$ & $...$ & $...$ & $...$ & $...$ & $...$ & $16.72$ & $0.02$ & $...$ & $...$ & $...$ & $...$ & $  P48$ \\ 
$   7.69$ & $18.31$ & $0.13$ & $...$ & $...$ & $...$ & $...$ & $...$ & $...$ & $...$ & $...$ & $...$ & $...$ & $Swift$ \\ 
$   7.96$ & $...$ & $...$ & $...$ & $...$ & $...$ & $...$ & $...$ & $...$ & $...$ & $...$ & $...$ & $...$ & $  P60$ \\ 
$  10.09$ & $17.95$ & $0.09$ & $...$ & $...$ & $...$ & $...$ & $...$ & $...$ & $...$ & $...$ & $...$ & $...$ & $Swift$ \\ 
$  11.93$ & $...$ & $...$ & $17.11$ & $0.02$ & $16.76$ & $0.02$ & $16.20$ & $0.02$ & $16.13$ & $0.04$ & $16.14$ & $0.07$ & $  P60$ \\ 
$  12.95$ & $...$ & $...$ & $...$ & $...$ & $16.65$ & $0.01$ & $16.14$ & $0.01$ & $16.03$ & $0.01$ & $...$ & $...$ & $  P60$ \\ 
$  14.03$ & $...$ & $...$ & $...$ & $...$ & $...$ & $...$ & $16.13$ & $0.01$ & $...$ & $...$ & $...$ & $...$ & $  P48$ \\ 
$  15.99$ & $...$ & $...$ & $16.93$ & $0.02$ & $16.53$ & $0.01$ & $15.98$ & $0.03$ & $15.88$ & $0.02$ & $15.84$ & $0.02$ & $  P60$ \\ 
$  18.17$ & $18.24$ & $0.10$ & $...$ & $...$ & $...$ & $...$ & $...$ & $...$ & $...$ & $...$ & $...$ & $...$ & $Swift$ \\ 
$  18.52$ & $18.66$ & $0.14$ & $...$ & $...$ & $...$ & $...$ & $...$ & $...$ & $...$ & $...$ & $...$ & $...$ & $Swift$ \\ 
$  18.93$ & $...$ & $...$ & $17.10$ & $0.03$ & $16.61$ & $0.02$ & $15.93$ & $0.01$ & $15.82$ & $0.02$ & $15.74$ & $0.04$ & $  P60$ \\ 
$  19.03$ & $...$ & $...$ & $...$ & $...$ & $...$ & $...$ & $15.96$ & $0.01$ & $...$ & $...$ & $...$ & $...$ & $  P48$ \\ 
$  20.03$ & $...$ & $...$ & $17.18$ & $0.02$ & $16.68$ & $0.01$ & $15.93$ & $0.00$ & $15.81$ & $0.01$ & $15.73$ & $0.02$ & $  P60$ \\ 
$  20.52$ & $18.73$ & $0.16$ & $...$ & $...$ & $...$ & $...$ & $...$ & $...$ & $...$ & $...$ & $...$ & $...$ & $Swift$ \\ 
$  20.93$ & $...$ & $...$ & $...$ & $...$ & $16.75$ & $0.01$ & $15.93$ & $0.01$ & $15.79$ & $0.02$ & $...$ & $...$ & $  P60$ \\ 
$  21.70$ & $...$ & $...$ & $...$ & $...$ & $16.86$ & $0.04$ & $15.90$ & $0.05$ & $15.82$ & $0.03$ & $15.73$ & $0.05$ & $   LT$ \\ 
$  21.92$ & $...$ & $...$ & $...$ & $...$ & $16.86$ & $0.05$ & $15.97$ & $0.09$ & $15.81$ & $0.03$ & $...$ & $...$ & $  P60$ \\ 
$  22.52$ & $18.97$ & $0.18$ & $...$ & $...$ & $...$ & $...$ & $...$ & $...$ & $...$ & $...$ & $...$ & $...$ & $Swift$ \\ 
$  24.92$ & $...$ & $...$ & $17.60$ & $0.06$ & $17.07$ & $0.02$ & $16.08$ & $0.01$ & $15.87$ & $0.02$ & $15.78$ & $0.03$ & $  P60$ \\ 
$  25.91$ & $...$ & $...$ & $...$ & $...$ & $17.16$ & $0.01$ & $16.12$ & $0.01$ & $15.89$ & $0.02$ & $...$ & $...$ & $  P60$ \\ 
$  26.00$ & $...$ & $...$ & $...$ & $...$ & $...$ & $...$ & $16.12$ & $0.01$ & $...$ & $...$ & $...$ & $...$ & $  P48$ \\ 
$  26.69$ & $...$ & $...$ & $...$ & $...$ & $17.24$ & $0.03$ & $16.11$ & $0.08$ & $15.92$ & $0.05$ & $15.81$ & $0.07$ & $   LT$ \\ 
$  26.91$ & $...$ & $...$ & $17.92$ & $0.05$ & $17.28$ & $0.02$ & $16.17$ & $0.01$ & $15.93$ & $0.03$ & $15.83$ & $0.05$ & $  P60$ \\ 
$  27.74$ & $...$ & $...$ & $...$ & $...$ & $17.34$ & $0.06$ & $16.19$ & $0.05$ & $15.96$ & $0.04$ & $15.82$ & $0.05$ & $   LT$ \\ 
$  27.91$ & $...$ & $...$ & $18.11$ & $0.07$ & $17.37$ & $0.02$ & $16.23$ & $0.01$ & $15.96$ & $0.04$ & $15.87$ & $0.02$ & $  P60$ \\ 
$  29.00$ & $...$ & $...$ & $...$ & $...$ & $...$ & $...$ & $16.27$ & $0.01$ & $...$ & $...$ & $...$ & $...$ & $  P48$ \\ 
$  29.01$ & $...$ & $...$ & $...$ & $...$ & $...$ & $...$ & $16.27$ & $0.01$ & $...$ & $...$ & $...$ & $...$ & $  P48$ \\ 
$  29.94$ & $...$ & $...$ & $...$ & $...$ & $...$ & $...$ & $...$ & $...$ & $...$ & $...$ & $15.91$ & $0.02$ & $  P60$ \\ 
$  30.66$ & $...$ & $...$ & $...$ & $...$ & $...$ & $...$ & $...$ & $...$ & $16.01$ & $0.10$ & $15.89$ & $0.06$ & $   LT$ \\ 
$  30.93$ & $...$ & $...$ & $...$ & $...$ & $...$ & $...$ & $...$ & $...$ & $...$ & $...$ & $15.96$ & $0.02$ & $  P60$ \\ 
$  31.66$ & $...$ & $...$ & $...$ & $...$ & $...$ & $...$ & $16.45$ & $0.06$ & $16.10$ & $0.05$ & $15.82$ & $0.07$ & $   LT$ \\ 
$  33.73$ & $...$ & $...$ & $...$ & $...$ & $17.80$ & $0.08$ & $...$ & $...$ & $16.19$ & $0.07$ & $16.07$ & $0.09$ & $   LT$ \\ 
$  33.89$ & $...$ & $...$ & $18.56$ & $0.14$ & $17.89$ & $0.05$ & $16.59$ & $0.03$ & $16.21$ & $0.03$ & $16.06$ & $0.04$ & $  P60$ \\ 
$  33.92$ & $...$ & $...$ & $18.59$ & $0.21$ & $...$ & $...$ & $...$ & $...$ & $...$ & $...$ & $...$ & $...$ & $  P60$ \\ 
$  35.96$ & $...$ & $...$ & $...$ & $...$ & $...$ & $...$ & $16.64$ & $0.01$ & $...$ & $...$ & $...$ & $...$ & $  P48$ \\ 
$  40.86$ & $...$ & $...$ & $...$ & $...$ & $...$ & $...$ & $...$ & $...$ & $16.45$ & $0.01$ & $...$ & $...$ & $  NTT$ \\ 
$  41.87$ & $...$ & $...$ & $19.02$ & $0.09$ & $18.38$ & $0.04$ & $16.92$ & $0.02$ & $16.49$ & $0.04$ & $16.25$ & $0.05$ & $  P60$ \\ 
$  41.92$ & $...$ & $...$ & $19.08$ & $0.08$ & $...$ & $...$ & $...$ & $...$ & $...$ & $...$ & $...$ & $...$ & $  P60$ \\ 
$  42.87$ & $...$ & $...$ & $19.12$ & $0.05$ & $18.40$ & $0.02$ & $16.96$ & $0.01$ & $16.51$ & $0.02$ & $...$ & $...$ & $  P60$ \\ 
$  42.92$ & $...$ & $...$ & $...$ & $...$ & $...$ & $...$ & $16.91$ & $0.01$ & $...$ & $...$ & $...$ & $...$ & $  P48$ \\ 
$  42.97$ & $...$ & $...$ & $...$ & $...$ & $...$ & $...$ & $16.89$ & $0.01$ & $...$ & $...$ & $...$ & $...$ & $  P48$ \\ 
$  44.93$ & $...$ & $...$ & $...$ & $...$ & $...$ & $...$ & $...$ & $...$ & $...$ & $...$ & $16.31$ & $0.02$ & $  P60$ \\ 
$  45.99$ & $...$ & $...$ & $...$ & $...$ & $...$ & $...$ & $17.01$ & $0.01$ & $...$ & $...$ & $...$ & $...$ & $  P48$ \\ 
$  46.05$ & $...$ & $...$ & $...$ & $...$ & $...$ & $...$ & $17.02$ & $0.01$ & $...$ & $...$ & $...$ & $...$ & $  P48$ \\ 
$  46.73$ & $...$ & $...$ & $...$ & $...$ & $18.56$ & $0.06$ & $17.14$ & $0.03$ & $16.65$ & $0.03$ & $16.34$ & $0.04$ & $   LT$ \\ 
$  48.99$ & $...$ & $...$ & $...$ & $...$ & $...$ & $...$ & $17.13$ & $0.01$ & $...$ & $...$ & $...$ & $...$ & $  P48$ \\ 
$  49.04$ & $...$ & $...$ & $...$ & $...$ & $...$ & $...$ & $17.11$ & $0.01$ & $...$ & $...$ & $...$ & $...$ & $  P48$ \\ 
$  50.85$ & $...$ & $...$ & $...$ & $...$ & $...$ & $...$ & $...$ & $...$ & $...$ & $...$ & $16.38$ & $0.02$ & $  P60$ \\ 
$  55.91$ & $...$ & $...$ & $19.40$ & $0.09$ & $18.71$ & $0.03$ & $17.34$ & $0.01$ & $16.90$ & $0.02$ & $16.46$ & $0.04$ & $  P60$ \\ 
$  56.01$ & $...$ & $...$ & $19.45$ & $0.06$ & $...$ & $...$ & $...$ & $...$ & $...$ & $...$ & $...$ & $...$ & $  P60$ \\ 
$  56.86$ & $...$ & $...$ & $...$ & $...$ & $...$ & $...$ & $17.31$ & $0.03$ & $...$ & $...$ & $...$ & $...$ & $  P48$ \\ 
$  56.90$ & $...$ & $...$ & $...$ & $...$ & $...$ & $...$ & $17.30$ & $0.02$ & $...$ & $...$ & $...$ & $...$ & $  P48$ \\ 
$  57.61$ & $...$ & $...$ & $...$ & $...$ & $...$ & $...$ & $17.32$ & $0.04$ & $...$ & $...$ & $16.52$ & $0.05$ & $   LT$ \\ 
$  59.73$ & $...$ & $...$ & $...$ & $...$ & $...$ & $...$ & $...$ & $0.06$ & $16.95$ & $0.08$ & $...$ & $...$ & $   LT$ \\ 
$  59.83$ & $...$ & $...$ & $19.61$ & $0.27$ & $18.79$ & $0.06$ & $17.45$ & $0.02$ & $16.99$ & $0.03$ & $16.52$ & $0.03$ & $  P60$ \\ 
$  65.72$ & $...$ & $...$ & $...$ & $...$ & $18.81$ & $0.06$ & $...$ & $...$ & $...$ & $...$ & $16.61$ & $0.07$ & $   LT$ \\ 
$  67.84$ & $...$ & $...$ & $...$ & $...$ & $...$ & $...$ & $17.50$ & $0.01$ & $...$ & $...$ & $...$ & $...$ & $  P48$ \\ 
$  67.89$ & $...$ & $...$ & $...$ & $...$ & $...$ & $...$ & $17.48$ & $0.01$ & $...$ & $...$ & $...$ & $...$ & $  P48$ \\ 
$  69.80$ & $...$ & $...$ & $...$ & $...$ & $...$ & $...$ & $17.52$ & $0.02$ & $...$ & $...$ & $...$ & $...$ & $  P48$ \\ 
$  71.76$ & $...$ & $...$ & $...$ & $...$ & $18.95$ & $0.04$ & $17.56$ & $0.03$ & $...$ & $...$ & $16.70$ & $0.03$ & $   LT$ \\ 
$  73.97$ & $...$ & $...$ & $...$ & $...$ & $...$ & $...$ & $17.62$ & $0.01$ & $...$ & $...$ & $...$ & $...$ & $  P48$ \\ 
$  74.99$ & $...$ & $...$ & $...$ & $...$ & $...$ & $...$ & $17.64$ & $0.01$ & $...$ & $...$ & $...$ & $...$ & $  P48$ \\ 
$  75.78$ & $...$ & $...$ & $...$ & $...$ & $...$ & $...$ & $17.64$ & $0.02$ & $...$ & $...$ & $...$ & $...$ & $  P48$ \\ 
$  77.72$ & $...$ & $...$ & $...$ & $...$ & $19.03$ & $0.08$ & $17.73$ & $0.02$ & $17.31$ & $0.03$ & $16.77$ & $0.05$ & $   LT$ \\ 
$  81.02$ & $...$ & $...$ & $...$ & $...$ & $...$ & $...$ & $17.75$ & $0.02$ & $...$ & $...$ & $...$ & $...$ & $  P48$ \\ 
$  82.85$ & $...$ & $...$ & $19.54$ & $0.07$ & $19.03$ & $0.02$ & $17.82$ & $0.01$ & $17.43$ & $0.03$ & $16.79$ & $0.03$ & $  P60$ \\ 
$  83.75$ & $...$ & $...$ & $...$ & $...$ & $19.05$ & $0.08$ & $17.81$ & $0.06$ & $17.43$ & $0.05$ & $...$ & $...$ & $  P60$ \\ 
$  84.75$ & $...$ & $...$ & $19.49$ & $0.19$ & $19.04$ & $0.05$ & $17.82$ & $0.03$ & $17.46$ & $0.04$ & $16.84$ & $0.04$ & $  P60$ \\ 
$  86.75$ & $...$ & $...$ & $...$ & $...$ & $19.06$ & $0.15$ & $17.87$ & $0.05$ & $17.48$ & $0.06$ & $16.88$ & $0.07$ & $  P60$ \\ 
$  87.74$ & $...$ & $...$ & $19.56$ & $0.22$ & $19.08$ & $0.08$ & $17.88$ & $0.04$ & $17.50$ & $0.04$ & $...$ & $...$ & $  P60$ \\ 
$  88.74$ & $...$ & $...$ & $...$ & $...$ & $...$ & $...$ & $...$ & $...$ & $17.52$ & $0.15$ & $16.83$ & $0.13$ & $  P60$ \\ 
$  89.74$ & $...$ & $...$ & $...$ & $...$ & $19.03$ & $0.13$ & $17.89$ & $0.05$ & $17.50$ & $0.04$ & $16.92$ & $0.05$ & $  P60$ \\ 
$  89.75$ & $...$ & $...$ & $...$ & $...$ & $...$ & $...$ & $...$ & $...$ & $...$ & $...$ & $...$ & $...$ & $  P60$ \\ 
$  90.74$ & $...$ & $...$ & $...$ & $...$ & $...$ & $...$ & $...$ & $...$ & $17.49$ & $0.18$ & $...$ & $...$ & $  P60$ \\ 
$  94.72$ & $...$ & $...$ & $19.50$ & $0.11$ & $19.11$ & $0.04$ & $17.98$ & $0.03$ & $17.60$ & $0.03$ & $16.99$ & $0.04$ & $  P60$ \\ 
$ 100.71$ & $...$ & $...$ & $...$ & $...$ & $19.18$ & $0.04$ & $18.05$ & $0.04$ & $17.67$ & $0.05$ & $17.05$ & $0.06$ & $  P60$ \\ 
$ 100.79$ & $...$ & $...$ & $19.68$ & $0.09$ & $...$ & $...$ & $...$ & $...$ & $...$ & $...$ & $17.03$ & $0.03$ & $  P60$ \\ 
$ 101.70$ & $...$ & $...$ & $...$ & $...$ & $19.20$ & $0.05$ & $18.08$ & $0.03$ & $17.70$ & $0.03$ & $...$ & $...$ & $  P60$ \\ 
$ 102.70$ & $...$ & $...$ & $...$ & $...$ & $19.19$ & $0.06$ & $18.07$ & $0.03$ & $17.70$ & $0.04$ & $...$ & $...$ & $  P60$ \\ 
$ 103.91$ & $...$ & $...$ & $...$ & $...$ & $19.21$ & $0.02$ & $18.12$ & $0.01$ & $17.74$ & $0.02$ & $...$ & $...$ & $  P60$ \\ 
$ 104.74$ & $...$ & $...$ & $...$ & $...$ & $19.23$ & $0.02$ & $18.12$ & $0.01$ & $17.72$ & $0.02$ & $...$ & $...$ & $  P60$ \\ 
$ 105.69$ & $...$ & $...$ & $...$ & $...$ & $19.25$ & $0.03$ & $18.10$ & $0.01$ & $17.76$ & $0.03$ & $...$ & $...$ & $  P60$ \\ 
$ 106.69$ & $...$ & $...$ & $...$ & $...$ & $19.24$ & $0.02$ & $18.14$ & $0.01$ & $17.75$ & $0.02$ & $...$ & $...$ & $  P60$ \\ 
$ 107.70$ & $...$ & $...$ & $...$ & $...$ & $19.25$ & $0.06$ & $...$ & $0.91$ & $17.76$ & $0.03$ & $...$ & $...$ & $  P60$ \\ 
$ 108.78$ & $...$ & $...$ & $...$ & $...$ & $19.26$ & $0.06$ & $18.13$ & $0.03$ & $17.80$ & $0.05$ & $...$ & $...$ & $  P60$ \\ 
$ 115.57$ & $...$ & $...$ & $...$ & $...$ & $19.33$ & $0.05$ & $18.24$ & $0.04$ & $...$ & $...$ & $...$ & $...$ & $  NOT$ \\ 
$ 115.59$ & $...$ & $...$ & $...$ & $...$ & $...$ & $...$ & $...$ & $...$ & $...$ & $...$ & $17.15$ & $0.09$ & $  NOT$ \\ 
$ 128.47$ & $...$ & $...$ & $...$ & $...$ & $19.49$ & $0.05$ & $18.40$ & $0.05$ & $18.00$ & $0.07$ & $17.56$ & $0.13$ & $  NOT$ \\ 
$ 139.64$ & $...$ & $...$ & $...$ & $...$ & $19.59$ & $0.03$ & $18.57$ & $0.03$ & $18.25$ & $0.02$ & $...$ & $...$ & $  GTC$ \\ 
$ 152.85$ & $...$ & $...$ & $...$ & $...$ & $19.78$ & $0.10$ & $18.72$ & $0.05$ & $18.50$ & $0.06$ & $...$ & $...$ & $  P60$ \\ 
$ 155.84$ & $...$ & $...$ & $20.53$ & $0.14$ & $19.81$ & $0.04$ & $18.79$ & $0.02$ & $18.54$ & $0.03$ & $...$ & $...$ & $  P60$ \\ 
$ 156.82$ & $...$ & $...$ & $20.44$ & $0.13$ & $19.85$ & $0.04$ & $18.78$ & $0.02$ & $18.55$ & $0.02$ & $...$ & $...$ & $  P60$ \\ 
$ 157.80$ & $...$ & $...$ & $...$ & $...$ & $...$ & $...$ & $...$ & $...$ & $18.57$ & $0.02$ & $...$ & $...$ & $  P60$ \\ 
$ 157.81$ & $...$ & $...$ & $20.52$ & $0.11$ & $19.85$ & $0.03$ & $18.83$ & $0.01$ & $...$ & $...$ & $...$ & $...$ & $  P60$ \\ 
$ 158.86$ & $...$ & $...$ & $20.42$ & $0.13$ & $19.88$ & $0.05$ & $18.83$ & $0.02$ & $18.57$ & $0.05$ & $...$ & $...$ & $  P60$ \\ 
$ 160.82$ & $...$ & $...$ & $20.45$ & $0.09$ & $19.91$ & $0.02$ & $18.86$ & $0.01$ & $18.60$ & $0.03$ & $...$ & $...$ & $  P60$ \\ 
$ 168.83$ & $...$ & $...$ & $20.51$ & $0.19$ & $20.04$ & $0.04$ & $18.95$ & $0.04$ & $18.76$ & $0.10$ & $...$ & $...$ & $  P60$ \\ 
$ 171.82$ & $...$ & $...$ & $20.71$ & $0.17$ & $20.10$ & $0.06$ & $18.99$ & $0.03$ & $18.79$ & $0.04$ & $...$ & $...$ & $  P60$ \\ 
$ 177.83$ & $...$ & $...$ & $...$ & $...$ & $...$ & $...$ & $...$ & $...$ & $18.90$ & $0.05$ & $...$ & $...$ & $  P60$ \\ 
$ 178.81$ & $...$ & $...$ & $...$ & $...$ & $...$ & $...$ & $19.01$ & $0.15$ & $18.81$ & $0.17$ & $...$ & $...$ & $  P60$ \\ 
$ 179.80$ & $...$ & $...$ & $...$ & $...$ & $20.18$ & $0.14$ & $19.17$ & $0.09$ & $18.91$ & $0.10$ & $...$ & $...$ & $  P60$ \\ 
$ 182.81$ & $...$ & $...$ & $...$ & $...$ & $...$ & $...$ & $19.15$ & $0.07$ & $18.95$ & $0.08$ & $...$ & $...$ & $  P60$ \\ 
$ 186.81$ & $...$ & $...$ & $...$ & $...$ & $...$ & $...$ & $19.16$ & $0.13$ & $19.00$ & $0.09$ & $...$ & $...$ & $  P60$ \\ 
$ 214.54$ & $...$ & $...$ & $...$ & $...$ & $20.81$ & $0.12$ & $19.79$ & $0.06$ & $...$ & $...$ & $...$ & $...$ & $  P60$ \\ 
$ 214.54$ & $...$ & $...$ & $...$ & $...$ & $...$ & $...$ & $...$ & $...$ & $19.61$ & $0.16$ & $...$ & $...$ & $  NTT$ \\ 
\enddata 
\tablecomments{The $B$-band magnitudes are given in the Vega system. The {\it Swift} $U$-band and the $griz$ magnitudes are given in the AB system.}
\end{deluxetable}

\begin{table}
\caption{{\it Swift}-UVOT Photometry of PTF12os/SN~2012P (AB magnitudes)}
\centering
\begin{tabular}{lllllll}
\hline \hline
Phase  & $U$ & $\sigma$   & $UVW1$  & $\sigma$   & $UVM2$  & $\sigma$ \\
JD $-$ 2,455,933.00  & [mag] & [mag] & [mag]      & [mag]      & [mag]      & [mag] \\
    \\ 
\hline 
 7.69   & 18.31 & 0.13 & 20.74 & 0.47 & 21.73 & 0.54    \\
  10.09 & 17.95 & 0.09 & 20.08 & 0.27 & 21.29 & 0.39    \\ 
  18.17  & 18.24 & 0.10 & 20.12 & 0.24 & 21.70 & 0.47    \\
  18.52  & 18.66 & 0.14 & 20.76 & 0.40 & 21.68 & 0.46    \\
  20.52 & 18.73 & 0.16 & 20.92 & 0.48 & 21.76 & 0.52    \\
  22.52   & 18.97 & 0.18 & 20.79 & 0.41 & ... & ...    \\
\hline
\end{tabular}
\label{tab:12osswift}
\end{table}

\begin{table}
\caption{{\it Swift}-UVOT Photometry of iPTF13bvn (AB magnitudes)}
\centering
\begin{tabular}{lllllll}
\hline \hline
Phase  & $U$ & $\sigma$   & $UVW1$  & $\sigma$   & $UVM2$  & $\sigma$ \\
 JD $-$ 2,456,459.17  & [mag] & [mag] & [mag]      & [mag]      & [mag]      & [mag] \\
    \\ 
\hline 
2.00 & ... & ... & 20.95 & 0.41 & ... & ... \\ 
3.54 & 18.83 & 0.17 & ... & ... & ... & ... \\ 
5.88 & 17.49 & 0.07 & 19.88 & 0.18 & ... & ... \\ 
11.36 & 16.77 & 0.04 & 19.00 & 0.08 & 22.25 & 0.71 \\ 
14.56 & 16.98 & 0.05 & 19.37 & 0.11 & 22.15 & 0.66 \\ 
18.91 & 17.53 & 0.06 & 20.20 & 0.22 & ... & ... \\ 
22.91 & 18.20 & 0.12 & 20.23 & 0.27 & ... & ... \\ 
23.56 & ... & ... & ... & ... & ... & ... \\ 
26.59 & 19.24 & 0.22 & ... & ... & ... & ... \\ 
37.78 & 20.29 & 0.38 & ... & ... & ... & ... \\ 
\hline

\end{tabular}
\label{tab:13bvnswift}
\end{table}

\clearpage
\begin{deluxetable}{lcccc}
\tabletypesize{\scriptsize}
\tablewidth{0pt}
\tablecaption{Log of the spectral observations of PTF12os\label{tab:spec12os}}
\tablehead{
\colhead{Date} &
\colhead{Phase}&
\colhead{Telescope+Instrument}&
\colhead{Range}\\
\colhead{[2012]}&
\colhead{JD $-$ 2,455,933.00}&
\colhead{[days]}&
\colhead{[\AA]}}
\startdata
14 Jan. &  +8.5         &  HET+LRS     & 4148-10403  \\
17 Jan. &  +11.5        &  GEMINI N.+GMOS    & 3500-9654   \\
18 Jan. &  +12.5        &  P200+DBSP     & 3470-10200  \\
22 Jan. &  +16.66       &  ASIAGO 1.82m+AFOSC &   3882-8180   \\
25 Jan. &  +19.75       &  NOT+ALFOSC & 3560-9086  \\
26 Jan. &  +20.5        &  KPNO 4m+RC Spec     & 3500-8450  \\
29 Jan. &  +23.63       &  ASIAGO 1.82m+AFOSC  &  3892-8180  \\
02 Feb. &  +27.5        &  Lick 3m+Kast         &   3500-10000 \\
10 Feb. &  +35.69       &  CA-2.2+CAFOS       & 3877-10120   \\
15 Feb.  &  +40.88       &  NTT+EFOSC  & 3910-9295  \\
20 Feb.  &  +45.03       &  Keck~1+LRIS       & 3830-10250  \\
23 Feb. &  +48.5        &  Lick 3m+Kast    & 3504-10152 \\
24 Feb. &  +49.77       &  NOT+ALFOSC     & 3974-9153   \\
25 Feb.   &  +50.5        &  NOT+ALFOSC     & 3402-9129   \\
28 Feb.  &  +53.63       &  CA-2.2+CAFOS      & 3877-10120  \\
29 Apr.  &  +1145       &  Keck~1+LRIS         & 3299-10250   \\
30 Apr.  &  +115.57      &  NOT+ALFOSC    & 3974-8999   \\
23 May   &  +138.66      &  GTC+OSIRIS  &   3600-10000   \\
08 Aug.  &  +214.5       &  NTT+EFOSC  & 3910-9295 \\ 
 \enddata
 \end{deluxetable}

 \begin{deluxetable}{lccccc}
\tabletypesize{\scriptsize}
\tablewidth{0pt}
\tablecaption{Log of the spectral observations of iPTF13bvn\label{tab:spec13bvn}}
\tablehead{
\colhead{Date} &
\colhead{Phase}&
\colhead{Telescope+Instrument}&
\colhead{Range}&
\colhead{Reference\tablenotemark{a}}\\
\colhead{}&
\colhead{JD $-$ 2,456,459.17}&
\colhead{[days]}&
\colhead{[\AA]}&
\colhead{}}
\startdata
17 June 2013  &  +2.15         &       SALT+Spectrograph    & 3500-9000  & This work\\
17 June 2013  &  +2.3        &       TNG+DOLORES     & 3164-7987& 1 \\
18 June 2013   &  +2.6        &      HET+LRS     & 4000-11000& 1 \\
18 June 2013   &  +3.4       &  Magellan Baade+FIRE & 9000-25000 & 1 \\
19 June 2013    &  +3.7        &   FTN+FLOYDS      & 3179-11038 &  1 \\
20 June 2013      &  +4.6       & FTS+FLOYDS & 3179-11038 & 1 \\
21 June 2013     &  +5.6        &     FTN+FLOYDS       & 3179-11038 &  1 \\
22 June 2013   &  +6.7       & FTN+FLOYDS    & 3179-11038 & 1 \\
24 June 2013    &  +8.7       &   FTN+FLOYDS     & 3179-11038   & 1 \\
24 June 2013  &  +8.5       &      Magellan Baade+FIRE     & 9000-25000 & 1 \\
26 June 2013     &  +10.4        &      HET+LRS       & 4172-10800  & 1 \\
27 June 2013    &  +11.5       &   APO+DIS     & 3338-10039 & 1 \\
27 June 2013  &  +11.7        &  FTN+FLOYDS      & 3179-11038     & 1 \\
01 July 2013      &  +15.5       &   HET+LRS    & 4172-10800 & 1 \\
02 July 2013    &  +16.7      &      IRTF+SpeX      & 9000-25000 & 1 \\
03 July 2013     &  +17.7       & FTN+FLOYDS    & 3179-11038 & This work \\
05 July 2013    &  +19.6       & P200+DBSP            & 3200-10100 & 2\\
09 July 2013    &  +23.7       & FTN+FLOYDS   & 3179-11038  & This work \\
11 July 2013    &  +25.7      &  Keck~2+DEIMOS       & 4450-9625  & 2\\
18 July 2013    &  +32.6       & FTN+FLOYDS   & 3179-11038  & This work \\
21 July 2013    &  +36.3      &  NOT+ALFOSC      &      3601-9141      & 2 \\
02 Aug. 2013      &  +47.8       & FTS+FLOYDS     & 3179-11038   & This work  \\
04 Aug. 2013     &  +49.6      &  P200+DBSP       &    3200-10100        & 2 \\
05 Aug. 2013    &  +50.7       & FTS+FLOYDS   & 3179-11038   &  This work \\
08 Aug. 2013     &  +53.7       & FTS+FLOYDS   & 3179-11038   & This work \\
02 Sep. 2013  &  +78.7      &  Magellan Baade+FIRE      &      9000-25000       & This work \\ 
04 Sep. 2013   &  +80.5      &  P200+DBSP    &     3200-10100       & 2\\
09 Sep. 2013     &  +85.6      &  Keck~1+LRIS      &     3200-10000       & 2\\
21 Feb. 2014  &  +250.6      &  NOT+ALFOSC      &      3301-9142      &  This work \\
26 May 2014  &  +344.8      &  Keck~2+DEIMOS      &     4915-10134       & This work \\
28 May 2014 &  +346.4      &  VLT+FORS2      &    3300-11000       & This work \\
26 June 2014  &  +376.3      &  WHT+ISIS    &     5400-9000      & This work  \\
\enddata 
\tablenotetext{a}{1. \cite{2013ApJ...775L...7C}, 2. \cite{2014A&amp;A...565A.114F} }
 \end{deluxetable}

\end{document}